\documentclass[12pt,preprint]{aastex}

\newcommand{\et}{et al.}
\newcommand{\kms}{km s$^{-1}$}
\newcommand{\ha}{H$\alpha$}
\newcommand{\solar}{\ifmmode_{\sun}\;\else$_{\sun}\;$\fi}

\newcommand{\HII}{H$\,${\sc ii}}
\newcommand{\HI}{H$\,${\sc i}}
\newcommand{\wtw}{W$_{20}$}
\newcommand{\lha}{L$_{H\alpha}$}
\newcommand{\x}{\enspace}
\newcommand{\xx}{\enspace\enspace}

\newcommand{\sfr}{$\dot{M}$}
\newcommand{\sfrtf}{$\dot{M}_{25}$}
\newcommand{\sfrd}{$\dot{M}_{D}$}
\newcommand{\logsfr}{$\log \dot{M}$}
\newcommand{\logsfrtf}{$\log \dot{M}_{25}$}
\newcommand{\logsfrd}{$\log \dot{M}_{D}$}
\newcommand{\rtf}{R$_{25}$}
\newcommand{\rd}{R$_D$}
\newcommand{\rh}{R$_H$}
\newcommand{\rha}{R$_{H\alpha}$}
\newcommand{\rhii}{R$_{HII}$}

\newcommand{\rhah}{R$_{H\alpha}$/R$_H$}
\newcommand{\rhad}{R$_{H\alpha}$/R$_D$}
\newcommand{\rhiitf}{R$_{HII}$/R$_{25}$}
\newcommand{\rhiih}{R$_{HII}$/R$_H$}
\newcommand{\rhiid}{R$_{HII}$/R$_D$}

\newcommand{\sfrunit}{M\solar\ yr$^{-1}$ kpc$^{-2}$}
\newcommand{\sigha}{$\Sigma_{H\alpha}$}
\newcommand{\mhilb}{M$_{HI}$/L$_B$}

\begin{document}

\title{Star Formation Properties of a Large Sample of Irregular Galaxies
}

\author{Deidre A. Hunter}
\affil{Lowell Observatory, 1400 West Mars Hill Road, Flagstaff, Arizona
86001 USA}
\email{dah@lowell.edu}

\and

\author{Bruce G. Elmegreen}
\affil{IBM T. J. Watson Research Center, PO Box 218, Yorktown Heights,
New York 10598 USA}
\email{bge@watson.ibm.com}

\begin{abstract}

We present the results of \ha\ imaging of a large sample of irregular galaxies.
Our sample includes 94 galaxies with morphological classifications of
Im, 26 Blue Compact Dwarfs (BCDs), and 20 Sm systems.
The sample spans a large range in galactic parameters including
integrated absolute magnitude (M$_V$ of $-$9 to $-$19),
average surface brightness (20 to 27 mag/arcsec$^2$),
current star formation activity (0 to 1.3 M\solar\ yr$^{-1}$ kpc$^{-2}$),
and relative gas content (0.02 to 5 M\solar/L$_{B}$).

The \ha\ images were used to measure the integrated star formation rates,
determine the extents of star formation in the disks, and compare
azimuthally-averaged radial profiles of current star formation
to older starlight.
The integrated star formation rates of Im galaxies
normalized to the physical size of
the galaxy span a range of a factor of 10$^4$ with 10\% Im galaxies
and one Sm system having no measureable star formation at the present time.
The BCDs fall, on average,
at the high star formation rate end of the range. We find no
correlation between star formation activity and proximity to other
catalogued galaxies.
Two galaxies located in voids are similar in properties to the Sm group
in our sample.

The \HII\ regions in these galaxies are most often found within the Holmberg
radius \rh, although in a few systems \HII\ regions are traced as far as 1.7\rh.
Similarly, most of the star formation is found within 3 disk scale-lengths
\rd, but in some galaxies \HII\ regions are traced as far as 6\rd.

A comparison of \ha\ surface photometry with V-band surface
photometry shows that the two approximately follow each other
with radius in Sm galaxies, but in most BCDs there is an excess of
\ha\ emission in the centers that drops with radius. In
approximately half of the Im galaxies \ha\ and V correspond
well, and in the rest there are small to large differences in the
relative rate of fall-off with radius.

The cases with strong gradients in the L$_{H\alpha}$/L$_V$ ratios
and with high central star formation rate densities, which include
most of the BCDs, require a significant fraction of their gas to
migrate to the center in the last Gyr. We discuss possible torques
that could have caused this without leaving an obvious signature,
including dark matter bars and past interactions or mergers with
small galaxies or HI clouds.  There is now a substantial amount of
evidence for these processes among many surveys of BCDs.  We note
that such gas migration will also increase the local pressure and
possibly enhance the formation of massive dense clusters, but
conclude that the star formation process itself does not appear to
differ much among BCD, Im and Sm types.  In particular, there is
evidence in the distribution function for H$\alpha$ surface
brightness that the turbulent Mach numbers are all about the same
in these systems. This follows from the H$\alpha$ distribution
functions corrected for exponential disk gradients, which are
log-normal with a nearly constant dispersion. Thus the influence
of shock triggered star formation is apparently no greater in BCDs
than in Im and Sm types.

\end{abstract}


\section{Introduction}

Irregular galaxies are interesting systems for many reasons.
Not only are they smaller, less evolved, and have lower amounts
of dust and shear in their interstellar mediums (ISM) than most spirals,
but they are also the most common type of galaxy in
the universe. In some models dwarf irregulars (dIm) are the galaxies that
formed first and became the building blocks of larger galaxies.
Because these tiny galaxies have evolved relatively slowly over time,
they chemically resemble the outer parts of present-day spirals.
In fact, ultra-low surface brightness dIms may represent the slowest
evolving galaxies in the universe (van den Hoek \et\ 2000).

For purposes of star formation studies, the lack of spiral
density waves in Im galaxies means that all of the non-spiral star formation
processes can be examined more clearly.
In fact Hunter, Gallagher, \& Rautenkranz (1982)
and Hunter \& Gallagher (1986) have shown that some Im galaxies
have star formation rates normalized to their size that are comparable even
to those of spiral galaxies.
Thus, spiral density waves are not necessary to a vigorous production
of stars (see also Elmegreen \& Elmegreen 1986).
Furthermore, Hunter and Gallagher
showed that irregular galaxies span a large range
in integrated star formation rates. What then
regulates star formation in tiny galaxies and
determines how fast an Im galaxy evolves?

To address this question, we have conducted a multi-wavelength
survey of a large sample of reasonably normal, relatively nearby,
non-interacting galaxies without spiral arms. The data consist of
UBV and \ha\ images for the entire sample, and JHK images, \HI\
maps, CO observations, and \HII\ region spectrophotometry for a
sub-sample. The \ha, UBV, and JHK image sets act as probes of star
formation on three different times scales: \ha\ images trace the
most recent star formation ($\leq$10 Myrs) through the ionization
of natal clouds by the short-lived massive stars; UBV, while a
more complicated clue, is dominated by the stars formed over the
past Gyr for on-going star formation (Gallagher \et\ 1984); and
JHK integrates over the lifetime of the galaxy where even in Im
galaxies global JHK colors are characteristic of old stellar
populations (Hunter \& Gallagher 1985b). We have already used
these data to conduct a case study of the Im galaxy NGC 2366
(Hunter, Elmegreen, \& van Woerden 2001) as well as studies of the
\HII\ region luminosity function and distributions (Youngblood \&
Hunter 1999, Roye \& Hunter 2000), gas abundances (Hunter \&
Hoffman 1999), and pressures of \HII\ regions relative to the
background galactic disk (Elmegreen \& Hunter 2000).

Here we present the \ha\ data---integrated star
formation rates, azimuthally-averaged \ha\ surface brightnesses,
and extents of star formation---of the full sample for the first time,
and we discuss the current star formation activity in these
galaxies.
Because this is the first presentation of the full survey sample,
in the next section we will discuss the sample and some of its
properties although results from broad-band imaging will
be reported in detail in a seperate paper. In Section \ref{sec-obs}
we present the \ha\ observations, and in Section \ref{sec-results}
we discuss star formation rates, the radial extents of star formation,
and \ha\ surface photometry.
A discussion of selected aspects is given in Section \ref{sec-disc}.

\section{The Sample} \label{sec-sample}

\begin{figure}
\plotone{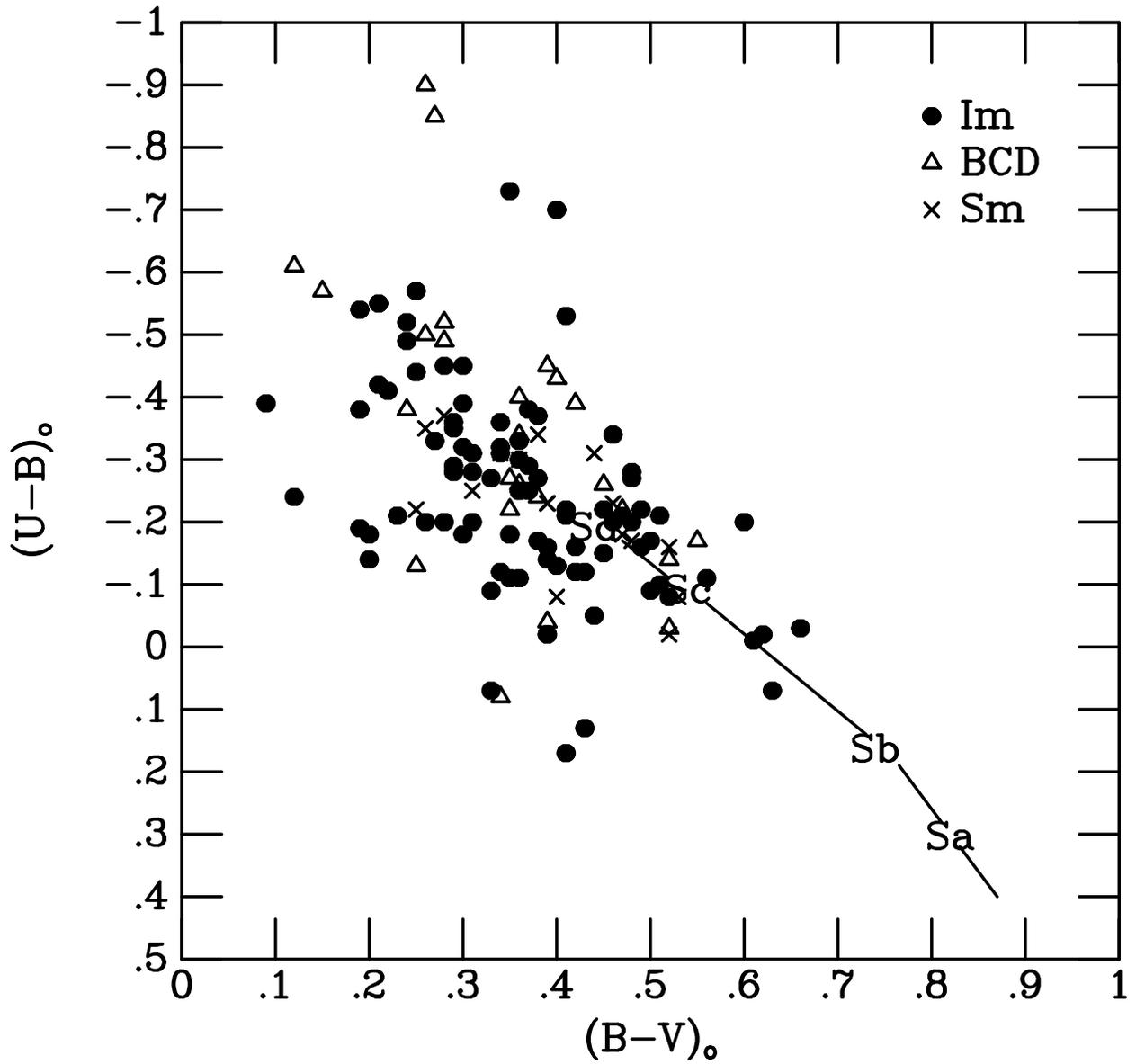}
\caption{
Integrated UBV colors of our survey galaxies.
Average colors are shown for spiral galaxies (de Vaucouleurs
\& de Vaucouleurs 1972). The UBV colors are corrected for reddening
using the foreground reddening of Burstein \& Heiles (1984), an
assumed internal reddening of E(B$-$V)$_s=$0.05, and the reddening
law of Cardelli \protect\et\ (1989).
\label{fig-ubv}}
\end{figure}

\begin{figure}
\epsscale{0.9}
\plotone{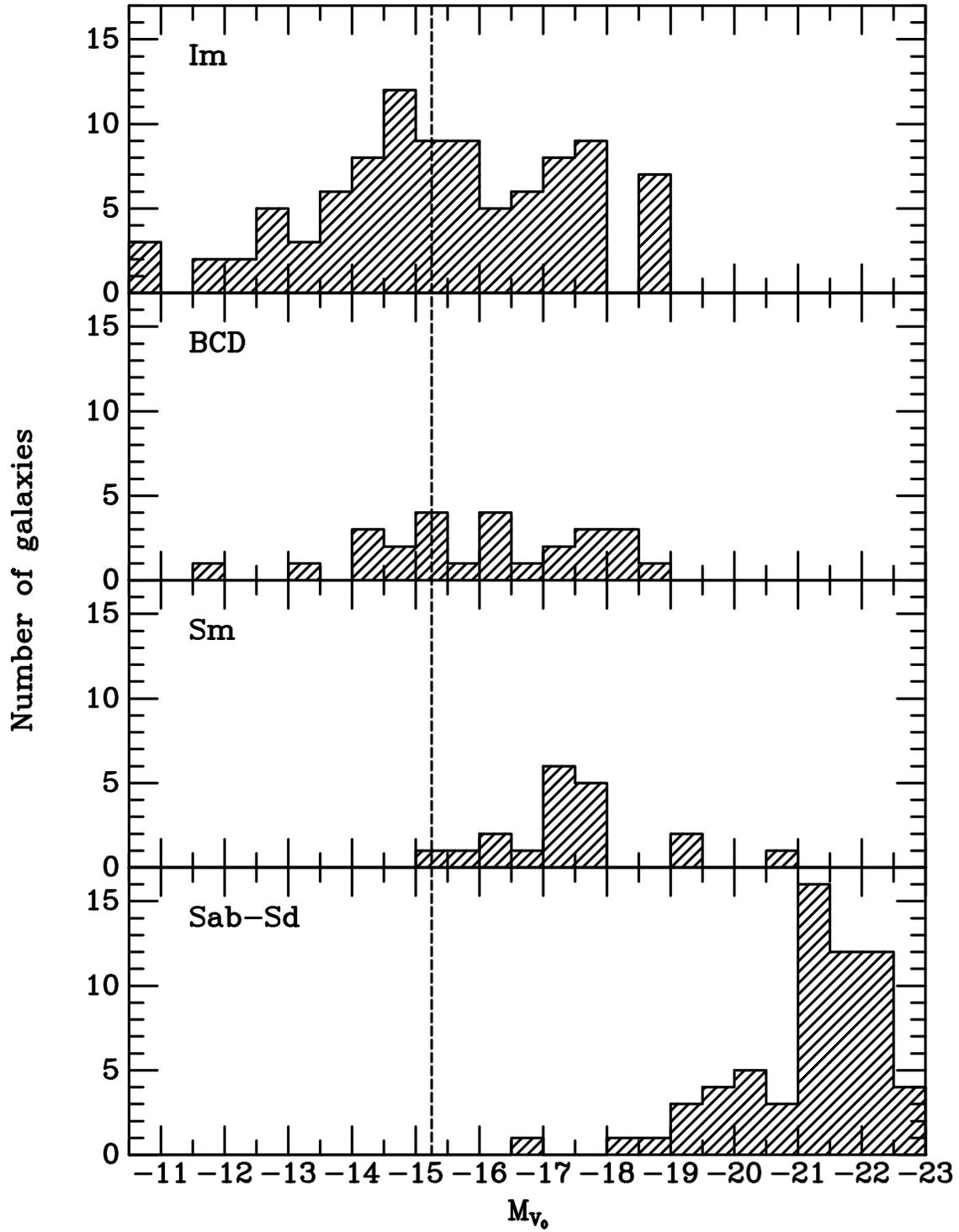}
\caption{
Number distribution of the survey galaxies in integrated M$_V$,
corrected for reddening. The vertical dashed line marks the median
value in M$_{V_0}$ for the Im galaxies.
\label{fig-mv}}
\end{figure}

\begin{figure}
\epsscale{0.9}
\plotone{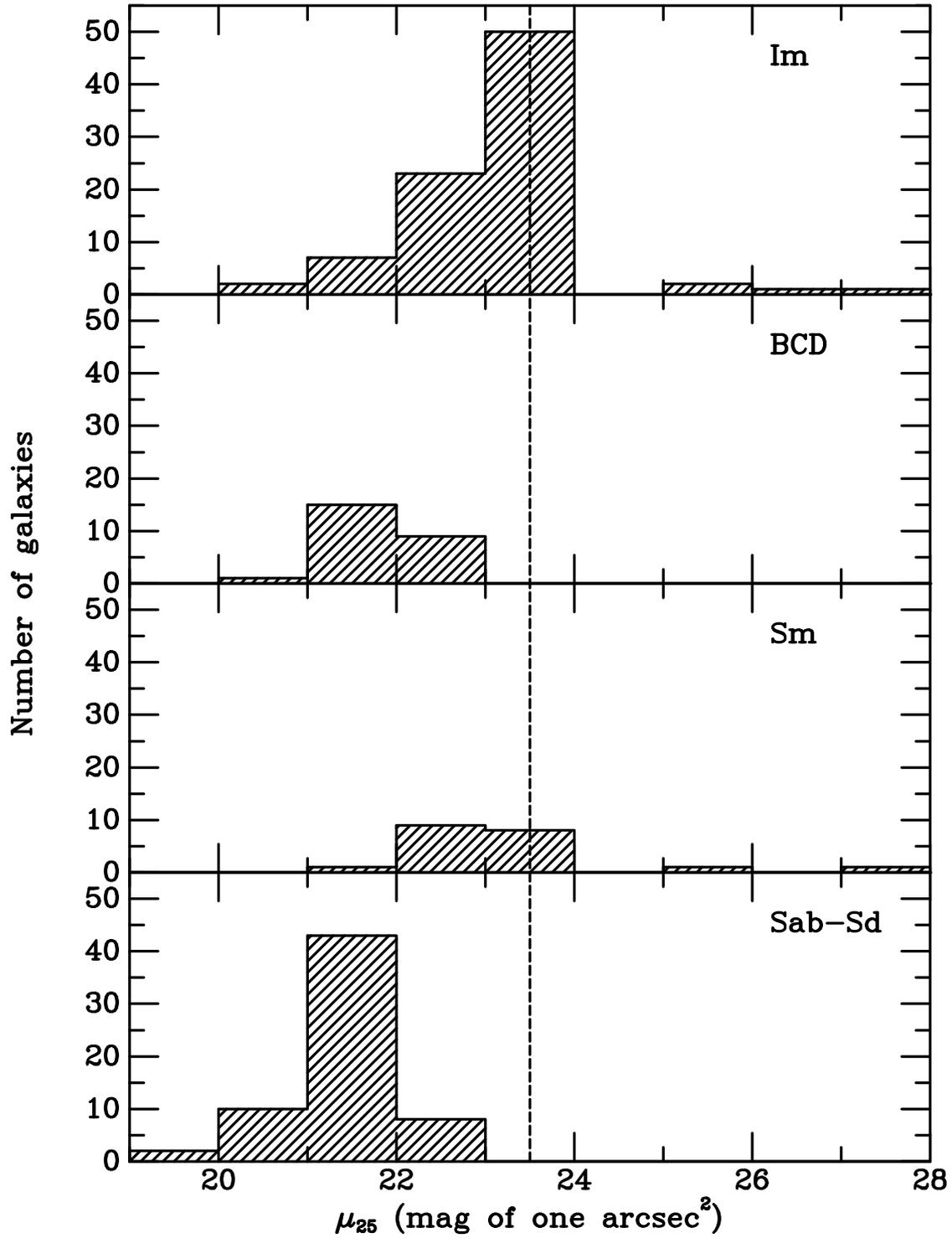}
\caption{
Number distribution of the survey galaxies in $\mu_{25}$,
the average surface brightness within a B-band isophote of
25 mag of one arcsec$^2$, corrected for reddening.
The dashed vertical line marks the median value of $\mu_{25}$
for the Im group.
\label{fig-mu25}}
\end{figure}

\begin{figure}
\epsscale{0.9}
\plotone{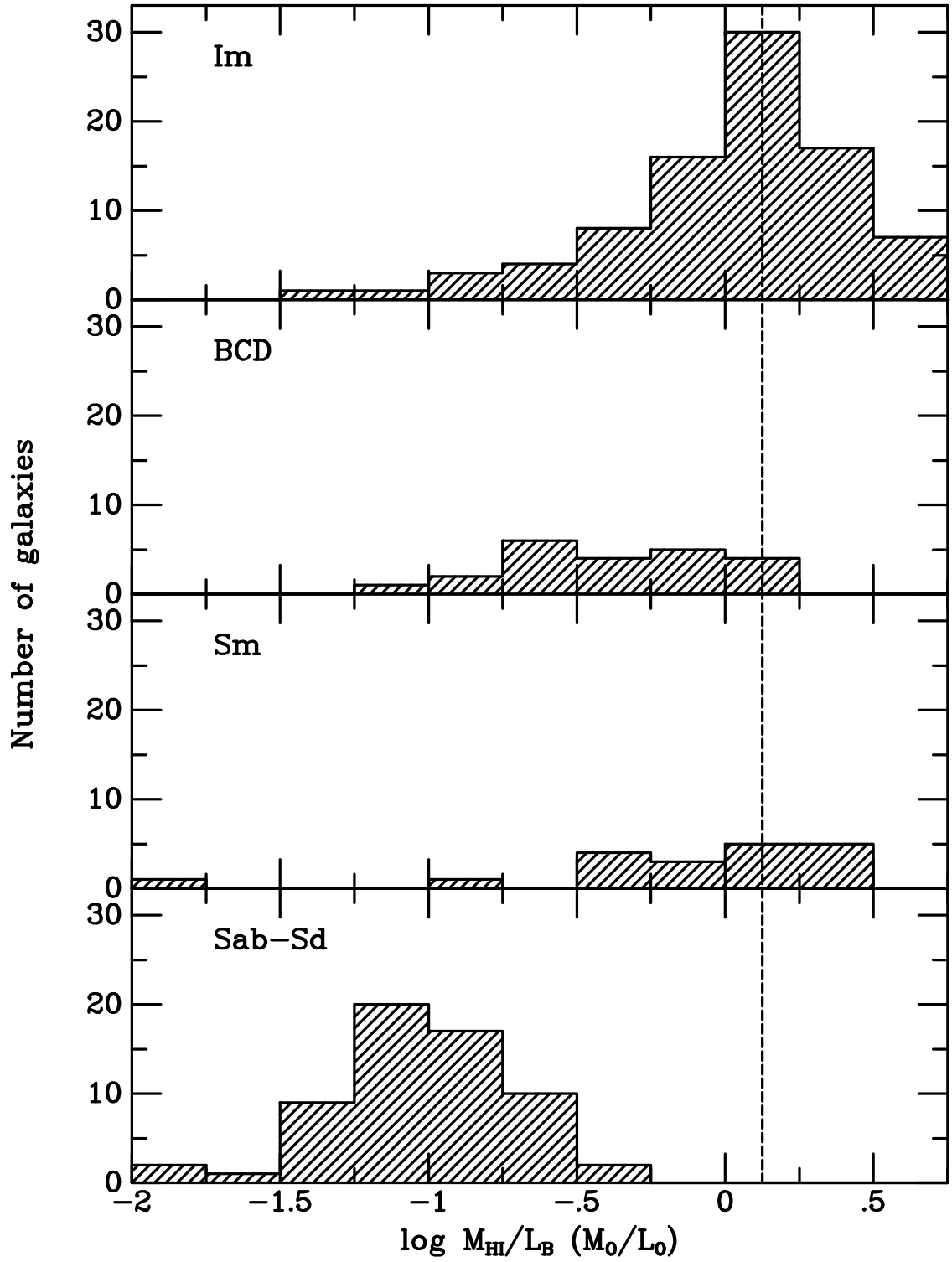}
\caption{
Number distribution of the survey galaxies in integrated \protect\mhilb.
The vertical dashed line marks the median value of $\log$ M$_{HI}$/L$_B$
for the Im galaxies to aid in
comparison to the other galaxy types.
\label{fig-mhilb}}
\end{figure}

The sample galaxies are listed in Table \ref{tab-sample} where
we have grouped the galaxies into three categories: Im, Blue Compact
Dwarf (BCD), and Sm.
Our sample includes 140 galaxies:
94 Im systems,
26 BCDs, and 20 Sms.
Our broad-band images will be presented in a separate paper, but
in order to characterize the sample we include here several
plots of integrated properties taken from that data set:
a UBV color-color diagram in Figure \ref{fig-ubv}, a histogram
of reddening-corrected M$_{V_0}$ values in Figure \ref{fig-mv},
a histogram of $\mu_{25}$ (average surface brightness
within a B-band isophote of 25 magnitudes of one arcsec$^2$)
in Figure \ref{fig-mu25},
and a histogram of \mhilb\ in Figure \ref{fig-mhilb}.

The morphological classifications were taken from de Vaucouleurs
\et\ (1991$=$RC3).
Most are simply classed as Im or Sm with minor modifications.
However, the morphological
classifications
of the BCDs are complex and varied,
reflecting the difficulty in determining
just where these systems fit in. One galaxy, NGC 3413, is classed
as an ``S0sp'' but we include it in the Im galaxy sample where our
imaging suggests that it belongs.
In addition to morphological classification,
most of the galaxies in this study were chosen to be relatively nearby;
most are closer than 30 Mpc.

The original sample of galaxies was chosen to be those classed
as Im in
the HI catalog of Fisher \& Tully (1975), and so the sample
is biased to systems
containing gas.
A few Sm systems were added from this catalog for comparison.
To this collection we later added galaxies that promised to probe the
extremes in various properties of the dwarf galaxies
since the extremes may hold important clues to a complete
picture of the evolution of these galaxies.
One such property is surface brightness. We added low surface brightness
systems cataloged from the surveys of Schombert \& Bothun (1988) and
Schombert \et\ (1992), particularly those with HI maps obtained by de Blok
\et\ (1996). These galaxies are more distant than our core sample,
the most distant being 86 Mpc. There were 15 Im galaxies added in this way
that form 16\%
of our total Im sample, and 3 Sm systems that form 15\% of the Sm sample.
In Figure \ref{fig-mu25}
all but 2 of these Im systems turn out
to have typical $\mu_{25}$ surface brightnesses and fall at the peak
in the number distribution of the entire sample, but they are
not responsible for defining the peak. The other two systems
form the two extreme bins at the faint end. Of the Sm systems,
one has a typical $\mu_{25}$ and the other two form the extreme
bins at the faint end of the distribution.
Thus, the augmentation of the original sample with these systems only produced
two Im (2\% of the total sample) and two Sm (10\%) galaxies that
really do form an extreme.

Another potential property of interest is the relative gas content.
We added 5 systems (5\% of our total Im sample) designated as having high
M$_{HI}$/L$_B$ ratios from the study by van Zee, Haynes, \& Giovanelli (1995).
Of these one turned out to have an average \mhilb, as seen in
Figure \ref{fig-mhilb}, 3 have modestly high \mhilb\
($\log$ \mhilb$>0.25$)
but are not extreme, and one is in the highest bin, although
there are 6 other galaxies in that bin too.
Thus, this augmentation of the original sample produced one extreme
Im galaxy, 1\% of the entire Im sample.

In addition to Im and a small sample of Sm galaxies,
we include a sample of BCD systems, kept as a seperate group,
in order to see how their
star-forming properties compare to those of Im galaxies.
BCDs are generally characterized
by intense emission-lines from \HII\ regions.
Some, although not all,
BCDs have luminosities, sizes, and other properties that make
them likely to be related
to Im systems (Kunth 1987).
Furthermore, detailed examination of one BCD with the {\it Hubble Space
Telescope} ({\it HST}), VIIZw403, has shown that it resembles an Im galaxy
but with an unusual star-formation past (Lynds \et\ 1998).
Yet other integrated properties of BCDs are statistically
different from those of Im galaxies (Chamaraux 1977, Papaderos
\et\ 1996).

Most BCDs were chosen from
the list of Thuan \& Martin (1981). The BCD or \HII\ galaxy class
appears to be a mixed bag of objects. Indeed some BCD galaxies
are actually star-forming pieces of other galaxies
(for example, NGC 2363 in NGC 2366).
In order to select BCDs that are most likely comparable systems to the
Im galaxies we have imposed an additional selection criteria not
applied to the rest of the sample.
We used \wtw, the width
at 20\% intensity of the integrated HI profile, since it is related
to the dynamical mass, to select systems that are
comparable in mass to Im systems. We chose systems with \wtw$\leq$175 \kms\
since this value corresponds to the upper envelope for the Im systems in
our sample.

Our sample, thus, is not complete in any sense, but it
spans a large range in galactic parameters
including integrated
luminosity (M$_V$ of $-$9 to $-$19),
average surface brightness (20 to 27 mag/arcsec$^2$),
current star formation activity (0 to 1.3 M\solar\ yr$^{-1}$ kpc$^{-2}$),
and relative gas content (0.02 to 5 M\solar/L$_{B}$).
A large sample is necessary in order to probe
the full range in galactic parameters
and to allow a meaningful probe of galaxies
that are proving to be very complex and diverse.

The original sample of Im and Sm galaxies
was taken from the \HI\ survey of Fisher \& Tully
(1975) that was based on the catalog of dwarfs compiled by
van den Bergh (1959, 1966) from a search of the National Geographic
Society and Palomar Sky Survey plates. Fisher \& Tully (1981) then
went on to produce a more complete survey of galaxies, dwarfs
and non-dwarfs of all Hubble types. They argue that their 1981
study is reasonably complete to radial velocities less than 1000 \kms,
sizes greater than 1.5\arcmin, declinations higher than $-35$\arcdeg,
and Galactic latitutes more than 30\arcdeg\ from the plane.
Their 1981 sample included 99 Im systems and 36 Sm systems that
satisfy these criteria. 37 galaxies from our Im sample (39\% of our sample)
represent 37\% of the Fisher and Tully sample, and 7 galaxies
from our Sm sample (35\% of our sample) represent 19\% of their sample.

There are 14 Im galaxies (15\% of our sample) and two Sm galaxies
(10\% of our sample) that are not part of the Fisher and Tully 1981
\HI\ survey or the augmentation samples discussed above.
In the figures we see that most of the added Im galaxies have typical
integrated properties.
The exceptions are as follows: 3 fall near the blue
end and 2 near the red end of UBV colors;
8 are at the high end of M$_V$ (M$_V<-17$), two are at the low end
(M$_V>-12.6$), and one is the lowest (M$_V=-9.4$);
7 are high in surface brightness ($\mu_{25}<22$ mag arcsec$^{-2}$) and none are low
($\mu_{25}>24$ mag arcsec$^{-2}$);
3 are low in \mhilb\ ($\log$ \mhilb$<-0.5$) and none are high
($\log$ \mhilb$>0.5$);
and
6 have high star formation rates ($\log$ \sfr$\geq-1$),
two have low rates ($\log$ \sfr$\leq-3$),
and two have no current measureable star formation.
The two Sm galaxies are typical in all properties.
Thus, while our sample of galaxies is not complete, it is
representative.

The distances to the galaxies in our sample
are given in Table \ref{tab-sample}.
We used distances determined from stellar population studies,
often from variable stars or the tip of the Red Giant Branch,
if they were available.
Other distances
were determined from the radial velocity relative to the Galactic
standard of rest V$_{GSR}$ (RC3) and a ``Hubble
constant'' of 65 \kms Mpc$^{-1}$.
References are given in Table \ref{tab-sample}.

Any system known to be interacting
was excluded because we are interested only in internal processes.
However, inclusion in this sample is not a guarantee that a galaxy has
not been affected by other galaxies at some time in the past. Surprises
sometimes await us when we learn more about individual galaxies
(for example, NGC 4449: Hunter \et\ 1998b).
NED was used to search
for galaxies within 1 Mpc
and 150 \kms\ of our sample galaxies.
The distance
to the nearest galaxy and their radial velocity difference
are given in Table \ref{tab-sample}.

For comparison to spiral galaxies we have used the sample compiled by
Kennicutt (1983). This sample spans the range of
morphologies from Sab to Sd with 74 systems. Kennicutt has measured
\ha\ fluxes for these galaxies. Other properties were obtained
from RC3 and Fisher \& Tully (1981).

Foreground reddening was determined from Burstein \& Heiles (1978)
and values are given in Table \ref{tab-sample}.
An internal reddening
correction of E(B$-$V)$_s$=0.05 was added for the stars and
of E(B$-$V)$_g$=0.1 for the gas.
The reddening law of Cardelli, Clayton, \& Mathis (1989) was
adopted.
For the spirals, internal reddening was taken to be 0.3, after Kennicutt (1983).

Reddenings from Schlegel \et\ (1998) were considered early on, but
we were uneasy that the Schlegel \et\ values are systematically higher
than those of Burstein and Heiles. For 49 Im and BCD galaxies, the
Schlegel \et\ E(B$-$V) values are on average a factor of two times
those of Burstein and Heiles. For most galaxies the difference is
nevertheless insignificant, especially compared to the error we must
make by adopting the same internal reddening for all galaxies and all \HII\
regions within each galaxy.
However, the higher values of Schlegel \et\ do not make sense in some cases.
For example, for the heavily reddened galaxy NGC 1569 optical spectroscopy
of its \HII\ regions yields an average total E(B$-$V) of 0.70$\pm$0.05
(see compilation in Hunter \& Hoffman 1999). The Burstein and Heiles value
for foreground reddening is 0.51 and the Schlegel \et\ value, obtained
from NED, is 0.70. Thus, the Schlegel \et\ value would mean that the
\HII\ regions in NGC 1569 have no internal reddening while the
Burstein and Heiles valule allows a reasonable 0.2 mag of internal
reddening.
Willick (1999), as quoted by NED, argues that both methods of determining
foreground reddening are likely to have systematic errors. It is not
clear to us at this point that one method is significantly superior to the
other, and we have prefered to use the values of Burstein and Heiles for
the reasons given.

\section{The \ha\ images} \label{sec-obs}

The \ha\ images
were obtained in 22 observing runs from 1988 to 1998.
The telescopes, instruments, and exposure times are given in
Table \ref{tab-obs}.
Some of these images were discussed by Hunter, Hawley, \& Gallagher
(1993) in their search for extra-HII region ionized gas structures.
Images are available from Hunter upon request.

The galaxies were imaged through narrow-band
\ha\ filters, usually with a FWHM$\sim$30 \AA.
We had a series of 5 filters forming a red-shift set with
some wavelength overlap between filters.
The continuum-only off-band filter was centered at 6440 \AA\ with a
FWHM of 95 \AA.
Usually multiple images through the \ha\ filter and sometimes
also through the off-band filter were obtained
and combined to remove cosmic rays.
The off-band image was shifted, scaled, and subtracted
from the \ha\ image to remove the stellar continuum.
The \ha\ emission was calibrated using the known \ha\ flux
from the \HII\ regions NGC 2363 (Kennicutt, Balick, \& Heckman 1980)
and NGC 604 (Appendix \ref{app-hii})
and from
spectrophotometric standard stars.
Calibrations determined from nebulae and standard stars
observed on the same nights agreed to within 4\%.
The \ha\ flux has been corrected for the shift of the bandpass with
temperature and the contribution from [NII]. The latter contribution
was usually of order 1\%.
Sky was subtracted from the image most often using a two-dimensional
fit to the background.

Detection limits in \ha\ photometry and comparison to measurements
in common with others
are discussed by Youngblood \& Hunter (1999).
They determined detection limits by artificially adding \HII\ regions
and determining the limit for recovering them.
In Table \ref{tab-obs} we also include the rms deviations in the background
of the \ha\ images
given in terms of the corresponding \ha\ surface brightness of a single pixel.
Note, however, that these limits are for a single pixel;
azimuthally-averaged surface photometry from ellipse photometry
can be lower than these limits if a detectable \HII\ region
is located in an annulus with a large area.

We mention here a few studies with integrated \ha\
measurements published after Youngblood \& Hunter (1999). There
are 12 galaxies in common with the sample of van Zee (2000).
Differences in the integrated \ha\ fluxes for these galaxies
are on average 2.8$\sigma$, with three galaxies having a 0.2--0.3$\sigma$
difference and two having a 7--9$\sigma$ difference.
From an alternative perspective, the differences are 21\% of
the flux on average except for two galaxies that differ by
a factor of 2--3.
Kniazen \et\ (2000) have estimated the integrated \ha\ flux
of HS0822$+$3542 from slit spectroscopy. Our flux agrees with theirs
to 6\%.
We have 34 galaxies in common with the \ha\ survey of James \et\ (2004).
The average difference in the \ha\ flux is 2.5$\sigma$ with one galaxy
having a difference of 6$\sigma$.

For doing azimuthally-averaged \ha\ surface photometry,
we have geometrically matched the
\ha\ image to the V-band image. The V-band images, to be presented
in a separate paper, were used to determine the position angle,
ellipticity, and center of the galaxy. The same ellipses used
for photometry of the V-band images were used for the \ha\ images.
Thus, the \ha\ surface photometry can be directly compared to that
from the V image. The position angle PA, minor-to-major axis
ratio $b/a$, step size, and galaxy center that were used for
the ellipse photometry are given in Table \ref{tab-obs}.
In a few cases V-band images were not available and we used instead
\ha\ off-band images or Digitized Sky Survey (DSS) extracts.

\section{Results} \label{sec-results}

\subsection{Integrated star formation rates} \label{sec-sfr}

We have used \ha\ as a tracer of the most recent star formation in
Im galaxies and \lha\ as a measure of the star formation rate.
Massive stars are short-lived and ionize the gas left over from
the star formation process. The \ha\ luminosity is a measure of
the flux of ionizing photons bathing the left over gas and hence a
measure of the number of massive stars that are present in the
region. With the assumption of a stellar initial mass function we
infer the presence of lower mass stars and then deduce the star
formation rate. The stellar initial mass function has been found
to be approximately constant among systems where it could be
directly measured, so this is a reasonable assumption. Extinction
due to dust is low, so we feel confident that \ha\ is revealing
the bulk of the star formation in these galaxies (Hopkins \et\
2001). One potential issue occurs for galaxies that are undergoing
a burst of star formation. Weilbacher \& Fritze-v. Alvensleben
(2001) have shown that the \ha\ luminosity does not translate into
\sfr\ as simply in galaxies where the star formation rate is
changing on short timescales as in galaxies where it is relatively
constant. Here we have treated all galaxies in the same manner,
but note that the \sfr\ deduced for galaxies with an unusually
high star formation rate compared to that in the past may not be
as well determined. Brinchmann \et\ (2004) find that the
relationship between star formation rate and \lha\ is also a
function of the stellar mass of the galaxy (driven primarily by
the metallicity of the galaxy), the ratio of \lha\ to star
formation rate increasing to less massive galaxies. However, their
models show that the simple method of converting \lha\ to \sfr\
using a single constant that we adopt here agrees with other
methods for the low-mass galaxies of our survey (their Figure 8).

The reddening-corrected \lha\ and star formation rates
\sfr\ are listed in Table \ref{tab-results}, and the number
distribution of \sfr\ is shown in Figure \ref{fig-sfr}.
We converted \lha\ to \sfr\ using
the formula derived in Appendix \ref{app-sfrformula}. This
assumes that stars 0.1 M\solar\ to 100 M\solar\ form in
the proportions first observed by Salpeter (1955). We also assume
an efficiency factor $\eta$ of 2/3 for the absorption
of ionizing photons by the nebulae (Gallagher, Hunter,
\& Tutukov 1984).

To compare star formation rates in galaxies we must
normalize the star formation rate to the ``size'' of the galaxy.
An obvious quantity for normalization would be the mass of the galaxy.
However, determining the mass is frought with uncertainties: There are varying
amounts of dark matter, large uncertainties in determining the
total stellar mass which increases with time, and a gas mass that is depleted with time.
Therefore, operationally mass does not make a good normalizer.
We have instead chosen to use the physical area of the galaxy.
In the past we used \rtf\ from which to calculate the area of the galaxy
$\pi$\rtf$^2$.
We did this because \rtf\ was readily available from RC3 for
a large number of galaxies.
However, \rtf\ is a surface-brightness defined size, and therefore,
it will depend to some extent on the location and level of the
star formation itself. Furthermore, there are some galaxies that
fall everywhere below a B-band surface brightness
of 25 mag of one arcsec$^2$ and yet are forming stars.

To address these problems, we have chosen here to use the scale-length
\rd\ measured from V-band images
to define the area of the galaxy. Of course, the scale-length
assumes a reasonably well-behaved exponential disk, and most
Im disks are (Hodge 1971, Hunter \& Gallagher 1985b, van Zee 2000).
Determining the scale-length
has its own uncertainties, but we believe that \rd\
is a more meaningful measure of the size of a galaxy than \rtf.
There are some galaxies that are not characterized by a single
fit to the surface brightness profile. These are 8 of the BCDs:
Haro 20, Haro 23, Mrk 32, Mrk 408, Mrk 600, Mrk 757, UGCA 290, and
VIIZw403.
All but one of these shows a steeper profile in the center than in the
outer parts.
Since we would expect intense central star formation to bias the
surface brightness profile,
we have used the fit to the outer
galaxy in these cases (see also Cair\'os \et\ 2001, Noeske \et\ 2003).

To ease the transition and for comparison to spirals for which
we only have \rtf, we include
the star formation rate normalized to \rtf\ \sfrtf, as well as
the star formation rate normalized to \rd\ \sfrd. These are
given in Table \ref{tab-results}. \sfrtf\ and \sfrd\ are
compared in Figure \ref{fig-comparesfr}, and the number distributions
of these two normalized star formation rates are
shown in Figures \ref{fig-sfr25} and
\ref{fig-sfrd}.
In order to represent galaxies with zero star formation rates on
these log plots, we have plotted them at \logsfrtf\ and \logsfrd\ of $-6$
to $-6.5$.
The \rtf\ and \rd\ are given in Table \ref{tab-params}.

Although we use the area $\pi$\rd$^2$ as the area for normalizing
\sfrd, that does not imply that all of the star formation
is found within \rd. In fact below we will see that most of the
star formation is found within 3--7\rd. We use $\pi$\rd$^2$
as the area simply as a way of normalizing by the relative physical
size of the galaxy.

We see in Figure \ref{fig-comparesfr} that the lower star
formation rate systems cluster around the line of
\sfrtf$\sim$\sfrd, while the higher star formation rate systems
approximately follow a line in which \sfrd$\sim10\times$\sfrtf,
which corresponds to \rtf$=\sqrt{10}$\rd. We can understand this
as due to lower surface brightness systems having shorter \rtf\
compared to \rd; they reach 25 mag arcsec$^{-2}$ faster with
radius than galaxies that start out brighter.  Higher surface
brightness galaxies have larger relative \rtf, approximately 3\rd\
(see also Heller et al. 2000). This is a consequence of the higher
star formation rate systems having higher central surface
brightnesses, $I_0$, because
$R_{25}/R_D=\log\left(I_0/I_{25}\right)$ for fixed surface
brightness $I_{25}$ at 25th mag arcsec$^{-2}$.

From Figures \ref{fig-sfr25} and \ref{fig-sfrd} one can see that
star formation rates in Im galaxies forming stars
span a factor of 300 in \sfrtf\
and $10^4$ in \sfrd.
BCDs, as expected, do generally have high
star formation rates although there is also a range
(see also Hopkins, Schulte-Ladbeck, \& Drozdovsky 2002).
Sm galaxies have about the same distribution in rates as Im galaxies.
In \sfrtf\ the irregular galaxies at the high end of the distribution
have normalized rates that are comparable
to those of spirals.
Not all galaxies are forming stars, however:
10\% of our Im (9 of 94) and 5\% of our Sm galaxies (1 of 20) are
not currently forming stars at a level that we can detect.

The typical \sfrtf\ star formation rate for the Im galaxies
is $10^{-3}$ \sfrunit. The typical spiral has an integrated star formation
rate that is about 3 times higher.
We have compared this to a
sample of 11 Sc galaxies whose radial
\ha\ surface brightnesses
are shown by Kennicutt (1989).
We see that the Sc spirals reach this star formation rate per
unit area between 0.35\rtf\ and 0.81\rtf. The average is
0.5\rtf. Thus, the Im galaxies more closely resemble the outer
parts of spirals in star formation rates.

The timescale $\tau$ to exhaust the current gas supply at the current
star formation rate is given in Table \ref{tab-results}; the
gas mass is given in Table \ref{tab-params}.
The number distribution
is shown in Figure \ref{fig-tau}.
We include all of the gas associated with the galaxy in the current
gas supply. Gas at large radii is problematical; we do not know what
role that gas will play in the future nor can we say with confidence
what role it is currently playing in star formation. In one Im galaxy,
the bulk population in the outer parts
is the same as everywhere else in the galaxy suggesting that star
formation must be taking place out there at some level (Hunter, in preparation).
Recycling of gas from dying stars
is not included in determining $\tau$,
but would add another factor of 1.5--4 to the timescale
(Kennicutt, Tamblyn, \& Congdon 1994).

Detailed studies of the stellar
populations in dIm galaxies that have shown that most evolve with a star
formation rate that varies in amplitude by a factor of only a few,
as would be expected in small galaxies (Ferraro \et\ 1989; Tosi
\et\ 1991; Greggio \et\ 1993; Marconi \et\ 1995; Gallart \et\
1996a,b; Aparicio \et\ 1997a,b; Dohm-Palmer \et\ 1998; Gallagher
\et\ 1998; Cole \et\ 1999; Gallart \et\ 1999; Kennicutt \& Skillman 2001;
Miller \et\ 2001; Wyder 2001). This is even true of some
BCD systems
(Schulte-Ladbeck \et\ 2000;
Crone \et\ 2002).
Thus, one can see that most of the Im and Sm galaxies
can sustain this rate for at least another
Hubble time and some much longer (see also van Zee 2001).

However, this is not true of most BCDs and a few Im galaxies
that have unusually high star formation rates.
The galaxies in our sample with unusually high \sfr\
(\sfrd$\geq-1$ in Table \ref{tab-results} and Figure \ref{fig-sfrd})
include IC 4662, NGC 1156, NGC 1569 among the Im systems
and Haro 3, Haro 4, Haro29, HS0822$+$3542, Mrk 67, NGC 1705, and
Zw2335 among the BCDs.
Most of these systems are also the ones with low $\tau$
(Table \ref{tab-results} and Figure \ref{fig-tau}),
indicating that this level of star formation is unsustainable.

This is what brought people early on to the belief that BCDs
as a class are
undergoing an unusually enhanced episode of star formation at the
present time (Thuan 1991).
Studies of the resolved stellar population of NGC 1569
find that the current star formation rate is unusual compared
to what it has been in the past (Greggio \et\ 1998).
In the case of NGC 1569, interaction with a nearby \HI\ gas
cloud is believed to be responsible for the current elevated
level of star formation (Stil \& Israel 1998).
Resolved stellar population studies of a few other Im galaxies
also show evidence of higher
amplitude variations either currently or in the past
(Tolstoy 1996; Israel 1988; Tolstoy \et\ 1998;
Karachentsev, Aparicio, \& Makarova 1999).
But, what causes statistically large variations to occur in
dwarfs that appear to be isolated is not known.

The current star formation rate does not correlate with many
integrated properties of the galaxies (see also Hunter \et\ 1982,
Hunter \& Gallagher 1985b, van Zee 2001). An expected correlation
is between star formation activity and average surface brightness
of the galaxy. This is shown in Figure \ref{fig-sfrsb}. Galaxies
with higher star formation rates are higher in surface brightness.
However, there is no correlation between surface brightness and
(U$-$B) or (B$-$V), which are longer term measures of the star
formation state of a galaxy (see also McGaugh \& Bothun 1994). In
terms of other quantities, there is a slight trend with lower
luminosity galaxies having lower star formation rates, shown in
Figure \ref{fig-sfrmv}, particularly when spiral galaxies are
included to extend the range in M$_V$. However, at any given
galactic magnitude, the scatter in \sfr\ is very large. There is a
stronger correlation between star formation rate and \mhilb\ and
between star formation and the fraction of the galactic mass (gas
plus stars) that is gas. These are shown in Figures
\ref{fig-sfrmhilb} and \ref{fig-sfrgasfrac}. The trends are in the
sense that galaxies with higher ratios of gas relative to their
luminosity or total baryonic mass have lower current star
formation rates. This makes sense if star formation rates are approximately
constant: galaxies forming stars at lower
rates have locked less of their gas up in stars. Other studies
have also shown that star formation rates are correlated with the
gas surface density in late-type galaxies (Buat, Deharveng, \&
Donas 1989) and total gas mass (P\'erez-Gonz\'alez \et\ 2003).

We were concerned about the effects that neighboring galaxies
might have on the star formation processes in our survey galaxies
(see, for example, Hashimoto \et\ 1998, Noeske \et\ 2001). In
Figure \ref{fig-sfrdist} we plot the current star formation rate
against the distance to the nearest catalogued galaxy, given in Table
\ref{tab-sample}. Galaxies for which no obvious neighbor was found
within our search parameters are plotted at a distance of 10000
kpc. One can see that galaxies with no neighbors span the
range from very high \sfrd\ to low \sfrd, most of the observed
range in our sample. Also, galaxies with a star formation rate of
zero span the same range in distance to nearest neighbors. So, it
would seem that a lack of neighbors has not been the cause
of no star formation. However, three galaxies at the high end of
the star formation rate range do fall outside the envelope defined
by the other galaxies towards smaller distances as might be
expected if gravitational interactions sometime in the past
contributed to the current star formation activity.

Among the BCDs, we do not see evidence of systematically closer
companions than for Im systems. This is consistent with the
studies by Campos-Aguilar \& Moles (1991) and Telles \& Terlevich
(1995) who also searched for bright companions. Recent studies
have found significant evidence for an excess of faint companions
to BCD galaxies, however. We return to this issue in Section
\ref{sect:bcd}.
At the other extreme, Bothun \et\ (1993) suggested that low
surface brightness galaxies have a deficit of companions. A plot
of the V-band surface brightness in one scale length $\mu_D$
against distance to the nearest galaxy shows only a somewhat
broader range in distances for the lower surface brightness
systems in this sample.

How do the two galaxies in voids
(0467$-$074, 1397$-$049) compare with the other irregular
galaxies? Their M$_V$ place them at the bright end of the range
for Im or BCD galaxies, more like Sm systems. Similarly, their
\rd\ are larger than is typical for Im or BCD galaxies. However,
except for 1397$-$049's U$-$B, the colors, surface brightnesses,
and star formation rates are typical of Im galaxies. These two
galaxies do not have the higher surface brightness or star
formation rates more typical of BCDs and, therefore, are not
similar to the void dwarfs studied by Popescu, Hopp, \& Rosa
(1999). However, like Popescu \et\ we find that the galaxies in
the voids are similar to systems in the field.

There have been several surveys published recently dealing
with star-forming properties of large numbers of galaxies over
the range of galaxy Hubble types, including Im and BCD systems.
Their observational approaches
are very different from ours, but it is worth commenting on
where our survey fits in with theirs. The first is that of
Gil de Paz \et\ (2000) who use \ha\ fluxes to determine star
formation rates for a sample of galaxies found
on objective-prism plates. Because of their method of detection,
this survey is limited to galaxies that are currently forming stars.
Their Figure 1(f) shows a histogram of \ha\ luminsity. We see that
the bright end of our survey begins just where their survey reaches
its faint limit. Gil de Paz \et\ normalize their star formation
rates to the stellar mass of the galaxy, a quantity that changes
as the galaxy evolves. If we compute the star formation rate for
our galaxies using their Equation 4 and divide by the integrated
galactic stellar mass, we find that our median log \sfr\ M$_{stars}^{-1}$
($10^{-11}$ yr$^{-1}$) is 1.2 for the Im galaxies with a range from
0.2 to 2.0 and 1.4 for
the BCDs with a range from 0.4 to 2.4.
Their mean value of log \sfr\ M$_{stars}^{-1}$
($10^{-11}$ yr$^{-1}$) for dwarf galaxies (based on 5 systems)
is 1.9$\pm$0.1 and for \HII\ galaxies (based on 18 systems) is 2.1$\pm$0.1.
Thus, the galaxies in our survey have
lower average specific star formation rates.
Gil de Paz \et\ also find a correlation between integrated galactic
stellar mass and the star formation per unit stellar mass
(their Figure 7(b)). Our sample alone shows no correlation, but
the relationship is quite broad and our range of parameter space
more limited than theirs.

A second survey is that of Brinchmann \et\ (2004) which uses
Sloan Digital Sky Survey (SDSS) images and spectra.
They also normalize to the integrated mass of the galaxy and find
that the value of log \sfr\ M$_{stars}^{-1}$ (yr$^{-1}$)
increases on average as the mass of the galaxy increases,
similar to Figure 7(b) of Gil de Paz \et\ (2000). To compare
to their Figure 22, we constructed histograms of
log \sfr\ M$_{stars}^{-1}$ (yr$^{-1}$) for our sample divided
into two sets: those with integrated galactic stellar masses
less than 10$^9$ M\solar\ and those more massive than this.
The median values in the two sub-samples are the same for the Im
galaxies, only 0.2 dex different for the BCDs and only 0.1 dex
different for the Sms. Brinchmann \et, on the other hand,
would predict a shift to higher log \sfr\ M$_{stars}^{-1}$ (yr$^{-1}$)
of 0.4 dex, using their middle mass bins in these ranges.
Thus, we do not see the correlation between \sfr\ M$_{stars}^{-1}$
and M$_{stars}$ that they see. However, our data do show the
correlation between \sfr\ and M$_{stars}$ that they find
(their Figure 17).

\subsection{Star formation extents} \label{sec-extents}

The radius of the last annulus in which \ha\ emission is detected
in our images \rha\
is given in Table \ref{tab-results}.
For most Im and Sm galaxies \rha\
is the same, or nearly the same,
as the radius of the furthest detectable discrete \HII\ region
\rhii.
However, in
some BCDs and starburst galaxies the furthest detectable
\ha\ emission is diffuse ionized gas
from the outer parts of a very large \HII\ region or extra-\HII\ region ionized gas.
In about 11 of our BCDs \rha\ is somewhat greater than the location of the
center of its furthest \HII\ region because of extended ionized gas around the
supergiant \HII\ region. In the two starburst galaxies NGC 1569 and NGC 1705, extra-\HII\
region ionized gas extends well beyond the main body of the optical galaxy
and any star-forming regions.
Therefore, in Table \ref{tab-results} we also include \rhii\ for each galaxy.
This is the ellipse, from the ellipse photometry, that includes the center of
the \HII\ region which is most distant from the center.
Also in Table \ref{tab-results} we give the ratio of \rhii\ to various
measures of the size of the galaxy: \rtf, \rh, and \rd. Histograms
of \rhiih\ and \rhiid\ are shown in Figures \ref{fig-histrharh}
and \ref{fig-histrhard}.

We see that most of the detectable \HII\ regions in our survey galaxies are
found within the Holmberg radius \rh. However, in some, \ha\ emission
is found beyond this, even to 1.7 times \rh.
Similarly, most of the star-formation is found within 3 disk scale-lengths,
but in some, \ha\ emission is found even to 6\rd.
This is just as Parodi \& Binggeli (2003) found for bright
lumps in B-band images of Im galaxies.
In BCDs, the \HII\ regions are significantly more concentrated
toward the center than in Im and Sm galaxies, consistent with
the starburst nature of the BCDs.

The location of \HII\ regions, as a tracer of recent star formation,
is consistent with most studies of resolved stellar populations in
Im galaxies. Aparicio, Tikhonov, \& Karachentsev  (2000) found
that there has been no recent star formation beyond \rh\ in DDO 187
and only old stars are found in the outer part of DDO 190
(Aparicio \& Tikhonov 2000).
WLM also has an extended halo of old stars (Minniti \& Zijlstra 1997),
as do DDO 216 (Gallagher \et\ 1998) and NGC 3109 (Minniti, Zijlstra,
\& Alonso 1999).
By contrast, Komiyama \et\ (2003) have found an extended distribution
of blue stars that follows the \HI\ distribution in NGC 6822,
well beyond the \ha.
They estimate the age of these stars at 180 Myrs.

\subsection{\ha\ surface photometry}  \label{sec-mu}

The \ha\ surface photometry was done in the same way as the
V-band or equivalent surface photometry, which will be discussed
in a separate paper. The surface brightness plot
of the first galaxy in our sample is shown in Figure \ref{fig-sb}.
Plots for the rest of the
galaxies in the sample will be published in the paper presenting
the broad-band images.
The \ha\ image is measuring the star formation activity over
the past 10 Myr or so, while the V-band light is dominated by
stars with lifetimes of order 1 Gyr if star-formation is
on-going (a convolution of the numbers of stars formed and their
brightness in the V-band, Gallagher, Hunter, \& Tutukov 1984).
Thus, the V-band starlight is a tracer of a
somewhat older stellar population.

Hunter \& Gallagher (1985b) and
Hunter \et\ (1998a) found that, for many Im galaxies in their
sample, there was a correlation between the surface density
of stars and the azimuthally-averaged star formation activity
(see also Ryder \& Dopita 1994; Brosch, Heller, \& Almoznino 1998;
Parodi \& Binggeli 2003).
This implies that star formation has been approximately constant with radius.
They noted that this could be a causal connection if stellar
energy input to the ISM acts as a feedback process to star
formation (see also Dopita \& Ryder 1994).

Our much larger sample here shows that the relationship between
stars and star formation is more complex.
We have examined the correspondence of surface photometry $\mu_V$ and \sigha\ by
looking at the ratio of L$_{H\alpha}$ to L$_V$ in the surface
photometry annuli.
The \ha\ surface photometry is expected to be less smooth
than the starlight since it represents a shorter-lived evolutionary stage.
So, exact correspondence between $\mu_V$ and \sigha\ is
not expected. The ratios L$_{H\alpha}$/L$_V$ as a function
of radius are shown in Figure \ref{fig-hadivv}.
We see that among the Im galaxies roughly half
of the Im galaxies have a good
correspondence between
stellar density and star formation surface photometry.
In the BCDs, on the other hand, the ratio L$_{H\alpha}$/L$_V$
most often drops steeply with radius.
For Sm galaxies,
the majority show a fairly constant ratio with radius.

We have used Figure \ref{fig-hadivv} to assign the relationship
between L$_{H\alpha}$ and L$_V$ for each galaxy to one of three
bins. The three bins are defined as 1) \ha\ falls off
approximately like V. That is, the ratio L$_{H\alpha}$/L$_V$ is
roughly constant, with statistical variations and not large
trends. 2) \ha\ and V show some differences and some similarity,
and 3) \ha\ and V are dissimilar with large-scale trends. Some
galaxies have too little star formation to be a part of this
exercise. Of the Im galaxies, we find that approximately 44\% (34
of 77) have \sigha\ that fall off with radius similar to $\mu_V$,
40\% (31 of 77) are quite different, and 16\% (12 of 77) are in
between. Of the BCD galaxies, 8\% (2 of 25) are similar, 76\% (19
of 25) are different, and 16\% (4 of 25) are in between. Of the Sm
galaxies, 55\% (10 of 18) are similar, 28\% (5 of 18) are
different, and 17\% (3 of 18) are in between. There is no
correlation between presence of a stellar bar or the value of
W$_{20}$, the FWHM of the integrated HI profile, of an Im galaxy
and its bin type. There is also no correlation between bin type
and distance to the nearest bright companion.
However, there is a small preference for Im galaxies in bin 1 to
be brighter overall in the V-band, and this is shown in Figure
\ref{fig-havclass_mv}.

\section{Discussion} \label{sec-disc}

\subsection{Sm galaxies}

In keeping with their classification as late-type spirals, the Sm
group falls at the high end of the range of absolute magnitude
of our sample although there is considerable overlap. Our Sm
sample ranges from $-15.9$ to $-19.5$ with a median of $-17$.
Because they tend to be more luminous, the Sm galaxies also
are physically larger (see also Hodge 1971) and, as a group,
tend to higher \sfr.
In other properties, however, the Sm group covers the same range
as the Im galaxies.

Most of the Sm galaxies have a roughly constant L$_{H\alpha}$/L$_V$
ratio with radius. This implies that the Sm galaxies are in a
sense more stable. They are large enough and have enough
star formation going on at all times that star formation
as a function of radius is approximately constant (Hunter \& Gallagher 1985b).
This is also consistent with the
trend for the brighter Im galaxies to also have roughly constant
L$_{H\alpha}$/L$_V$ with radius. The statistical variations
with location within
smaller galaxies simply make this correlation harder to
maintain. However, smaller galaxies may still maintain a correspondence
integrated over a longer period of time, as ``star formation
bubbles around a galaxy'' (Hunter \& Gallagher 1986) propelled
by changes in local characteristics.

\subsection{BCD galaxies}
\label{sect:bcd}

In contrast to the Sm group, the BCD group covers the same range
in M$_V$ as the Im galaxies, but falls at the extremes in other
properties. While the absolute star formation rate \sfr\ is only
modestly high in BCDs compared to typical Ims, the normalized star
formation rates \sfrtf\ and \sfrd\ are significantly higher. In
keeping with the higher star formation rates, the BCDs also have a
smaller time scale to exhaust the gas $\tau$ and higher surface
brightnesses.

Part of the contribution to higher \sfrd\ is the smaller \rd\ of
BCDs (see also Papaderos \et\ 1996, Patterson \& Thuan 1996, van
Zee \et\ 1998). This is illustrated in Figure \ref{fig-mvrd} where
one can see that the BCDs have smaller R$_D$ relative to their
absolute magnitude, implying that they have systematically steeper
profiles. This is in spite of the fact that in 8 systems we
have fit the V-band azimuthal profile only in the outer parts
where the profile becomes less steep and that we have reached as
far out as \rh\ in 13 of the 26 systems. Thus, to the extent that
the V-band images are affected by the intense star formation
taking place in these galaxies, they appear to be affected over a
fairly large portion of the optical galaxy or the galaxy beyond
the affected region must be quite low in surface brightness
(see also Noeske \et\ 2003).

In addition, the star formation in BCDs is often
concentrated to the central regions (see also Davies, Sugai, \&
Ward 1998). One can see this in Figures \ref{fig-histrharh}
and \ref{fig-histrhard}. However, \rhii\ does not take the intensity
of the \HII\ region into account, and in some galaxies small regions
are located outside the dominant central star-forming region.
The central concentration is even clearer in the plots of $\log$
L$_{H\alpha}$/L$_V$ as a function of radius (Figure
\ref{fig-hadivv}). There the ratios often fall steeply with
radius, implying unusually high L$_{H\alpha}$ relative to L$_V$
near the centers.

The central concentration of star formation may be accompanied by
a similar concentration of neutral gas. While only 3 of our BCDs
have published HI maps, other samples of BCDs have been mapped in
HI. Taylor \et\ (1995), van Zee \et\ (1998), and Simpson \&
Gottesman (2000) have stated that the HI gas is centrally
concentrated in their BCDs. Van Zee, Salzer, \& Skillman (2001)
further state that BCDs have systematically steeper rotation
curves, although the rotation curves are not given. This is in
contrast to galaxies at the other extreme; very low surface
brightness galaxies have more slowly rising rotation curves (de
Blok, McGaugh, \& van der Hulst 1996; but see Swaters, Madore, \&
Trewhella 2000).

The radial variation of L$_{H\alpha}$/L$_V$ in a galaxy implies
that the position of star formation has migrated to the center
within the last Gyr, which is approximately the timescale for stars
that dominate the light in the V-band for on-going star
formation. A major shift in the
location of star formation requires a significant amount of
angular momentum loss from the gas.  Such a loss is usually the
result of bar torques or companion galaxy tidal torques. In the
case of a merger, the torque also comes from gas pressure and
shocks as the two interstellar media merge. For our BCD sample,
and for the Im and Sm cases that also have L$_{H\alpha}$/L$_V$
gradients, there is no systematic evidence for strong bars or
bright companions.  Still, most of these low mass galaxies are
clearly axisymmetric and irregular, and any slight shear in the
presence of such irregularities can lead to bar-like torques and
mass inflow.

Low mass galaxies also have dark matter halos with velocity
dispersions comparable to the dispersion in the stars and gas, as
may be inferred from their systematically thicker shapes
(Hodge \& Hitchcock 1966; van den Bergh 1988;
Staveley-Smith, Davies, \& Kinman 1992; Binggeli \& Popescu 1995;
Sung \et\ 1998).
In these
cases, two-fluid or three-fluid instabilities involving the
visible galaxy gravitationally coupled to a rotating dark matter
halo could in principle lead to bar formation in the halo in much
the same way as a stellar disk forms a bar in its center. The
presence of such halo bars (e.g., Bekki \& Freeman 2003) could be
inferred from the rotation pattern of gas in the disk, which would
show the characteristic central twist in the iso-velocity lines of
a velocity contour diagram (Bosma 1981).  We note that the HI
velocity distribution in the nearby BCD-like galaxy IC 10 (Wilcots
\& Miller 1998; see their Fig. 10) has twists and distortions in
the center, and their tilted ring model for the rotation curve has
sharp changes in the fitted inclinations and position angles of
the inner disk when they force a circular rotation pattern (see
their Fig. 5). Wilcots \& Miller point out that a counter-rotating
stream of HI in the outer part of IC 10 is a likely cause of the
starburst, but the inner irregularities are interesting also as
possible drivers of gas migration.

One tempting interpretation of the central concentration of gas in
BCDs is that they have been affected by interactions with other
galaxies sometime in the past.  An interaction can strip angular
momentum from the gas and stars and lead to overall disk
contraction.  The high gas density in the center then causes the
starburst. For example, the ratio of R$_{25}$ to $R_D$ increases
from $\xi_1$ to $\xi_2$ if $\xi_2-\xi_1=2\ln(S)$ where $S$ is the
shrinkage factor of an exponential disk that conserves mass. Our
observations are consistent with $\xi_1\sim1.5$ for Im and
$\xi_2=3$ for BCDs (cf. Fig. \ref{fig-comparesfr}), in which case
$S=1.3$ requires only a modest disk shrinkage.  Theoretical models
by Noguchi (1988) were among the first to show how amorphous
galaxies with mass inflow can result from the interaction of two
galaxies of comparable mass passing on parabolic orbits.  This
might also be consistent with the otherwise smooth distribution of
the HI in BCDs (Simpson \& Gottesman 2000).

Recent observations are consistent with this picture. Taylor,
Brinks \& Skillman (1993) and Taylor et al. (1995) found
previously unknown companions or HI clouds around half of the HII
galaxies they surveyed compared to a companion fraction of 1/4
among comparable LSB dwarfs (Taylor et al. 1996). Taylor (1997)
concluded from this that the incidence of companions for HII
region galaxies is twice that for LSB dwarfs, implying that the
companions triggered starbursts in some cases. The Fornax Cluster
galaxy FCC 35 is another example of a BCD with a nearby
intergalactic HI cloud (Putman et al. 1998), and NGC 1569 and IC 10
are examples of starbursts with large HI clouds in their outer
parts (Stil \& Israel 1998, Wilcots \& Miller 1998).
M\'endez \& Esteban
(2000) observed a high fraction of irregular WR galaxies with
uncatalogued low surface brightness companions and suggested these
companions were responsible for the starbursts. Similar
conclusions were reached by Pustilnik's group: Pustilnik et al.
(2001a) suggested that $>80$\% of the 86 BCDs in their survey had
star formation triggered by faint visible companions or previous
mergers. Pustilnik et al. (2001b) found that the BCD galaxy SBS
0335-052 is at a gas concentration in a giant HI cloud that also
contains another BCD galaxy, and they suggested that the
starbursts were triggered either by mutual interactions or by
disturbances from a nearby giant galaxy. Pustilnik et al. (2003)
also observed a faint, blue, low surface brightness dwarf within
11 kpc of the BCD galaxy HS 0822+3542, which was formerly thought
to be isolated. Similarly, Bravo-Alfaro et al. (2004) suggested
that the BCD galaxy Haro 4 was interacting with the nearby spiral
Haro 26. Sometimes the evidence for an interaction is indirect.
For example, the warp and disturbed velocity field
(C${\hat{\rm o}}$t\'e, Carignan \& Freeman 2000) of the BCD galaxy
NGC 625 suggested a recent interaction to Cannon et al. (2003).
\"Ostlin et al. (2001) also found irregular H$\alpha$ velocity
fields and secondary dynamical components in the 6 BCDs they
studied, and suggested these galaxies had mergers with gas-rich
dwarfs or massive gas clouds.

Considering these recent discoveries, it is not surprising that BCD
companions were hard to observe: most turn out to be very faint.
For example, Telles \& Maddox (2000) did not find a significant
incidence of companions to HII galaxies in a plate-scanned survey
of galaxies down to 20.5 mag in B-band, where the catalog was
90\%-95\% complete. The companions that are now implicated in BCD
interactions are usually fainter than this limit, and fainter than
our limit in Fig. \ref{fig-sfrdist}. The faintness of triggering
companions is probably reasonable, though. Most galaxies are
dwarfs, so the most likely object to perturb any given galaxy is a
dwarf. For large galaxies, these dwarf perturbations are indeed
common but they do very little to enhance the star formation rate
and gas concentration in the large galaxy. But for small galaxies,
even a dwarf or HI cloud interaction can be significant, leading
to angular momentum redistribution, mass inflow, and rapid star
formation in a dense core.  Thus the most likely significant
disturbers of small galaxies are other small galaxies or small gas
clouds.

Adding to the ambiguity of BCD origins is the likelihood that some
interactions occurred long ago and the BCD has been bursting on
and off regularly ever since. In a study of color-magnitude
diagrams of BCDs, Schulte-Ladbeck, et al.(2001) found that there
were no significant star formation gaps lasting longer than a few
times 100 Myr over the last Gyr of evolution. Since the gas
consumption time is $\sim1$ Gyr, the interactions in some cases
could have started this long ago (Lynds \et\ 1998; Crone et al.
2002; \"Ostlin et al. 2003). The primary remnant of these former
interactions would simply be the low specific angular momentum of
the present day disk; this would continue to drive starburst
episodes as long as there is gas (e.g., van Zee, Salzer, \&
Skillman 2001). The low specific angular momentum of BCDs could
also have resulted from the last big merger (Vitvitska et al.
2002), or from an old interaction that deformed the dark matter
halo into a prolate spheroid, which then drove a disk response
like a bar in a barred galaxy (Bekki \& Freeman 2003).

Another possibility is an evolutionary scheme between dIm and
BCDs. This has been proposed by Simpson \& Gottesman (2000) in an
optical and HI comparison of low surface brightness dIm and BCDs.
The HI in numerous dIm takes on a ring-like morphology. The
formation of these giant, sometimes galaxy-sized, rings would be
easiest to explain as due to a starburst in the galaxy center that
blew a very large hole in the gas. The energy requirements are
reasonable for this explanation. Thus, an evolution of BCD to dIm
as the gas reservoir expands and dissipates has some appeal.
However, there is a problem with the lack of evidence for the
presence of an aging starburst population in the center of the HI
ring in at least one such system (see Simpson, Hunter, \& Knezek
2004). Also, differences in the scale-length suggest that there
must be some redistribution of the stellar component as well, and
this leads us once again back to gravitational interactions.

\subsection{Probability Distribution Functions}

The high star formation rates and dense core structures of BCD
galaxies (see Figures \ref{fig-histrharh}, \ref{fig-histrhard},
and \ref{fig-hadivv}) do not necessarily imply that the star
formation processes differ much from those in Im and Sm types --
only that the star formation regions crowd closer together in the
BCD centers, producing smaller $R_D/R_{25}$ and higher luminosity
densities. To compare the star formation processes in a global
sense, we examined the probability density functions (pdf) of
H$\alpha$ emission for each galaxy.  In a turbulent medium with
isothermal density and pressure variations, as in an HII region,
the gas density should have a pdf that is approximately log-normal
(Ostriker, Gammie \& Stone 1999; Klessen 2000; Ostriker et al.
2001; Li et al. 2003) and the column density should have a pdf
that is log-normal too, although somewhat diluted by the average
of unrelated density elements on the lines of sight (Padoan et al.
2000; Ostriker et al. 2001; V\'azquez-Semadeni \& Garc\'ia 2001).
The width of the pdf increases with the Mach number of the
turbulence (Padoan et al. 2000). For star formation triggered
globally in turbulent shocks (Elmegreen 1993; see review in Mac
Low \& Klessen 2004), the H$\alpha$ emission throughout a whole
galaxy might be expected to follow this turbulent pdf too. We find
this to be the case for all of our galaxies.

The pdfs for the \ha\ images of our galaxies are shown in Figure
\ref{fig-pdf}. We constructed them from a histogram of the \ha\
surface brightness of individual pixels in the \ha\ image within
the region used in the ellipse photometry (Table \ref{tab-obs}).
Pixels without \ha\ emission were rejected, but no cutoff was made
for positive values of noise at the low count end. The surface
brightness of each pixel was normalized to the
azimuthally-averaged \ha\ surface brightness at that radius (for
example, Figure \ref{fig-sb}) in order to remove the exponential
disk.

Figure \ref{fig-pdf} is plotted on log-log axes so that a
log-normal is an inverted parabola. All of the pdfs in fact have
this shape, and they all have about the same FWHM too, although
their absolute brightnesses vary from galaxy to galaxy. In some
cases there is a lower cutoff in the surface brightness that
limits the pdf to the upper half, but even then most pdfs resemble
a log-normal in the plotted parts.  Some have a slight excess at
the highest surface brightness, giving the pdfs a flair on the
lower right.

There is apparently no systematic difference in the H$\alpha$
surface brightness pdfs for the various types of galaxies. The BCD
pdfs look about the same as the Im and Sm pdfs.  This implies that
the Mach numbers of the turbulence in the ionized gas components
are about the same, even though the areal densities of star
formation are much higher in the BCDs.  We conclude that the
individual star formation regions are not intrinsically different
in the various galaxy types, only that the regions are crowded
more closely together in the BCDs. This crowding is not without
consequences -- it can lead to higher pressures and perhaps more
massive clusters in proportion to the total star formation rate
by the size of sample effect. But the expansion and turbulent
speeds in BCD HII regions are not significantly higher than in
other galaxies, and therefore the balance between pressurized
triggering and spontaneous instabilities as causes of star
formation is probably not much different either.

\acknowledgments

We are deeply grateful to Ralph Nye who did a great job
making and keeping the Hall 1.1 m and
Perkins 1.8 m telescopes and their instruments operating.
We appreciate the TI CCD provided to Lowell Observatory by the National Science
Foundation and another on loan
from the U.\ S.\ Naval Observatory in Flagstaff.
The \ha\ imaging would not have been possible without filters purchased through
funds provided by a Small Research Grant from the American Astronomical Society,
National Science Foundation grant AST-9022046,
and grant 960355 from JPL.
Funding for this work was provided by the Lowell Research Fund
and by the National Science Foundation through grants AST-0204922 to DAH
and AST-0205097 to BGE.

Some of the observations presented here benefited from the Fabry-Perot
focal reducer
that was built by Ohio State University,
and we appreciate the effort of M.\ Wagner to keep it
working.
We also wish to thank Emily Bowsher for geomtrically matching the
\ha\ images to the V-band images for some of the galaxies
as part of the 2003 Research Experiences for Undergraduates
program of Northern Arizona University funded by the National
Science Foundation through grant 9988007.
This research has made use of the NASA/IPAC Extragalactic Database
(NED) which is operated by the Jet Propulsion Laboratory, California
Institute of Technology, under contract with the National Aeronautics
and Space Administration.

\appendix

\section{Spectrophotometry of nebulae} \label{app-hii}

Spectrophotometric observations with large apertures
were obtained of several \HII\ regions in nearby galaxies
in 1981 October and December. We used the cooled dual-beam
Intensified Reticon Scanner (IRS) mounted on the No. 1 0.9 m
telescope at Kitt Peak National Observatory. We used three
gratings to cover from 3500 \AA\ to 7450 \AA\ with 4.5 \AA\
spectral resolution.
These observations
were part of a program to observe stars for
a stellar library, and the instrumentation,
observations, and data reductions are discussed in more detail by
Jacoby, Hunter, \& Christian (1984). In Table \ref{tab-hii}
we present integrated spectrophotometry
of the major emission-lines measured from spectra of
\HII\ regions observed at that time.
The \ha\ emission from NGC 604
was used to calibrate the \ha\ images presented in this paper.

\section{Determining the Star Formation Rate from \lha} \label{app-sfrformula}

Here we outline the derivation of the formula that we used to
convert \lha\ to the star formation rate in solar-masses per year.
There are three quantitative assumptions that enter the derivation.
First, we assume that when stars form from a gas cloud, the number
of stars is a power law that depends on the mass of the star.
That is, the number of stars formed of mass $m$ is given by
\begin{equation}
\Phi(m) = A m^\gamma ,
\label{eq:imf}
\end{equation}
where the power law index $\gamma$ is
related to the ``slope of the stellar initial mass function''
$\Gamma$ as $\Gamma=\gamma+1$. Here we assume the Salpeter (1955)
slope $\Gamma$ of $-1.35$.

Second, we assume an upper stellar mass limit of 100 M\solar.
We know that stars as massive as 150 M\solar\ are found in very
rich star-forming regions (Massey \& Hunter 1998), but they
are relatively rare, and integrating the stellar IMF to 100 M\solar\
is consistent with the vast majority of star-forming regions found
in galaxies. We also assume that the lower stellar mass limit is 0.1
M\solar.

Third, we assume an efficiency factor $\eta$ of 2/3 for the absorption
of ionizing photons by the nebula (Gallagher, Hunter,
\& Tutukov 1984). That is, 1/3
of the ionizing photons emitted by massive stars in \HII\ regions
are absorbed by dust or escape the nebula, and 2/3 of the emitted
photons result in \HII.

The number of ionizing photons $N_L$ per second implied by the flux from a nebula
is
\begin{equation}
N_L = \frac{\rm L_{H\beta}}{h \nu_{H\beta}} \left( \frac{\alpha_B}{\alpha_{H\beta}} \right),
\label{eq:nl}
\end{equation}
where $h \nu_{H\beta}$ is the energy of the H$\beta$ photon and
$\alpha_B$ and $\alpha_{H\beta}$ are the recombination coefficients for Case B
recombination. $\alpha_B$ is given in Table 2.1 and $\alpha_{H\beta}$ in Table 4.4
of Osterbrock (1974). Table 4.4 also gives the ratio of the line intensity of
\ha\ to H$\beta$. Taking values for a nebular temperature of 10$^4$ K and
density of 10$^2$ cm$^{-3}$, we have
\begin{equation}
N_L = 7.37\times10^{11} {\rm L_{H\alpha}} \ {\rm photons\ s^{-1}}.
\label{eq:nlha}
\end{equation}
Therefore, from the integrated \lha\ of the galaxy,
we know the total number of ionizing photons
required to ionize all of the \HII\ regions,
and, if we account for losses, we know the total number of ionizing photons
$N_L^G$, in photons per second, being produced by massive stars in the galaxy:
\begin{equation}
N_L^G = \frac{7.37\times10^{11}}{\eta} {\rm L_{H\alpha}} \ {\rm photons\ s^{-1}}.
\label{eq:nlg}
\end{equation}

The ionized gas is produced by stars with masses 18 M\solar\ to 100 M\solar.
The number of stars between 18 M\solar\ and 100 M\solar\ is, therefore,
\begin{equation}
n_{18}^{100} = \frac{N_L^G}{\langle N_L^* \rangle}
           = \frac{7.37\times10^{11} {\rm L_{H\alpha}} / \eta}{\langle N_L^* \rangle},
\label{eq:n18100}
\end{equation}
where $\langle N_L^* \rangle$ is the average number of ionizing photons produced
per star.
The number of stars between 18 M\solar\ and 100 M\solar\ is
also given by the IMF:
\begin{equation}
n_{18}^{100} = \int_{18}^{100} A m^\gamma dm
\label{eq:n18100imf}
\end{equation}
and the total mass in stars $M_*$ is given by
\begin{equation}
M_* = {\int_{0.1}^{100} A m^\gamma m dm}
      = \frac{n_{18}^{100}}{100^\Gamma-18^\Gamma} \left(\frac{\Gamma}{\Gamma+1}\right) {(100^{\Gamma+1}-0.1^{\Gamma+1})}.
\label{eq:mass}
\end{equation}
Thus,
\begin{equation}
M_* = 432 n_{18}^{100} {\rm M_{\sun}}.
\label{eq:massn18100}
\end{equation}

To put this altogether, we need the
weighted average $\langle N_L^*\rangle$ output per star, where
we weight by the IMF and by the Hydrogen-burning lifetime of the star. $N_L^*$
is higher the more massive the star is, but the number of stars go down
and the lifetime is shorter as the mass goes up.
We have used the data compiled by Panagia (1973) to determine
$N_L^*$ as a function of the mass of the star. Below a mass of 18 M\solar\
$N_L^*$ drops precipitously, so we consider only stars with masses
greater than or equal to 18 M\solar\ as contributing to the ionization
of the \HII\ regions. A fit to the data of Panagia yields
\[ \log N_L^* =  \left\{ \begin{array}{ll}
                 $ 44.129 + 3.276 {\rm log} m $ & \mbox{$ 18-41 {\rm M}_{\sun}$} \\
                 $ 47.170 + 1.417 {\rm log} m $ & \mbox{$ \geq 41 {\rm M}_{\sun}$}
                 \end{array}
                 \right. \]
We use the data of Schaerer \et\ (1993) for the Hydrogen-burning lifetimes
of massive stars at a $Z=0.008$, a metallicity representative of many
Im galaxies. A parameterization of the lifetime $t_{ms}(m)$ as function of the mass
of the star is given by
\[ \log t_{ms}(m) =  \left\{ \begin{array}{ll}
                 $ 8.475 - 1.160 {\rm log} m $ & \mbox{$ 15-33 {\rm M}_{\sun}$} \\
                 $ 7.472 - 0.501 {\rm log} m $ & \mbox{$ \geq 33 {\rm M}_{\sun}$}
                 \end{array}
                 \right. \]
where $t_{ms}(m)$ is in years.
The IMF- and $t_{ms}(m)$-weighted average $\langle N_L^*\rangle$ is given by
\begin{equation}
\langle N_L^* \rangle = {\frac{\int_{18}^{100} N_L^*(m) t_{ms}(m) A m^\gamma dm}
    {\int_{18}^{100} A m^\gamma dm}} = 2.52\times10^{63} {\rm \ photons,}
\label{eq:avenlstar}
\end{equation}
now integrated over the lifetimes of these stars.
If we convert $N_L^G$ from photons s$^{-1}$ to photons yr$^{-1}$, we have the
star formation rate $\dot{M}$
\begin{equation}
\dot{M} = 5.96\times10^{-42} {\rm L_{H\alpha}} \ {\rm M_{\sun} yr^{-1}}.
\label{eq:sfr}
\end{equation}

This star formation rate formula differs from others we have used
in the past. It yields \sfr\ that are a factor of 1.2 lower than
rates given by the formula of Hunter \& Gallagher (1986) and 1.8
times lower than those of Gallagher \et\ (1984). This formulation
yields star formation rates that are 14\% higher than those used
by Kennicutt (1998), a factor of 10 lower than those determined by
Gil de Paz \et\ (2000), and 40\% higher than numbers determined by
Brinchmann \et\ (2004) for the integrated galactic mass of a
typical galaxy in our sample (their Figure 7).

\clearpage


%


\begin{deluxetable}{lllrrrrrl}
\tabletypesize{\scriptsize}
\tablewidth{0pt}
\tablecaption{Galaxy Sample. \label{tab-sample}}
\tablehead{
\colhead{} & \colhead{} & \colhead{} & \colhead{} & \colhead{}
& \colhead{} & \multicolumn{3}{c}{Nearest Neighbor\tablenotemark{f}} \\
\cline{7-9}
\colhead{} & \colhead{} & \colhead{}
& \colhead{D} & \colhead{} & \colhead{}
& \colhead{$\Delta$d} & \colhead{$\Delta$V$_r$} & \colhead{} \\
\colhead{Galaxy} & \colhead{Other Names\tablenotemark{a}}
& \colhead{Type\tablenotemark{b}}
& \colhead{(Mpc)} & \colhead{Ref\tablenotemark{c}}
& \colhead{E(B$-$V)$_f$\tablenotemark{d}}
& \colhead{(kpc)} & \colhead{(km/s)} & \colhead{Neighbor}
}
\startdata
\vspace{4mm}
{\bf Im Galaxies:}&&&&&&&&\\
\vspace{4mm}
A1004$+$10 & PGC 29428, UGC 5456,        &    I0?      &  6.5\x & \nodata & 0.01\xx &  570 &  $+$13\xx & UGC 5288 (Sdm:)        \\
           & IRASF10046$+$1036           &             &        & \nodata &         &      &           &                        \\
A2228$+$33 & PGC 69019, UGC 12060,       &    IBm      & 16.9\x & \nodata & 0.01\xx & 1545 &  $-$98\xx & NGC 7320 (Sd)          \\
           & IRASF22282$+$3334           &             &        &         &         &      &           &                        \\
CVnIdwA  & UGCA 292                      &    Im?      &  4.1\x & 15      & 0.01\xx &  260 &   $+$2\xx & UGC 7605 (Im)          \\
D508-2   & LSBC F508-V01                 &    Im       & 29.9\x & \nodata & 0.00\xx & 2260 & $-$103\xx & KISSB 109 (-)          \\
D575-5   & LSBC F575-03                  &    dI       &  6.2\x & \nodata & 0.03\xx &  730 &   $-$5\xx & IC 798 (E0)            \\
\enddata
\tablenotetext{a}{This does not include all possible identifications.
Identifications were obtained from NED.}
\tablenotetext{b}{Morphological Hubble types are from de Vaucouleurs et al.\ (1991). If no type is
given there, we have used types given by NED.}
\tablenotetext{c}{Reference for the distance to the galaxy.
If no reference is given, the distance
was determined from V$_{GSR}$ given by de Vaucouleurs et al.\ (1991) and a Hubble constant
of 65 km s$^{-1}$ Mpc$^{-1}$.}
\tablenotetext{d}{Foreground reddening from Burstein \& Heiles (1984).}
\tablenotetext{f}{Nearest galaxy within 1 Mpc in the plane of the sky and $\pm$150 km s$^{-1}$.
Distance to the galaxy $\Delta d$ was taken as a combination of apparent distance on the sky and radial
distance with a Hubble constant
of 65 km s$^{-1}$ Mpc$^{-1}$. The search used NED.}
\tablerefs{
(1) Aparicio 1994;
(2) Aparicio, Tikhonov, \& Karachentsev 2000;
(3) Dohm-Palmer \et\ 1998;
(4) Dolphin 2000;
(5) Dolphin \et\ 2002;
(6) Dolphin \et\ 2003;
(7) Freedman \et\ 2001;
(8) Gallagher \et\ 1998;
(9) Gallart, Aparicio, \& Vilchez 1996;
(10) Hidalgo, Mar\'in-Franch, \& Aparicio 2003;
(11) Hunter 2001;
(12) Karachentsev, Aparicio, \& Makarova  1999;
(13) Karachentsev, Musella, \& Grimaldi 1996;
(14) Karachentsev \et\ 2002;
(15) Karachentsev \et\ 2003a;
(16) Karachentsev \et\ 2003b;
(17) Lee \et\ 1999;
(18) Lee \& Kim 2000;
(19) Maiz-Apellaniz, Cieza, \& Mackenty 2002;
(20) Makarova \et\ 1998;
(21) Massey \& Armandroff 1995;
(22) M\'endez \et\ 2002;
(23) Miller \et\ 2001;
(24) Minniti \& Zijlstra 1997;
(25) O'Connell \et\ 1994;
(26) Tolstoy \et\ 1995;
and
(27) Tosi \et\ 2001.
}
\end{deluxetable}

\clearpage


%


\begin{deluxetable}{lllccccrcrrrc}
\tabletypesize{\scriptsize}
\rotate
\tablewidth{0pt}
\tablecaption{Observations and photometry parameters. \label{tab-obs}}
\tablehead{
\colhead{} & \colhead{} & \colhead{} & \colhead{} & \colhead{} & \colhead{}
& \colhead{}
& \multicolumn{5}{c}{Ellipse photometry parameters\tablenotemark{a}}
& \colhead{} \\
\cline{8-12}
\colhead{} & \colhead{} & \colhead{} & \colhead{}
& \colhead{Exposures\tablenotemark{d}} & \colhead{Seeing\tablenotemark{e}}
& \colhead{Scale}
& \colhead{PA}
& \colhead{} & \colhead{Step} & \multicolumn{2}{c}{Center (2000)}
& \colhead{$\log$ rms$_{\Sigma_{H\alpha}}$\tablenotemark{f}} \\
\colhead{Galaxy}
& \colhead{Mos\tablenotemark{b}}
& \colhead{Date}
& \colhead{Instr\tablenotemark{c}}
& \colhead{(s)}
& \colhead{(\protect\arcsec)}
& \colhead{(\protect\arcsec)}
& \colhead{(\protect\arcdeg)}
& \colhead{{\it b/a}} & \colhead{(\protect\arcsec)}
& \colhead{RA} & \colhead{DEC}
& \colhead{(ergs/s/pc$^2$)}
}
\startdata
{\bf Im Galaxies:}&&&&&&&&&&&&\\
&&&&&&&&&&&&\\
A1004+10  & \nodata & 9703 & PT,4:1,NOTI & 1$\times$1800 & 3.4 & 0.43 &  -33.3 & 0.63 &  9.1 & 10:07:19.6 &  10:21:47 & 32.48 \\
A2228+33  & \nodata & 9410 & PT,FP,NCCD  & 2$\times$1800 & 2.3 & 0.49 &   38.5 & 0.92 & 11.3 & 22:30:34.0 &  33:49:14 & 32.16 \\
CVnIdwA   & \nodata & 0005 & HT,2048     & 3$\times$1800 & 3.1 & 1.13 &   79.5 & 0.78 & 11.3 & 12:38:40.2 &  32:45:40 & 32.19 \\
D508-2    & \nodata & 0004 & HT,2048     & 3$\times$1800 & 5.0 & 1.13 &   70.6 & 0.76 &  9.1 & 13:04:34.3 &  26:46:24 & 32.07 \\
D575-5    & \nodata & 9605 & PT,4:1,NCCD & 1$\times$2700 & 2.0 & 0.41 &   55.2 & 0.51 &  9.1 & 12:55:41.0 &  19:12:35 & 32.32 \\
D634-3    & \nodata & 0001 & HT,2048     & 3$\times$2700 & 3.7 & 1.13 &   -9.4 & 0.76 &  5.7 &  9:08:53.8 &  14:35:03 & 32.04 \\
\enddata
\tablenotetext{a}{Position angle PA, minor-to-major axis ratio
$b/a$, ellipse semi-major axis step size,
and position of center used to do photometry in concentric
ellipses. The PA given here has been corrected for the PA of the CCD camera
on the sky, usually about 0.5\arcdeg, but the PA
appropriate to the image itself was used for the photometry.}
\tablenotetext{b}{Galaxies that were, or were potentially, larger
than the field of view were mosaiced. This column designates the
relative pointing in the mosaic. A ``C'' designates the center.
No entry means that the galaxy
was not mosaiced.}
\tablenotetext{c}{Telescope, instrument, and detector used for the
\protect\ha\ observations.
Telescopes: PT$=$1.8 m Perkins Telescope at Lowell Observatory;
HT$=$1.1 m Hall Telescope at Lowell Observatory;
KPNO4m$=$Kitt Peak National Observatory 4 m telescope;
CTIO1.5m$=$Cerro Tololo Inter-American Observatory 1.5 m telescope.
Instruments: 4:1$=$4:1 focal reducer; FP$=$Ohio State University
Fabry Perot used as a
simple 3:1 focal reducer; those without a designation were direct
cameras.
Detectors: NCCD$=$a TI $800\times800$ CCD provided to Lowell Observatory
by the National Science Foundation;
NOTI$=$a TI $800\times800$ CCD on loan to Lowell Observatory from
the US Naval Observatory, Flagstaff Station;
2048$=$SITe $2048\times2048$ CCD;
Mos$=$KPNO Mosaic camera.}
\tablenotetext{d}{A () designates images taken under clear sky conditions
to calibrate the other frames taken under non-photometric conditions.}
\tablenotetext{e}{FWHM of a stellar profile on the final combined
\protect\ha\ image.}
\tablenotetext{f}{The H$\alpha$ surface brightness corresponding
to the rms of one pixel in the original H$\alpha$ image.}
\end{deluxetable}

\clearpage


%


\begin{deluxetable}{lrcccrrrccc}
\tabletypesize{\scriptsize}
\rotate
\tablewidth{0pt}
\tablecaption{H$\alpha$ photometry, star formation rates, H$\alpha$ extents
\label{tab-results}}
\tablehead{
\colhead{}
& \colhead{$\log$ L$_{H\alpha}$}
& \colhead{\protect\logsfr}
& \colhead{\protect\logsfrtf\tablenotemark{a}}
& \colhead{\protect\logsfrd\tablenotemark{b}}
& \colhead{$\log \tau$\tablenotemark{c}}
& \colhead{\protect\rha\tablenotemark{d}}
& \colhead{\protect\rhii\tablenotemark{e}}
& \colhead{}
& \colhead{}
& \colhead{}
\\
\colhead{Galaxy}
& \colhead{(ergs/s)}
& \colhead{(M\protect\solar/yr)}
& \colhead{(M\protect\solar/yr/kpc$^2$)}
& \colhead{(M\protect\solar/yr/kpc$^2$)}
& \colhead{(yr)}
& \colhead{(\protect\arcmin)}
& \colhead{(\protect\arcmin)}
& \colhead{\protect\rhiitf}
& \colhead{\protect\rhiih}
& \colhead{\protect\rhiid}
}
\startdata

{\bf Im Galaxies:}&&&&&&&&&&\\
&&&&&&&&&&\\
A1004+10  & 39.52$\pm$ 0.00 & -1.71 & -2.42 & -1.28 &  9.30 &  0.68 &  0.53 & 0.78 &  0.55 &  2.91\\
A2228+33  & 40.16$\pm$ 0.01 & -1.07 & -2.73 & -2.39 & 10.42 &  1.61 &  1.80 & 2.32 &  1.20 &  3.46\\
CVnIdwA   & 38.70$\pm$ 0.05 & -2.52 & \nodata & -2.64 & 10.44 &  0.66 & 0.47 & \nodata &  0.54 &  0.87\\
D508-2    & 39.02$\pm$ 0.32 & -2.20 & -3.27 & -3.21 & 11.38 &  0.53 &  0.53 & 2.39 &  0.96 &  2.56\\
D575-5    &  0.00 & \nodata & \nodata & \nodata & \nodata &  \nodata &  \nodata & \nodata &  \nodata &  \nodata\\
D634-3    &  0.00 & \nodata & \nodata & \nodata & \nodata &  \nodata &  \nodata & \nodata &  \nodata &  \nodata\\
D646-8    &  0.00 & \nodata & \nodata & \nodata & \nodata &  \nodata &  \nodata & \nodata &  \nodata &  \nodata\\
DDO 9     & 39.78$\pm$ 0.01 & -1.45 & -3.12 & -2.93 & 10.74 &  2.36 &  2.36 & 2.23 &  1.09 &  2.78\\
DDO 22    & 39.06$\pm$ 0.02 & -2.17 & -3.08 & -2.73 & 10.71 &  0.66 &  0.66 & 1.26 &  0.70 &  1.87\\
DDO 24    & 39.92$\pm$ 0.01 & -1.31 & -2.85 & -2.03 & 10.44 &  1.61 &  1.61 & 1.51 & \nodata &  3.93\\
DDO 25    & 40.17$\pm$ 0.01 & -1.06 & -2.69 & -1.87 &  9.90 &  1.42 &  1.42 & 1.22 &  0.82 &  3.16\\
DDO 26    & 39.26$\pm$ 0.01 & -1.96 & -3.49 & -2.91 & 11.39 &  1.33 &  1.33 & 2.06 &  1.25 &  3.99\\
DDO 27    & 39.23$\pm$ 0.03 & -2.00 & \nodata & -2.87 & 10.86 &  0.84 & 0.84 & \nodata &  1.35 &  2.93\\
DDO 33    & 40.54$\pm$ 0.01 & -0.69 & -2.92 & -2.68 & 10.49 &  1.04 &  1.04 & 1.13 &  0.88 &  1.48\\
DDO 34    & 39.39$\pm$ 0.03 & -1.84 & -2.49 & -2.40 & 10.52 &  1.60 &  1.60 & 3.61 &  1.57 &  3.98\\
DDO 35    & 40.78$\pm$ 0.00 & -0.45 & -2.53 & -1.48 & 10.02 &  0.85 &  0.85 & 0.85 & \nodata &  2.84\\
DDO 38    & 40.05$\pm$ 0.01 & -1.18 & -2.72 & -2.61 & 10.57 &  1.05 &  1.05 & 1.94 &  1.02 &  2.22\\
DDO 39    & 39.61$\pm$ 0.02 & -1.62 & -2.95 & -3.25 & 10.98 &  2.36 &  2.36 & 3.89 &  1.27 &  2.74\\
DDO 40    & 40.45$\pm$ 0.00 & -0.78 & -2.67 & -2.02 & 10.09 &  1.23 &  1.23 & 1.61 &  0.99 &  3.39\\
DDO 43    & 38.80$\pm$ 0.01 & -2.43 & -2.91 & -2.19 & 10.53 &  1.04 &  0.94 & 1.53 &  1.06 &  3.53\\
DDO 46    & 38.79$\pm$ 0.01 & -2.43 & -3.08 & -2.96 & 10.69 &  0.85 &  0.85 & 1.15 & \nodata &  1.32\\
DDO 47    & 39.25$\pm$ 0.01 & -1.98 & -2.55 & -2.75 & 10.71 &  3.69 &  3.69 & 5.15 &  1.65 &  4.08\\
DDO 50    & 39.98$\pm$ 0.00 & -1.25 & -2.68 & -1.83 & 10.24 & 99.99 & \nodata & \nodata & \nodata & \nodata\\
DDO 52    & 38.22$\pm$ 0.02 & -3.01 & -3.30 & -3.27 & 11.11 &  1.23 &  1.23 & 2.73 &  1.14 &  2.82\\
DDO 53    & 38.95$\pm$ 0.00 & -2.28 & -2.49 & -2.50 & 10.67 &  1.19 &  1.05 & 1.53 &  0.77 &  1.52\\
DDO 63    & 38.97$\pm$ 0.05 & -2.25 & -3.06 & -3.44 & 10.56 &  1.99 &  1.61 & 1.25 &  0.74 &  0.81\\
DDO 64    & 39.29$\pm$ 0.01 & -1.94 & -2.93 & -2.68 & 10.46 &  1.23 &  1.04 & 1.26 &  0.85 &  1.69\\
DDO 68    & 38.61$\pm$ 0.12 & -2.62 & -3.41 & -2.66 & 11.27 &  1.61 &  1.61 & 2.33 &  1.47 &  5.51\\
DDO 69    & 37.28$\pm$ 0.00 & -3.95 & -3.42 & -3.28 & 11.08 &  3.26 &  3.26 & 2.49 &  1.36 &  2.92\\
DDO 70    & 38.22$\pm$ 0.01 & -3.00 & -3.31 & -2.86 & 10.72 &  3.26 &  3.26 & 1.54 &  0.88 &  2.60\\
DDO 75    & 39.03$\pm$ 0.00 & -2.20 & -2.61 & -1.40 & 10.21 &  3.08 &  3.08 & 1.30 &  1.00 &  5.22\\
DDO 86    & 39.70$\pm$ 0.02 & -1.53 & -2.85 & -2.74 & 10.72 &  1.23 &  1.04 & 1.99 &  1.05 &  2.25\\
DDO 87    & 38.75$\pm$ 0.02 & -2.47 & \nodata & -3.16 & 10.84 &  1.42 & 1.23 & \nodata &  1.07 &  1.93\\
DDO 99    & 38.95$\pm$ 0.04 & -2.28 & -2.91 & -2.71 & 10.60 &  1.42 &  1.42 & 1.45 &  0.69 &  1.84\\
DDO 101   & 38.98$\pm$ 0.02 & -2.25 & -3.26 & -2.99 &  9.65 &  0.66 &  0.66 & 0.96 &  0.63 &  1.31\\
DDO 105   & 39.90$\pm$ 0.01 & -1.32 & -3.00 & -2.85 & 10.65 &  2.55 &  2.55 & 2.73 &  1.24 &  3.25\\
DDO 115   & 38.23$\pm$ 0.01 & -3.00 & -3.07 & -2.51 & 10.24 &  0.68 &  0.68 & 1.04 &  0.69 &  1.98\\
DDO 120   &  0.00 & \nodata & \nodata & \nodata & \nodata & \nodata  & \nodata & \nodata & \nodata  & \nodata \\
DDO 125   & 38.51$\pm$ 0.05 & -2.72 & \nodata & -2.58 & 10.48 & 1.65 & 1.65 & \nodata & \nodata & 2.49 \\
DDO 126   & 39.16$\pm$ 0.01 & -2.06 & -2.70 & -2.45 & 10.33 &  1.99 &  1.89 & 2.30 &  1.08 &  3.08\\
DDO 133   & 39.45$\pm$ 0.00 & -1.77 & -2.82 & -2.93 & 10.48 &  2.55 &  2.55 & 2.40 &  1.10 &  2.12\\
DDO 143   & 39.53$\pm$ 0.04 & -1.70 & \nodata & -2.82 & 10.52 &  1.61 & 1.61 & \nodata &  1.19 &  2.24\\
DDO 154   & 38.80$\pm$ 0.01 & -2.43 & -3.02 & -2.60 & 11.05 &  1.61 &  1.42 & 1.60 &  0.92 &  2.60\\
DDO 155   & 38.57$\pm$ 0.00 & -2.66 & -2.36 & -1.50 &  9.79 &  1.04 &  0.66 & 1.06 &  0.70 &  2.87\\
DDO 165   & 38.78$\pm$ 0.02 & -2.44 & -3.53 & -3.52 & 10.76 &  2.36 &  2.36 & 1.67 &  1.11 &  1.70\\
DDO 167   & 38.35$\pm$ 0.01 & -2.88 & -2.93 & -2.41 & 10.27 &  0.66 &  0.47 & 0.96 &  0.63 &  1.75\\
DDO 168   & 39.03$\pm$ 0.00 & -2.20 & -3.02 & -2.33 & 10.68 &  1.79 &  1.61 & 1.13 &  0.70 &  2.49\\
DDO 169   & 38.21$\pm$ 0.14 & -3.02 & -3.57 & -3.29 & 11.15 &  0.28 &  0.28 & 0.41 &  0.21 &  0.57\\
DDO 171   & 39.95$\pm$ 0.02 & -1.27 & -3.01 & -2.35 & 10.09 &  0.66 &  0.66 & 0.87 &  0.54 &  1.85\\
DDO 183   & 38.38$\pm$ 0.04 & -2.85 & -3.47 & -3.21 & 10.87 &  0.85 &  0.85 & 1.09 &  0.67 &  1.48\\
DDO 185   & 39.01$\pm$ 0.01 & -2.21 & -2.87 & -2.58 & 10.33 &  1.99 &  1.80 & 1.75 &  0.90 &  2.43\\
DDO 187   & 37.70$\pm$ 0.02 & -3.53 & -3.43 & -2.64 & 11.01 &  0.47 &  0.47 & 0.69 &  0.45 &  1.70\\
DDO 210   &  0.00 & \nodata & \nodata & \nodata & \nodata & \nodata  & \nodata  & \nodata & \nodata &  \\
DDO 215   & 39.12$\pm$ 0.04 & -2.10 & -3.05 & -2.72 & 10.83 &  1.60 &  1.60 & 4.09 &  1.79 &  5.94\\
DDO 216   & 36.86$\pm$ 0.03 & -4.36 & -4.08 & -4.15 & 10.38 &  1.31 &  1.31 & 0.85 &  0.33 &  0.78\\
DDO 220   & 39.39$\pm$ 0.01 & -1.83 & -3.13 & -2.91 & 10.67 &  0.68 &  0.68 & 1.19 &  0.77 &  1.52\\
F533-1    & 39.17$\pm$ 0.05 & -2.05 & -2.77 & -2.68 & \nodata &  0.53 &  0.53 & 2.73 &  1.16 &  3.02\\
F563-V1   & 39.31$\pm$ 0.05 & -1.91 & \nodata & -3.55 & 10.97 & 0.18 & 0.18 & \nodata & \nodata & 0.82\\
F563-V2   & 40.48$\pm$ 0.04 & -0.75 & -2.96 & -2.03 & 10.36 &  0.43 &  0.43 & 1.21 &  0.84 &  3.57\\
F564-V3   &  0.00 & \nodata & \nodata & \nodata & \nodata & \nodata  & \nodata  &\nodata  &\nodata   &\nodata  \\
F565-V1   &  0.00 & \nodata & \nodata & \nodata & \nodata & \nodata  & \nodata  &\nodata  & \nodata  & \nodata \\
F565-V2   & 38.65$\pm$ 0.05 & -2.57 & \nodata & -3.44 & 11.68 & 0.32 & 0.32 & \nodata & \nodata & 3.37\\
F608-1    & 38.79$\pm$ 0.04 & -2.43 & -2.63 & -3.07 & \nodata &  0.55 &  0.55 & 3.99 &  1.13 &  2.40\\
F615-1    & 36.97$\pm$ 0.33 & -4.25 & \nodata & -4.76 & \nodata &  0.55 & 0.55 & \nodata &  1.20 &  2.12\\
F651-2    & 39.71$\pm$ 0.04 & -1.52 & -2.81 & -2.91 & \nodata &  0.79 &  0.73 & 2.39 &  1.08 &  2.14\\
F721-V2   & 39.14$\pm$ 0.02 & -2.09 & -2.15 & -2.51 & \nodata &  0.25 &  0.15 & 1.34 &  0.43 &  0.89\\
F750-V1   & 38.54$\pm$ 0.05 & -2.69 & -3.07 & -2.95 & \nodata &  0.35 &  0.35 & 1.97 &  0.93 &  2.28\\
IC 10     & 39.92$\pm$ 0.00 & -1.31 & \nodata & -1.31 &  9.78 & 99.99 & \nodata & \nodata & \nodata & \nodata\\
IC 1613   & 38.62$\pm$ 0.00 & -2.61 & -3.13 & -2.64 & 10.27 & 99.99 & \nodata & \nodata & \nodata & \nodata\\
IC 4662   & 40.21$\pm$ 0.00 & -1.01 & \nodata & -0.48 &  9.65 &  1.62 & 1.62 & \nodata & \nodata &  4.66\\
LGS 3     &  0.00 & \nodata & \nodata & \nodata & \nodata & \nodata  & \nodata  & \nodata  &\nodata  & \nodata \\
M81dwA    &  0.00 & \nodata & \nodata & \nodata & \nodata & \nodata  & \nodata  & \nodata  & \nodata & \nodata \\
NGC 1156  & 40.68$\pm$ 0.00 & -0.54 & -2.20 & -0.87 &  9.70 &  1.79 &  1.79 & 1.07 &  0.84 &  4.96\\
NGC 1569  & 40.74$\pm$ 0.00 & -0.49 & \nodata &  0.11 &  8.61 & 99.99 & \nodata & \nodata & \nodata & \nodata\\
NGC 2101  & 40.46$\pm$ 0.05 & -0.76 & \nodata & -1.25 & 10.16 & 0.94 & 0.94 & \nodata & \nodata & 4.20\\
NGC 2366  & 40.20$\pm$ 0.00 & -1.02 & -2.40 & -1.73 &  9.98 &  5.64 &  5.64 & 1.91 &  1.19 &  4.10\\
NGC 3413  & 39.68$\pm$ 0.01 & -1.55 & -2.69 & -1.50 & 10.16 &  0.47 &  0.47 & 0.63 &  0.45 &  2.47\\
NGC 3738  & 39.77$\pm$ 0.00 & -1.45 & -2.63 & -1.72 &  9.77 &  1.04 &  0.66 & 0.43 &  0.28 &  1.23\\
NGC 3952  & 40.77$\pm$ 0.00 & -0.45 & -2.45 & -1.34 &  9.79 &  0.85 &  0.85 & 0.99 &  0.70 &  3.56\\
NGC 4163  & 38.19$\pm$ 0.08 & -3.04 & -3.26 & -2.43 & 10.35 &  1.04 &  1.04 & 1.16 &  0.71 &  3.03\\
NGC 4214  & 40.35$\pm$ 0.00 & -0.88 & -2.30 & -1.10 &  9.76 & 99.99 & \nodata & \nodata & \nodata & \nodata\\
NGC 6822  & 39.26$\pm$ 0.00 & -1.97 & \nodata & -1.96 & 10.21 & 99.99 & \nodata & \nodata & \nodata & \nodata\\
SagDIG    & 37.43$\pm$ 0.15 & -3.79 & -3.71 & -3.02 & 10.82 &  1.58 &  1.58 & 0.98 & \nodata &  2.17\\
UGC 199   & 39.23$\pm$ 0.15 & -2.00 & -2.86 & -3.40 & 11.04 &  0.68 &  0.68 & 3.90 &  0.97 &  2.09\\
UGC 5209  & 38.04$\pm$ 0.07 & -3.18 & -3.45 & -2.99 & 10.74 &  0.30 &  0.30 & 0.90 & \nodata &  1.52\\
UGC 8011  & 38.97$\pm$ 0.03 & -2.26 & \nodata & -3.14 & 10.81 &  0.66 & 0.66 & \nodata &  0.61 &  1.43\\
UGC 8055  & 38.62$\pm$ 0.04 & -2.60 & -2.71 & -2.67 & 10.90 &  0.56 &  0.56 & 2.17 & \nodata &  2.25\\
UGC 8276  & 38.28$\pm$ 0.07 & -2.95 & -3.54 & -3.42 & 11.29 &  0.56 &  0.56 & 1.93 &  1.06 &  2.21\\
UGC 8508  & 38.45$\pm$ 0.01 & -2.77 & -2.88 & -2.12 & 10.31 &  1.04 &  0.84 & 1.00 &  0.66 &  2.39\\
UGC 10140 & 39.61$\pm$ 0.04 & -1.61 & -3.01 & -2.78 & 10.63 &  0.25 &  0.25 & 1.10 & \nodata &  1.43\\
UGC 10281 & 39.02$\pm$ 0.04 & -2.21 & -2.73 & -3.19 & 11.02 &  0.68 &  0.68 & 3.54 &  0.94 &  2.10\\
WLM       & 38.39$\pm$ 0.00 & -2.84 & -3.33 & -2.85 & 10.74 &  4.25 &  4.25 & 1.25 &  0.73 &  2.15\\
0467-074  & 40.44$\pm$ 0.03 & -0.78 & -3.06 & -2.48 & 10.30 &  0.32 &  0.32 & 1.31 & \nodata &  2.52\\
1397-049  & 39.06$\pm$ 0.05 & -2.16 & -4.01 & -3.74 & 11.79 &  0.00 &  0.00 & 0.00 &  0.00 &  0.00\\

{\bf BCD Galaxies:}&&&&&&&&&&\\
&&&&&&&&&&\\
Haro 3    & 40.91$\pm$ 0.00 & -0.31 & \nodata & -0.16 &  9.26 &  0.51 & 0.51 & \nodata & \nodata &  4.86\\
Haro 4    & 39.97$\pm$ 0.00 & -1.26 & -1.50 & -0.29 &  8.91 &  0.46 & 0.15 & 0.56 &  0.37 &  2.24\\
Haro 8    & 40.58$\pm$ 0.00 & -0.65 & -2.21 & -1.05 &  9.22 &  0.46 & 0.15 & 0.28 &  0.20 &  1.08\\
Haro 14   & 40.12$\pm$ 0.00 & -1.11 & -2.40 & -1.03 &  9.73 &  0.46 & 0.25 & 0.44 &  0.35 &  2.14\\
Haro 20   & 40.18$\pm$ 0.01 & -1.04 & -2.74 & -1.91 &  9.92 &  0.25 & 0.05 & 0.10 & \nodata &  0.26\\
Haro 23   & 40.22$\pm$ 0.00 & -1.01 & -2.33 & -1.69 &  9.13 &  0.25 & 0.15 & 0.34 &  0.19 &  0.71\\
Haro 29   & 39.77$\pm$ 0.00 & -1.46 & -1.72 & -0.82 &  9.39 &  0.56 & 0.35 & 0.72 &  0.41 &  2.04\\
Haro 36   & 39.41$\pm$ 0.01 & -1.82 & -2.87 & -1.96 & 10.11 &  0.76 & 0.66 & 0.91 & \nodata & 2.60\\
Haro 38   & 39.64$\pm$ 0.01 & -1.59 & -2.40 & -1.78 &  9.82 &  0.56 & 0.25 & 0.69 &  0.40 &  1.42\\
Haro 43   & 40.15$\pm$ 0.00 & -1.08 & -2.33 & -1.33 & 10.09 &  0.28 & 0.20 & 0.75 &  0.48 &  2.35\\
HS0822+3542    & 39.09$\pm$ 0.01 & -2.14 & -1.86 & -0.73 &  9.53 &  0.13 & 0.05 & 0.39 & \nodata &  1.43\\
Mrk 5     & 39.88$\pm$ 0.00 & -1.35 & -2.17 & -1.14 &  9.57 &  0.40 &  0.17 & 0.49 & \nodata &  1.60\\
Mrk 16    & 40.49$\pm$ 0.01 & -0.74 & -2.58 & -1.55 &  9.93 &  0.35 &  0.15 & 0.35 & \nodata &  1.14\\
Mrk 32    & 39.20$\pm$ 0.01 & -2.02 & -2.65 & -2.10 & 10.29 &  0.25 &  0.15 & 0.52 &  0.31 &  0.99\\
Mrk 67    & 39.86$\pm$ 0.05 & -1.36 & \nodata & -0.55 &  8.93 & 0.25 & 0.04 & \nodata & \nodata & 0.82\\
Mrk 178   & 39.10$\pm$ 0.00 & -2.12 & -2.31 & -1.53 &  9.25 &  1.03 &  0.91 & 1.49 &  0.90 &  3.64\\
Mrk 408   & 40.23$\pm$ 0.00 & -1.00 & -2.26 & -1.09 &  9.68 &  0.28 &  0.06 & 0.17 &  0.12 &  0.65\\
Mrk 600   & 40.15$\pm$ 0.00 & -1.08 & -2.10 & -1.02 &  9.82 &  0.51 &  0.51 & 1.27 & \nodata &  4.43\\
Mrk 757   & 39.25$\pm$ 0.01 & -1.98 & -2.45 & -1.53 &  9.51 &  0.20 &  0.08 & 0.21 & \nodata &  0.62\\
NGC 1705  & 40.15$\pm$ 0.00 & -1.07 & \nodata & -0.82 &  9.29 &  1.54 & 0.45 & \nodata & \nodata &  1.58\\
NGC 6789  & 38.55$\pm$ 0.04 & -2.68 & -2.98 & -1.91 & \nodata &  0.47 &  0.34 & 0.48 & \nodata &  1.54\\
SBS1415+437   & 39.05$\pm$ 0.01 & -2.18 & -2.92 & -2.31 & \nodata &  0.60 &  0.33 & 0.79 & \nodata &  1.60\\
UGCA 290  & 37.92$\pm$ 0.06 & -3.31 & -2.83 & -1.77 & \nodata &  0.62 &  0.43 & 1.08 &  0.79 &  3.68\\
IZw115    & 38.52$\pm$ 0.05 & -2.71 & -3.75 & -2.87 & 11.10 &  0.35 &  0.15 & 0.29 &  0.20 &  0.79\\
VIIZw403  & 39.34$\pm$ 0.00 & -1.88 & -2.19 & -1.82 &  9.86 &  0.99 &  0.68 & 1.09 &  0.61 &  1.67\\
Zw2335    & 40.20$\pm$ 0.00 & -1.02 & -2.10 & -0.89 &  9.27 &  0.15 &  0.01 & 0.04 &  0.02 &  0.14\\

{\bf Sm Galaxies:}&&&&&&&&&&\\
&&&&&&&&&&\\
DDO 18    & 39.76$\pm$ 0.05 & -1.47 & -3.30 & -3.05 & 10.78 &  0.85 &  0.85 & 1.35 &  0.89 &  1.82\\
DDO 48    & 39.64$\pm$ 0.01 & -1.59 & -2.57 & -2.95 & 10.82 &  1.61 &  1.61 & 4.77 &  1.43 &  3.10\\
DDO 54    & 39.54$\pm$ 0.04 & -1.69 & -3.36 & -2.63 & 10.58 &  0.85 &  0.85 & 1.15 &  0.76 &  2.68\\
DDO 88    & 38.90$\pm$ 0.03 & -2.33 & -3.33 & -2.60 & 10.42 &  1.04 &  1.04 & 1.25 &  0.82 &  2.91\\
DDO 122   & 39.87$\pm$ 0.00 & -1.36 & -2.91 & -2.11 & 10.06 &  1.42 &  1.23 & 0.98 &  0.74 &  2.45\\
DDO 135   & 39.18$\pm$ 0.00 & -2.05 & -2.78 & -1.88 &  8.41 &  1.79 &  1.79 & 1.36 &  0.95 &  3.83\\
DDO 150   & 39.67$\pm$ 0.00 & -1.55 & -2.86 & -2.09 & 10.13 &  1.42 &  1.42 & 1.38 &  0.99 &  3.34\\
DDO 173   & 40.07$\pm$ 0.01 & -1.16 & -2.93 & -2.18 & 10.35 &  1.04 &  1.04 & 1.19 &  0.84 &  2.80\\
DDO 180   & 40.68$\pm$ 0.00 & -0.54 & -2.64 & -1.66 &  9.86 &  1.04 &  1.04 & 0.90 &  0.64 &  2.80\\
DDO 204   & 40.46$\pm$ 0.00 & -0.76 & -2.62 & -1.74 &  9.97 &  1.79 &  1.79 & 1.47 &  1.20 &  4.05\\
DDO 214   & 40.71$\pm$ 0.01 & -0.51 & -3.04 & -2.36 & 10.12 &  1.23 &  1.23 & 0.98 &  0.76 &  2.14\\
DDO 217   & 40.13$\pm$ 0.00 & -1.09 & -3.07 & -2.73 & 10.44 &  3.50 &  3.50 & 1.78 &  1.20 &  2.64\\
F561-1    & 40.29$\pm$ 0.05 & -0.93 & \nodata & -2.68 & 10.42 & 0.47 & 0.47 & \nodata & \nodata & 2.33 \\
F567-2    &  0.00 & \nodata & \nodata & \nodata & \nodata & \nodata & \nodata & \nodata & \nodata &\nodata \\
F583-1    & 39.96$\pm$ 0.02 & -1.26 & -2.72 & -2.57 & 10.75 &  0.99 &  0.91 & 3.25 &  1.29 &  3.87\\
NGC 2552  & 40.06$\pm$ 0.00 & -1.16 & -2.83 & -1.98 & 10.05 &  1.42 &  1.42 & 0.92 &  0.71 &  2.43\\
NGC 3109  & 39.52$\pm$ 0.00 & -1.71 & \nodata & -2.39 & 10.49 & 99.99 & \nodata & \nodata & \nodata & \nodata\\
NGC 3510  & 40.43$\pm$ 0.00 & -0.80 & -2.36 & -1.52 & 10.11 &  1.79 &  1.79 & 1.59 &  0.99 &  4.19\\
UGC 5716  & 39.51$\pm$ 0.03 & -1.72 & -2.93 & -2.65 & 10.75 &  0.68 &  0.68 & 1.65 &  0.77 &  2.28\\
UGC 11820 & 40.07$\pm$ 0.01 & -1.15 & -2.59 & -2.74 & 10.59 &  2.01 &  1.92 & 3.95 &  1.33 &  3.33\\
\enddata
\tablenotetext{a}{Integrated star formation rate normalized to the area
of the galaxy within \protect\rtf.}
\tablenotetext{b}{Integrated star formation rate normalized to the area
of the galaxy within \protect\rd.}
\tablenotetext{c}{Time scale to exhaust the total current gas content
of the galaxy at
the current star formation rate.}
\tablenotetext{d}{Semi-major axis of largest ellipse that contains
\protect\ha\ emission. An entry of 99.99 means that the \protect\ha\
emission exceeded the size of the region that we imaged.}
\tablenotetext{e}{Semi-major axis of largest ellipse that contains the center of a
discrete \protect\HII\ region. An entry of 99.99 means that the \protect\HII\ potentially extended
beyond the region that we imaged.}
\end{deluxetable}

\clearpage


%


\begin{deluxetable}{lccccr}
\tabletypesize{\scriptsize}
\tablewidth{0pt}
\tablecaption{Other parameters.
\label{tab-params}}
\tablehead{
\colhead{}
& \colhead{\protect\rtf}
& \colhead{\protect\rd}
& \colhead{}
& \colhead{$\log {\rm M}_{gas}$\tablenotemark{a}}
& \colhead{} \\
\colhead{Galaxy}
& \colhead{(kpc)}
& \colhead{(kpc)}
& \colhead{M$_V$}
& \colhead{(M\protect\solar)}
& \colhead{Ref\tablenotemark{b}}
}
\startdata

{\bf Im Galaxies:}&&&&&\\
&&&&&\\
A1004+10  &  1.29 &  0.35 & -15.93 &  7.59 & 12\\
A2228+33  &  3.83 &  2.57 & -17.75 &  9.35 & 11\\
CVnIdwA   & \nodata &  0.65 & -12.65 &  7.92 & 25\\
D508-2    &  1.94 &  1.81 & -15.74 &  9.18 & 17\\
D575-5    & \nodata &  0.85 & -12.85 &  7.52 & 17\\
D634-3    &  0.09 &  0.13 & -10.16 &  5.98 & 17\\
D646-8    & \nodata &  2.02 & -14.69 &  7.79 & 17\\
DDO 9     &  3.87 &  3.11 & -17.46 &  9.29 &  7\\
DDO 22    &  1.61 &  1.09 & -15.44 &  8.54 &  7\\
DDO 24    &  3.35 &  1.29 & -17.09 &  9.13 &  7\\
DDO 25    &  3.72 &  1.44 & -17.27 &  8.84 &  7\\
DDO 26    &  3.27 &  1.68 & -16.80 &  9.43 & 11\\
DDO 27    & \nodata &  1.54 & -15.28 &  8.86 &  7\\
DDO 33    &  7.36 &  5.62 & -18.63 &  9.80 &  7\\
DDO 34    &  1.19 &  1.08 & -15.36 &  8.68 &  7\\
DDO 35    &  6.18 &  1.84 & -18.74 &  9.57 &  7\\
DDO 38    &  3.36 &  2.93 & -17.11 &  9.39 &  7\\
DDO 39    &  2.61 &  3.71 & -17.42 &  9.36 &  7\\
DDO 40    &  4.97 &  2.36 & -17.88 &  9.31 &  7\\
DDO 43    &  0.99 &  0.43 & -14.31 &  8.10 & 11\\
DDO 46    &  1.19 &  1.03 & -14.45 &  8.26 &  7\\
DDO 47    &  1.09 &  1.37 & -15.46 &  8.73 & 11\\
DDO 50    &  2.93 &  1.11 & -16.61 &  8.99 &  3\\
DDO 52    &  0.79 &  0.76 & -14.27 &  8.10 &  7\\
DDO 53    &  0.72 &  0.73 & -13.84 &  8.39 & 16\\
DDO 63    &  1.43 &  2.22 & -14.73 &  8.31 & 18\\
DDO 64    &  1.78 &  1.33 & -15.42 &  8.52 & 11\\
DDO 68    &  1.41 &  0.60 & -15.17 &  8.65 & 11\\
DDO 69    &  0.31 &  0.26 & -11.67 &  7.13 &  1\\
DDO 70    &  0.80 &  0.48 & -14.10 &  7.72 & 11\\
DDO 75    &  0.90 &  0.22 & -13.91 &  8.01 & 11\\
DDO 86    &  2.58 &  2.28 & -16.51 &  9.19 &  7\\
DDO 87    & \nodata &  1.25 & -14.67 &  8.37 & 11\\
DDO 99    &  1.17 &  0.92 & -14.88 &  8.32 & 18\\
DDO 101   &  1.81 &  1.32 & -15.75 &  7.40 & 20\\
DDO 105   &  3.90 &  3.28 & -17.70 &  9.33 &  7\\
DDO 115   &  0.61 &  0.32 & -13.28 &  7.24 &  2\\
DDO 120   &  2.23 &  1.01 & -16.33 &  8.33 &  7\\
DDO 125   & \nodata & 0.48 & -14.70 &  7.76 &  7\\
DDO 126   &  1.18 &  0.88 & -14.85 &  8.27 &  7\\
DDO 133   &  1.89 &  2.15 & -15.96 &  8.71 &  7\\
DDO 143   & \nodata &  2.05 & -15.58 &  8.82 &  2\\
DDO 154   &  1.11 &  0.69 & -14.51 &  8.62 &  4\\
DDO 155   &  0.40 &  0.15 & -12.53 &  7.13 & 11\\
DDO 165   &  1.98 &  1.95 & -15.69 &  8.32 &  7\\
DDO 167   &  0.60 &  0.33 & -12.98 &  7.39 &  7\\
DDO 168   &  1.45 &  0.66 & -15.27 &  8.48 & 11\\
DDO 169   &  1.06 &  0.78 & -14.58 &  8.13 &  7\\
DDO 171   &  4.16 &  1.95 & -17.90 &  8.82 &  7\\
DDO 183   &  1.16 &  0.86 & -14.54 &  8.02 & 11\\
DDO 185   &  1.20 &  0.86 & -14.95 &  8.12 & 21\\
DDO 187   &  0.50 &  0.20 & -12.95 &  7.48 & 11\\
DDO 210   &  0.12 &  0.17 & -10.88 &  6.52 & 14\\
DDO 215   &  1.67 &  1.15 & -15.63 &  8.73 &  7\\
DDO 216   &  0.40 &  0.44 & -13.29 &  6.02 &  7\\
DDO 220   &  2.49 &  1.94 & -16.28 &  8.84 &  7\\
F533-1    &  1.29 &  1.17 & -15.34 & \nodata & \nodata\\
F563-V1   & \nodata & 3.71 & -17.22 &  9.05 &  5\\
F563-V2   &  7.26 &  2.47 & -18.63 &  9.61 &  5\\
F564-V3   &  0.59 &  0.38 & -13.24 &  7.27 &  5\\
F565-V1   &  0.34 &  0.32 & -12.45 & \nodata & \nodata\\
F565-V2   & \nodata & 1.52 & -16.17 &  9.10 &  5\\
F608-1    &  0.71 &  1.18 & -14.86 & \nodata & \nodata\\
F615-1    & \nodata &  1.02 & -13.89 & \nodata & \nodata\\
F651-2    &  2.50 &  2.80 & -16.42 & \nodata & \nodata\\
F721-V2   &  0.61 &  0.92 & -14.34 & \nodata & \nodata\\
F750-V1   &  0.88 &  0.76 & -14.34 & \nodata & \nodata\\
IC 10     & \nodata &  0.57 & -17.11 &  8.47 & 10\\
IC 1613   &  1.03 &  0.59 & -14.60 &  7.66 & 10\\
IC 4662   & \nodata &  0.30 & -16.56 &  8.64 &  2\\
LGS 3     & \nodata &  0.20 &  -9.41 &  5.18 & 23\\
M81dwA    & \nodata &  0.37 & -11.73 &  7.26 & 19\\
NGC 1156  &  3.81 &  0.82 & -18.67 &  9.16 & 11\\
NGC 1569  & \nodata &  0.28 & -17.57 &  8.12 & 12\\
NGC 2101  & \nodata & 0.99 & -17.69 &  9.40 &  2\\
NGC 2366  &  2.76 &  1.28 & -16.66 &  8.96 &  7\\
NGC 3413  &  2.10 &  0.54 & -17.16 &  8.61 &  2\\
NGC 3738  &  2.18 &  0.77 & -17.12 &  8.32 & 11\\
NGC 3952  &  5.61 &  1.56 & -18.91 &  9.34 &  2\\
NGC 4163  &  0.73 &  0.28 & -14.37 &  7.31 &  2\\
NGC 4214  &  2.90 &  0.72 & -17.56 &  8.88 &  2\\
NGC 6822  & \nodata &  0.56 & -15.12 &  8.24 &  6\\
SagDIG    &  0.52 &  0.23 & -12.45 &  7.03 & 14\\
UGC 199   &  1.52 &  2.85 & -16.48 &  9.04 & 24\\
UGC 5209  &  0.77 &  0.46 & -13.76 &  7.56 &  2\\
UGC 8011  & \nodata &  1.57 & -15.56 &  8.55 &  2\\
UGC 8055  &  0.64 &  0.61 & -13.62 &  8.30 &  2\\
UGC 8276  &  1.12 &  0.98 & -14.57 &  8.34 &  2\\
UGC 8508  &  0.64 &  0.27 & -13.59 &  7.54 &  7\\
UGC 10140 &  2.81 &  2.17 & -16.80 &  9.02 & 24\\
UGC 10281 &  1.03 &  1.74 & -15.64 &  8.81 & 24\\
WLM       &  0.99 &  0.58 & -14.39 &  7.90 & 10\\
0467-074  &  7.72 &  4.00 & -18.69 &  9.52 &  9\\
1397-049  &  4.75 &  3.47 & -17.73 &  9.63 &  9\\

{\bf BCD Galaxies:}&&&&&\\
&&&&&\\
Haro 3    & \nodata &  0.48 & -18.23 &  8.95 & 12\\
Haro 4    &  0.74 &  0.19 & -14.80 &  7.65 &  8\\
Haro 8    &  3.42 &  0.90 & -17.91 &  8.57 &  8\\
Haro 14   &  2.49 &  0.52 & -17.77 &  8.62 &  8\\
Haro 20   &  3.99 &  1.53 & -18.32 &  8.88 &  8\\
Haro 23   &  2.58 &  1.24 & -17.72 &  8.12 &  8\\
Haro 29   &  0.76 &  0.27 & -14.47 &  7.93 &  8\\
Haro 36   &  1.90 &  0.67 & -15.84 &  8.29 &  8\\
Haro 38   &  1.45 &  0.70 & -15.22 &  8.23 &  8\\
Haro 43   &  2.39 &  0.76 & -17.18 &  9.01 &  8\\
HS0822    &  0.41 &  0.11 & -13.15 &  7.39 & 13\\
Mrk 5     &  1.45 &  0.44 & -16.02 &  8.22 & 22\\
Mrk 16    &  4.72 &  1.44 & -18.74 &  9.19 & 22\\
Mrk 32    &  1.16 &  0.61 & -15.43 &  8.27 & 22\\
Mrk 67    & \nodata & 2.61 &  \nodata &  7.57 & 22\\
Mrk 178   &  0.70 &  0.28 & -14.11 &  7.13 & 22\\
Mrk 408   &  2.43 &  0.63 & -18.02 &  8.68 & 22\\
Mrk 600   &  1.83 &  0.53 & -16.30 &  8.74 & 22\\
Mrk 757   &  0.97 &  0.34 & -15.46 &  7.53 & 22\\
NGC 1705  & \nodata &  0.42 & -16.25 &  8.22 & 15\\
NGC 6789  &  0.80 &  0.23 & -14.77 & \nodata & \nodata\\
SBS1415   &  1.32 &  0.66 & -15.01 & \nodata & \nodata\\
UGCA 290  &  0.33 &  0.10 & -11.73 & \nodata & \nodata\\
IZw115    &  1.88 &  0.68 & -16.01 &  8.39 &  8\\
VIIZw403  &  0.80 &  0.52 & -14.27 &  7.98 & 22\\
Zw2335    &  1.95 &  0.49 & -17.47 &  8.25 &  8\\

{\bf Sm Galaxies:}&&&&&\\
&&&&&\\
DDO 18    &  4.66 &  3.47 & -17.52 &  9.31 &  7\\
DDO 48    &  1.75 &  2.70 & -16.67 &  9.23 &  7\\
DDO 54    &  3.89 &  1.67 & -17.39 &  8.89 &  7\\
DDO 88    &  1.80 &  0.77 & -15.87 &  8.09 &  7\\
DDO 122   &  3.37 &  1.34 & -17.15 &  8.70 &  7\\
DDO 135   &  1.31 &  0.47 & -15.06 &  6.36 &  7\\
DDO 150   &  2.54 &  1.05 & -16.31 &  8.58 &  7\\
DDO 173   &  4.32 &  1.84 & -17.56 &  9.19 &  7\\
DDO 180   &  6.33 &  2.04 & -19.19 &  9.32 &  7\\
DDO 204   &  4.80 &  1.75 & -17.58 &  9.21 & 11\\
DDO 214   & 10.33 &  4.75 & -19.48 &  9.61 &  7\\
DDO 217   &  5.52 &  3.71 & -17.94 &  9.35 &  7\\
F561-1    & \nodata & 4.21 & -18.66 &  9.49 &  5\\
F567-2    & \nodata & \nodata & -17.40 &  9.67 &  5\\
F583-1    &  3.03 &  2.54 & -17.17 &  9.49 &  5\\
NGC 2552  &  3.84 &  1.45 & -17.34 &  8.89 & 11\\
NGC 3109  & \nodata &  1.24 &  \nodata &  8.79 &  7\\
NGC 3510  &  3.44 &  1.30 & -17.27 &  9.31 & 11\\
UGC 5716  &  2.28 &  1.65 & -16.30 &  9.03 &  2\\
UGC 11820 &  2.96 &  3.51 & -17.56 &  9.44 &  2\\
\enddata
\tablenotetext{a}{Total galactic atomic gas mass. The HI mass has been
multiplied by 1.34 to account for He.}
\tablenotetext{b}{Reference from which the HI mass was taken.
Masses were modified to reflect the distances used here as necessary.}
\tablerefs{
(1) Allsopp 1978;
(2) Bottinelli \et\ 1990;
(3) Bureau \& Carignan 2002;
(4) Carignan \& Beaulieu 1989;
(5) de Blok, McGaugh, \& van der Hulst 1996;
(6) de Blok \& Walter 2000;
(7) Fisher \& Tully 1981;
(8) Gordon \& Gottesman 1981;
(9) Huchtmeier, Hopp, \& Kuhn 1997;
(10) Huchtmeier, Seiradakis, \& Materne 1981;
(11) Hunter \& Gallagher 1985a;
(12) Hunter, Gallagher, \& Rautenkranz 1982;
(13) Kniazen \et\ 2000;
(14) Lo, Sargent, \& Young 1993;
(15) Meurer, Staveley-Smith, \& Killeen 1998;
(16) Nordgren \et, private communication;
(17) Pildis, Schombert, \& Eder 1997;
(18) RC3;
(19) Sargent, Sancisi, \& Lo 1983;
(20) Stil \& Israel 2002;
(21) Swaters 1999;
(22) Thuan \& Martin 1981;
(23) Young \& Lo 1997;
(24) van Zee, Haynes, \& Giovanelli 1995;
and
(25) van Zee \et\ 1997.
}
\end{deluxetable}

\clearpage

\begin{deluxetable}{lccccccc}
\tabletypesize{\scriptsize}
\tablecaption{Spectrophometry of HII regions. \label{tab-hii}}
\tablewidth{0pt}
\tablehead{
\colhead{HII Region}
& \colhead{Aperture\tablenotemark{a}}
& \colhead{[OII]$\lambda$3727\tablenotemark{b}}
& \colhead{[OIII]$\lambda$5007\tablenotemark{b}}
& \colhead{H$\beta$\tablenotemark{b}}
& \colhead{H$\alpha$\tablenotemark{b}}
& \colhead{[NII]$\lambda$6584\tablenotemark{b}}
& \colhead{[SII]$\lambda$6717,6731\tablenotemark{b}}
}
\startdata
NGC 595 & 13.5 & 5.95 & 4.37 & 3.47 & 18.31 & 2.01 & 1.54 \\
        & 22.0 & 16.22 & 11.22 & 8.69 & \nodata & \nodata & \nodata \\
NGC 604 & 13.5 & 16.36 & 23.33 & 10.44 & 53.69 & 6.05 & 6.21 \\
        & 22.0 & 51.19 & 50.31 & 25.27 & \nodata & \nodata & \nodata \\
NGC 2363 & 13.5 & 8.60 & 85.04 & 12.06 & 42.30 & 0.75 & 1.33 \\
NGC 5461 & 13.5 & 4.41 & \nodata & 3.36 & 22.42 & 3.46 & 2.69 \\
NGC 5462 & 13.5 & \nodata & \nodata & \nodata & 4.25 & 0.37 & 1.04 \\
\enddata
\tablenotetext{a}{Aperture diameter in arcseconds.}
\tablenotetext{b}{Integrated emission in units of
$10^{-13}$ ergs cm$^{-2}$ s$^{-1}$.}
\end{deluxetable}

\clearpage

\begin{figure}
\epsscale{0.9}
\plotone{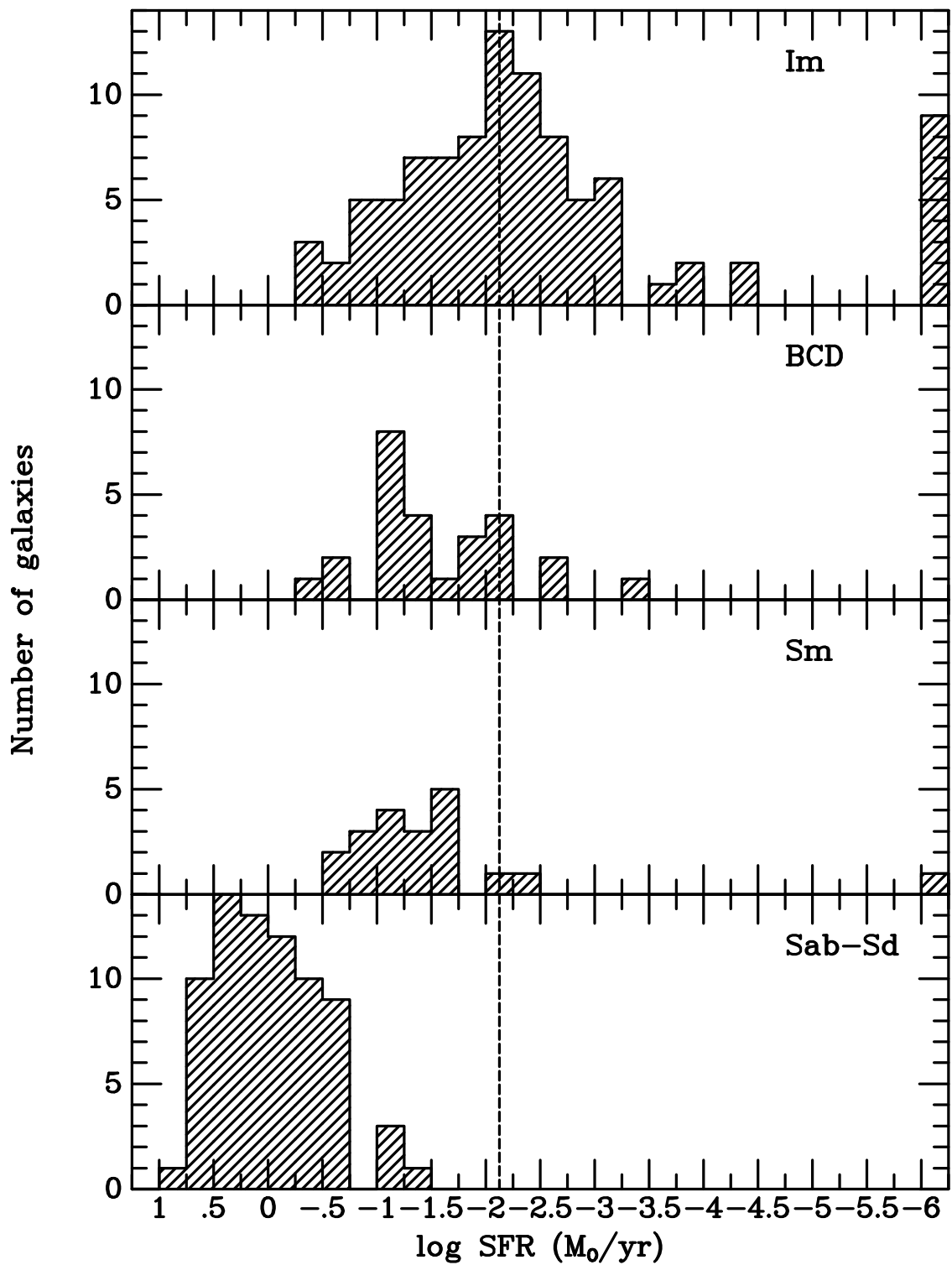}
\caption{
Number distribution of the survey galaxies and comparison
spirals in integrated star formation rate.
Galaxies at \protect\logsfr$\leq-6$ are those with zero
star formation rates.
The vertical dashed line marks the median value of \protect\logsfr\
for the Im galaxies.
\label{fig-sfr}}
\end{figure}

\begin{figure}
\plotone{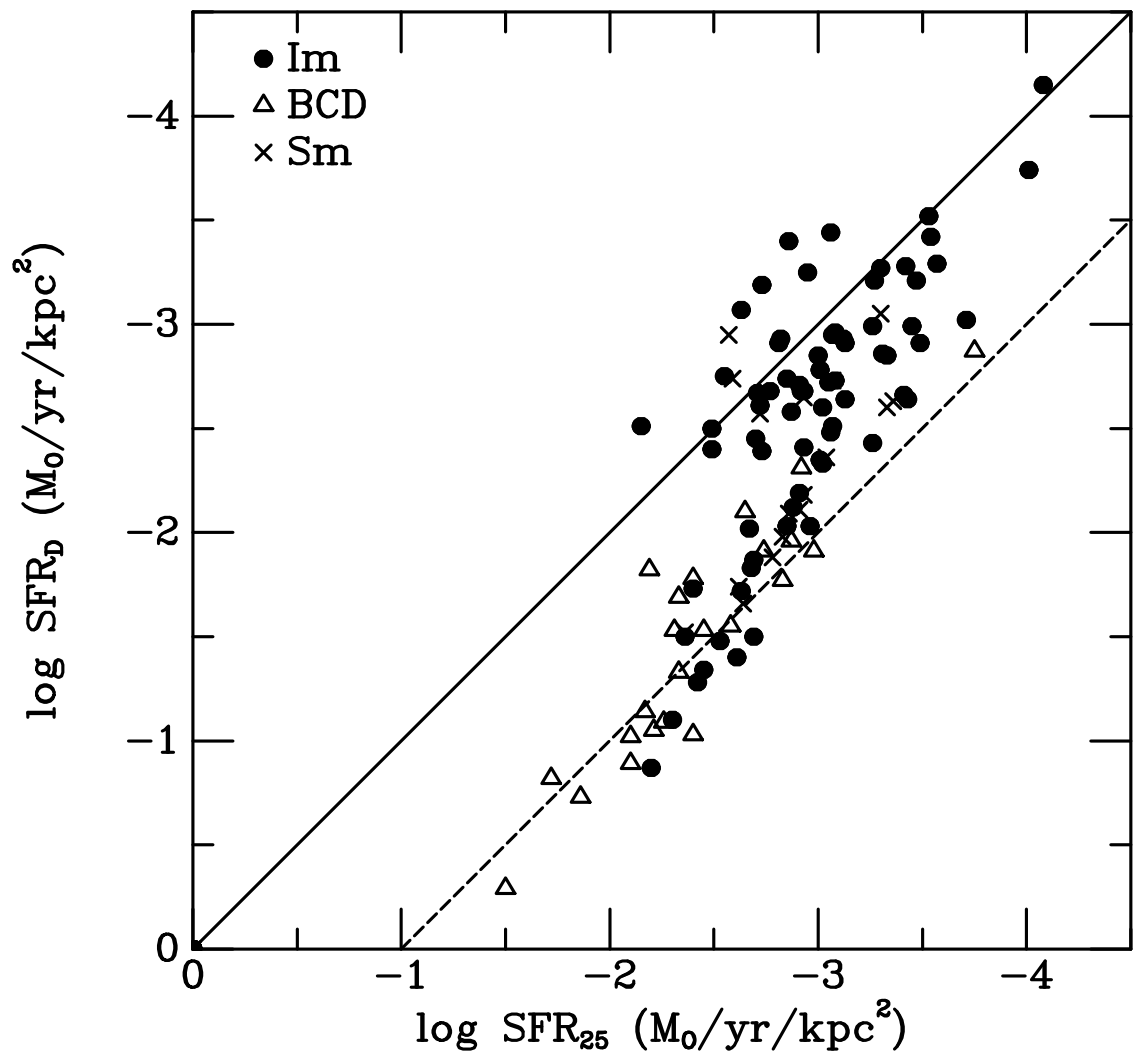}
\caption{ Comparison of integrated star
formation rates \protect\sfr\ normalized to the area of the galaxy
determined from \protect\rtf\ and from \protect\rd. The solid line
deliniates equal quantities. The dashed line is for
\protect\sfrd$=10\times$\protect\sfrtf, which corresponds to
R$_{25}=\sqrt{10}$R$_D$. The area of the galaxy is computed from
$\pi$R$^2$. \label{fig-comparesfr}}
\end{figure}

\begin{figure}
\epsscale{0.9}
\plotone{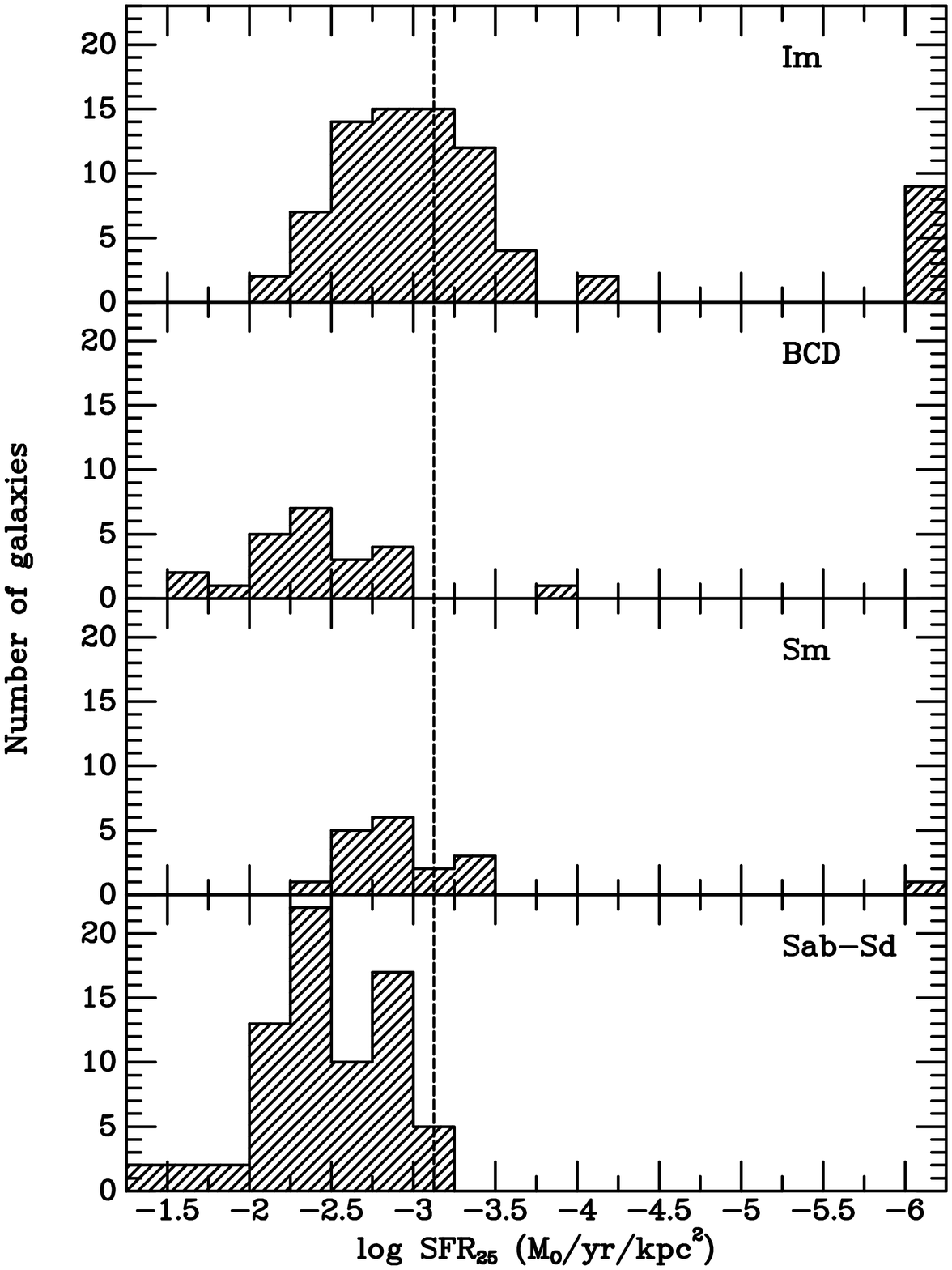}
\caption{
Number distribution of the survey galaxies and comparison
spirals in integrated star formation rate normalized to the
size of the galaxy within \protect\rtf.
Galaxies at \protect\logsfrtf$\leq-6$ are those with zero
star formation rates.
The vertical dashed line marks the median star formation rate of
an Im galaxy.
\label{fig-sfr25}}
\end{figure}

\begin{figure}
\plotone{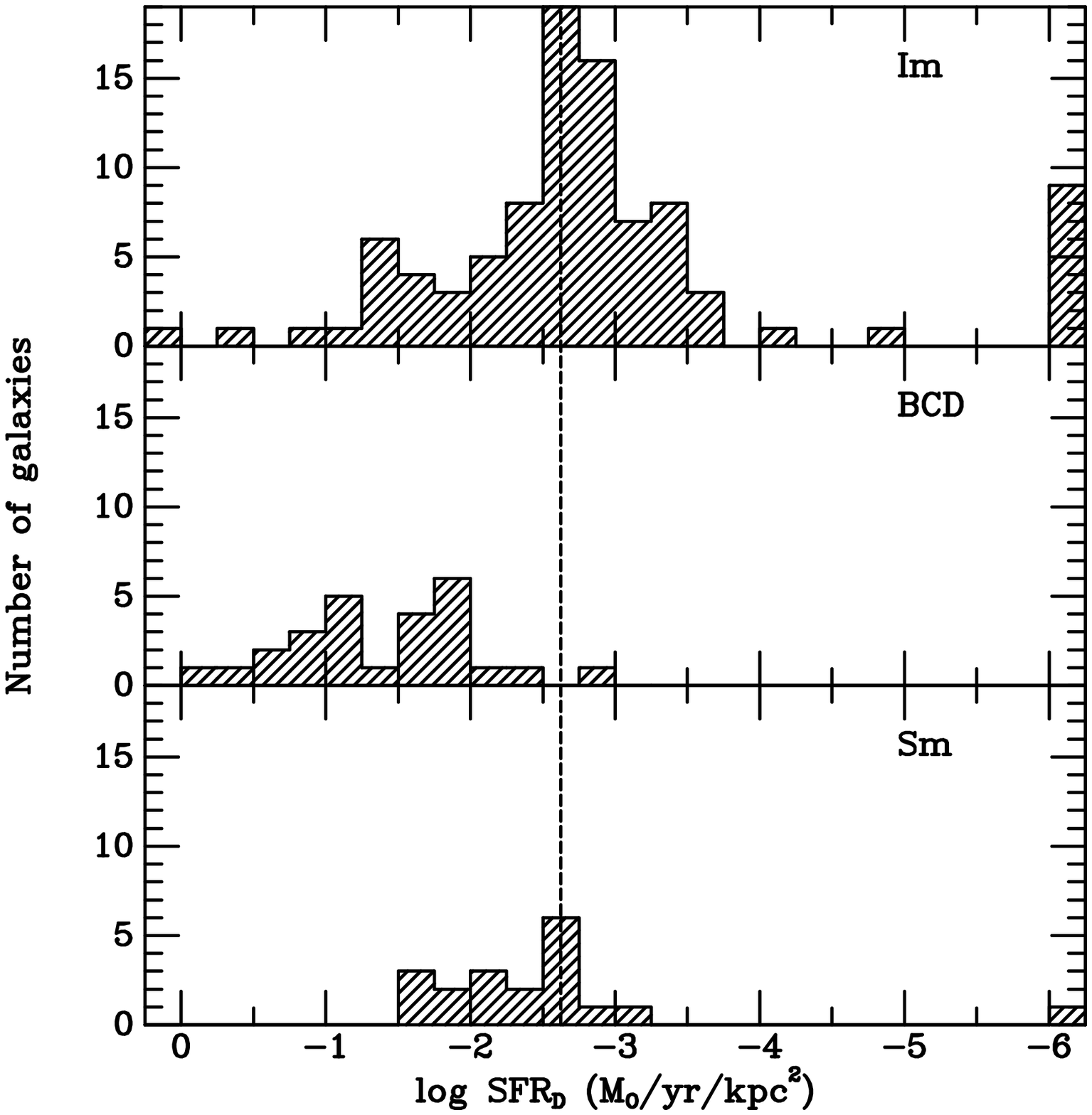}
\caption{
Number distribution of the survey galaxies and comparison
spirals in integrated star formation rate normalized to the
size of the galaxy within the scale-length \protect\rd.
Galaxies at \protect\logsfrtf$\leq-6$ are those with zero
star formation rates.
The vertical dashed line marks the median \protect\logsfrd\ for an Im
galaxy.
\label{fig-sfrd}}
\end{figure}

\begin{figure}
\epsscale{0.9}
\plotone{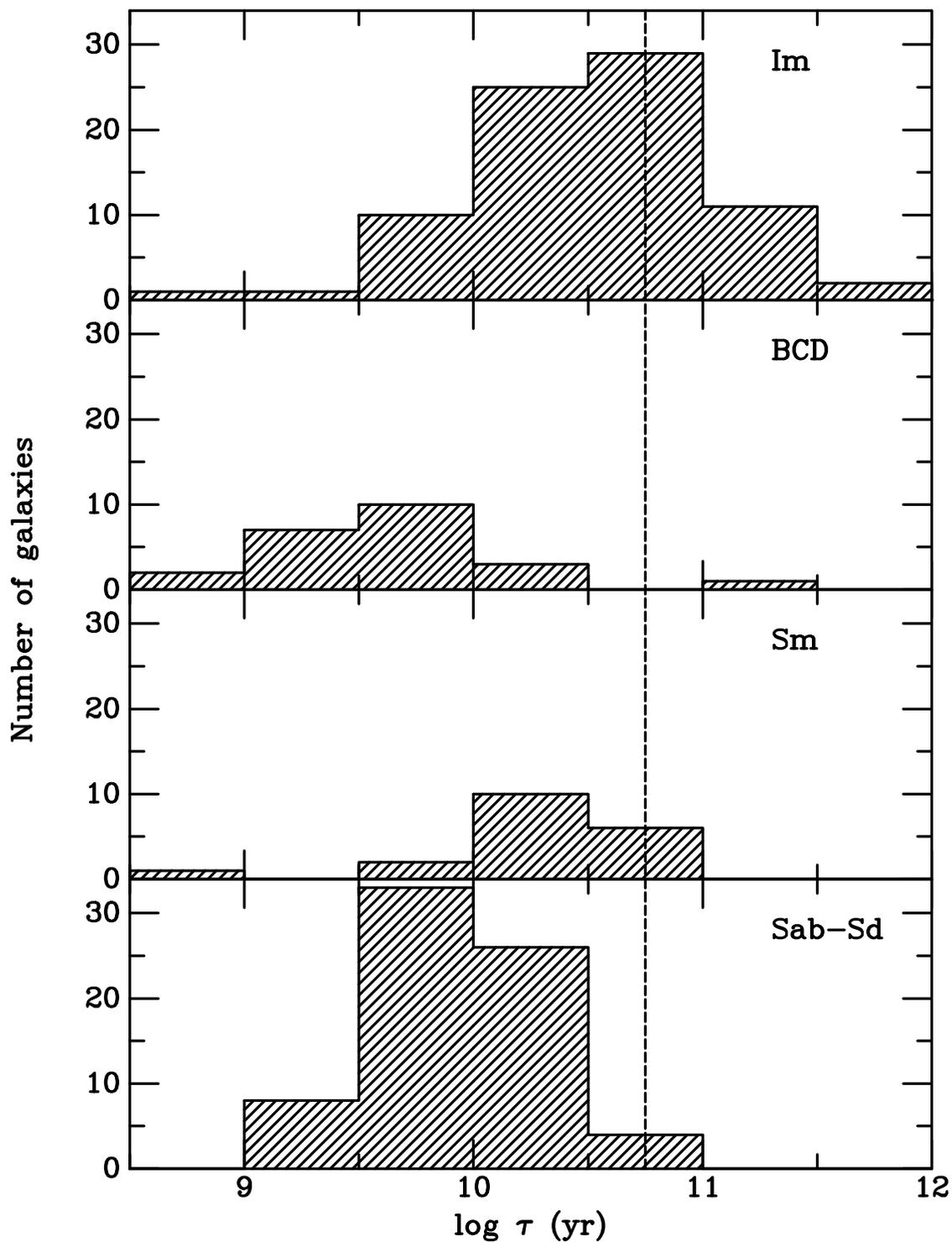}
\caption{
Number distribution of the survey galaxies and comparison
spirals in timescale $\tau$ to exhaust the current gas supply
at the current star formation rate. The gas supply is taken
as the total gas associated with the galaxy, including HI and He.
The vertical dashed line marks the median value of an Im galaxy.
\label{fig-tau}}
\end{figure}

\begin{figure}
\plotone{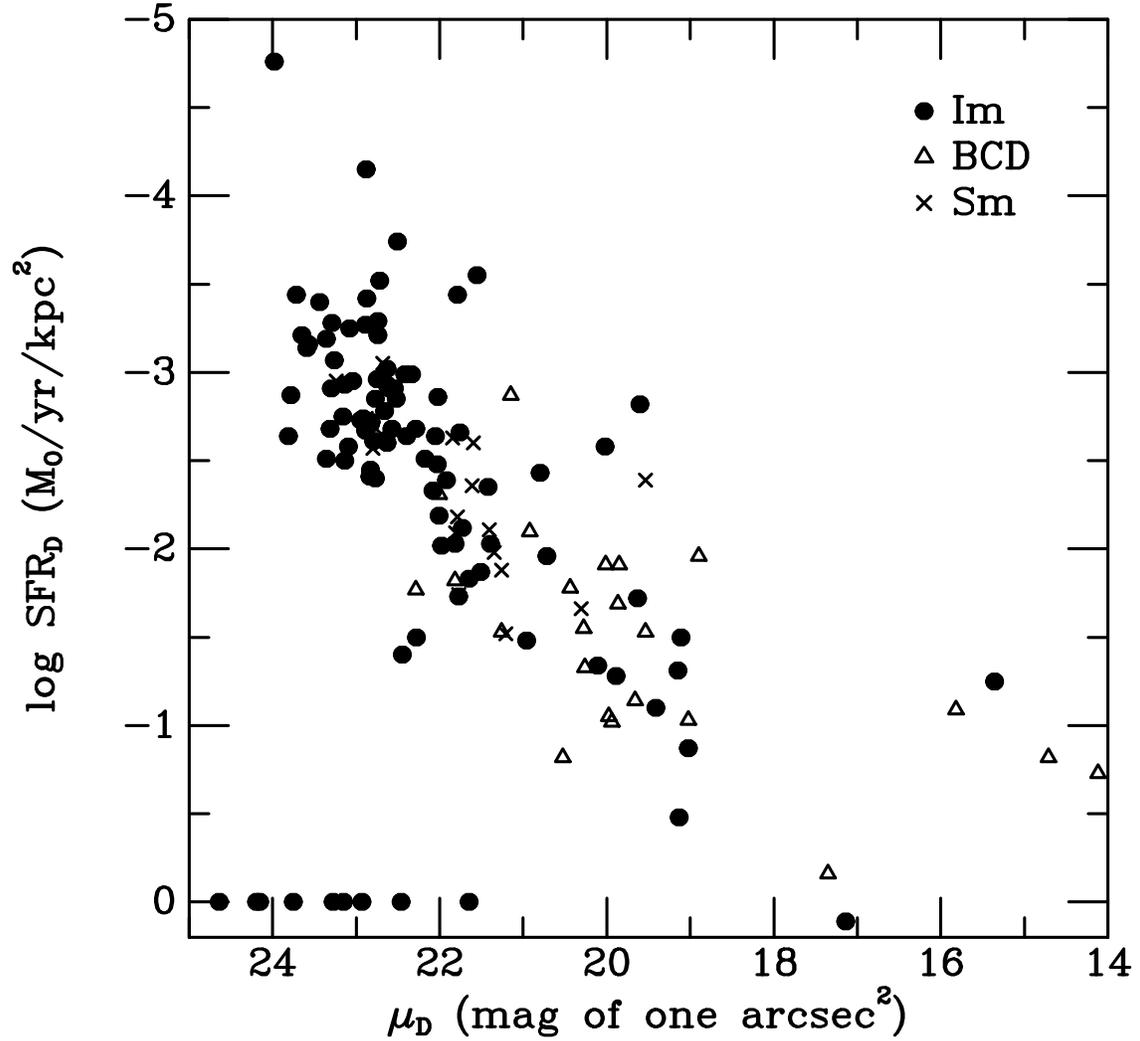}
\caption{
Normalized star-formation rate \protect\sfrd\ plotted against
the average V-band surface brightness within one scale length $\mu_D$.
Galaxies with star formation rates of zero are plotted along the
x axis at a log of 0.
The Im, BCD, and Sm samples
are shown with different symbols.
\label{fig-sfrsb}}
\end{figure}

\begin{figure}
\plotone{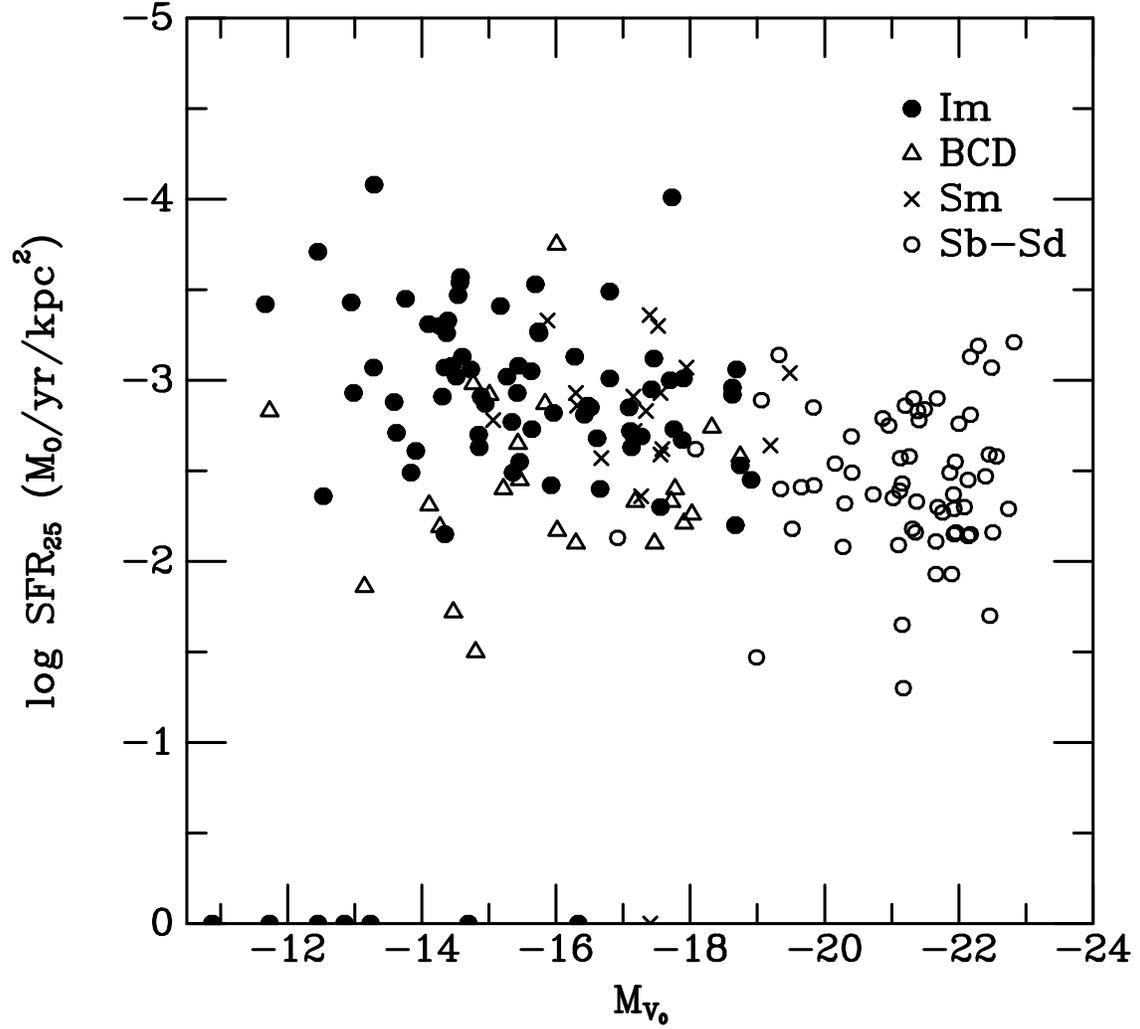}
\caption{
Normalized star-formation rate \protect\sfrtf\ plotted against
reddening-corrected galactic M$_V$. The Im, BCD, and Sm samples
are shown with different symbols. The Sb-Sd galaxies are taken
from the literature. Galaxies plotted at \protect\logsfrd$=0$
have a star formation rate of 0.
\label{fig-sfrmv}}
\end{figure}

\begin{figure}
\plotone{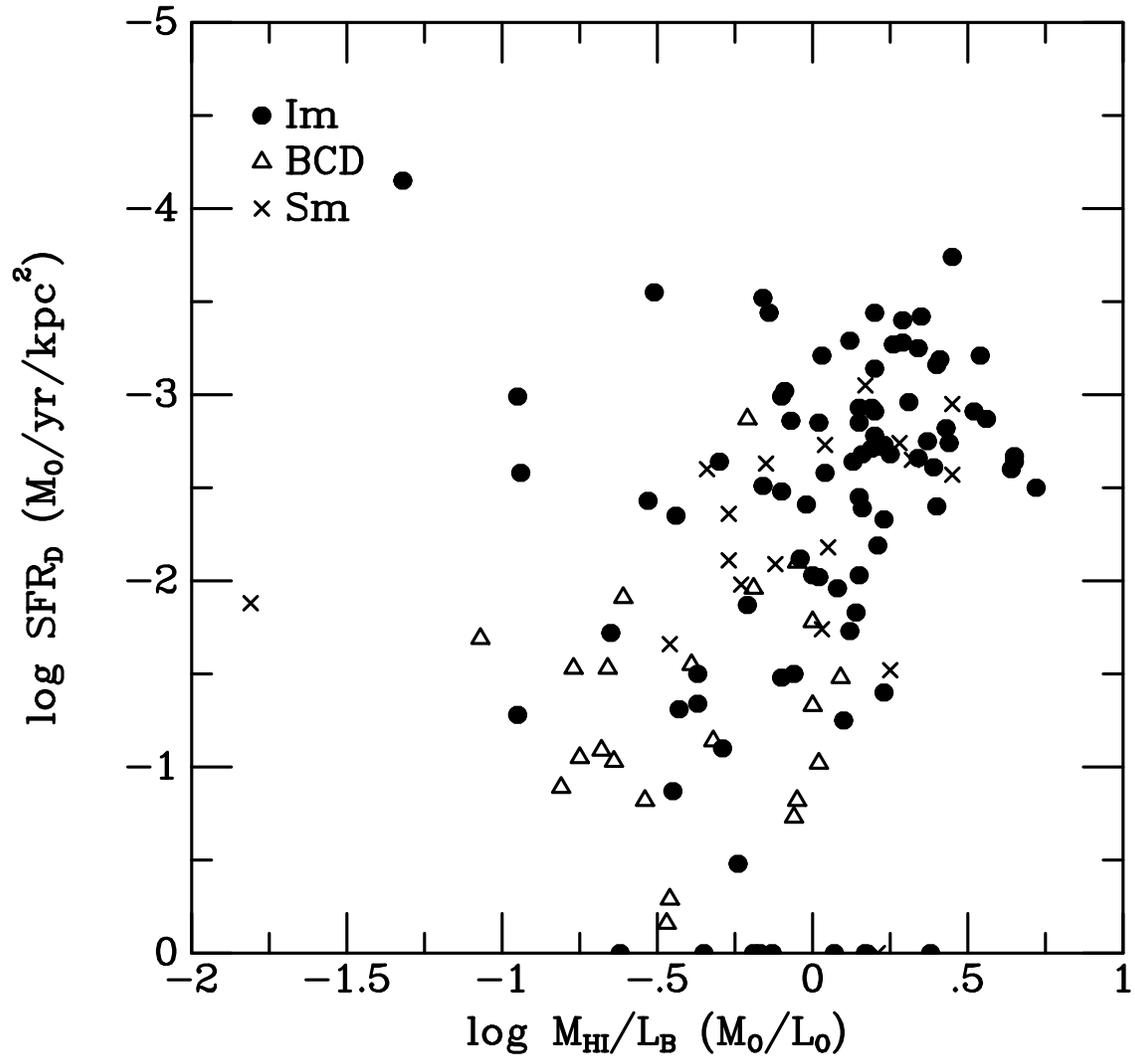}
\caption{
Normalized star-formation rate \protect\sfrd\ plotted against
galactic \protect\mhilb. The Im, BCD, and Sm samples
are shown with different symbols.
Galaxies plotted at \protect\logsfrd$=0$ have a star formation
rate of 0.
\label{fig-sfrmhilb}}
\end{figure}

\begin{figure}
\plotone{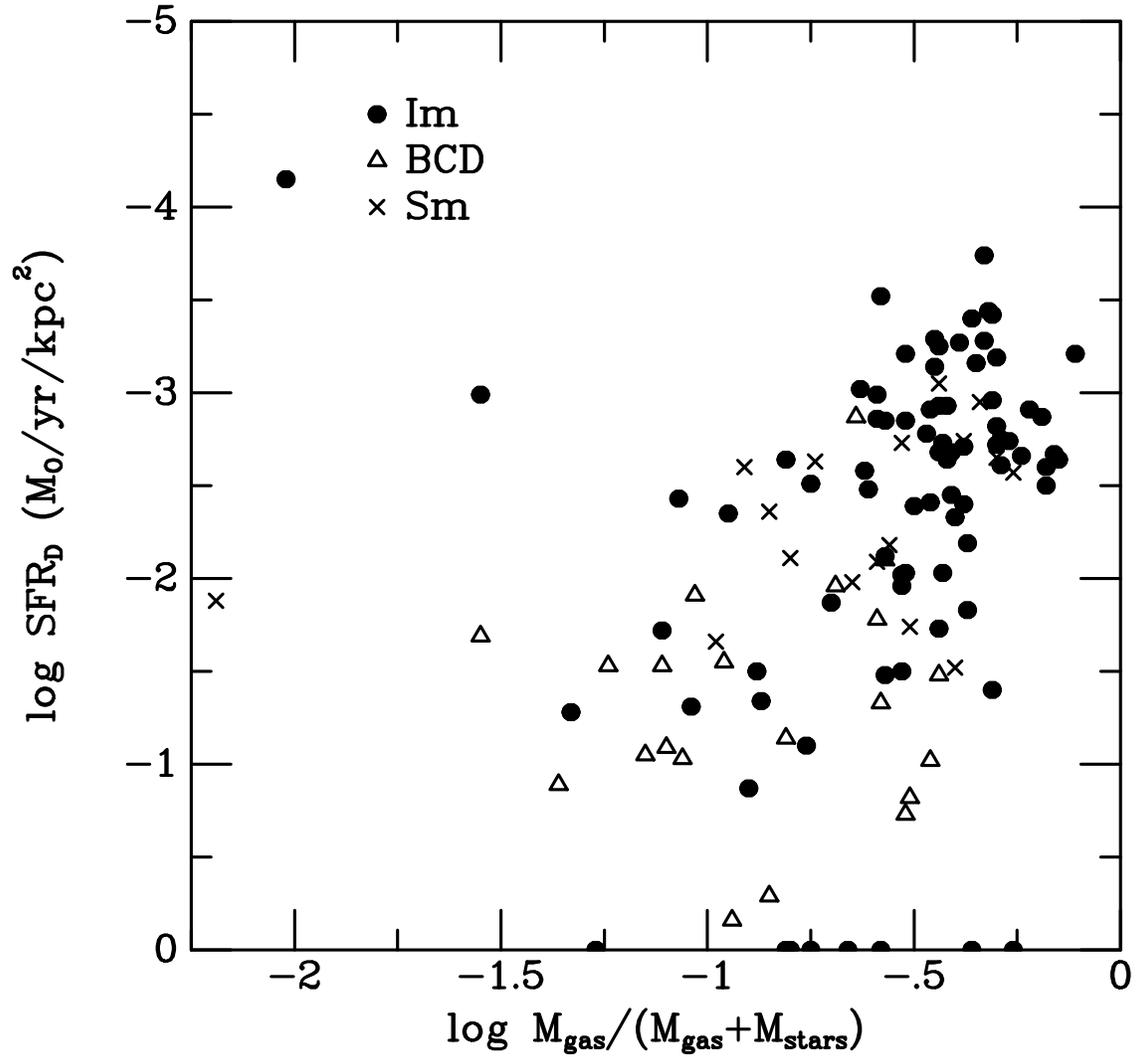}
\caption{
Normalized star-formation rate \protect\sfrd\ plotted against
the gas fraction of the galaxy.
The gas mass includes \protect\HI\ and He.
The Im, BCD, and Sm samples
are shown with different symbols.
Galaxies without any current star formation are plotted at \protect\logsfrd$=0$.
\label{fig-sfrgasfrac}}
\end{figure}

\begin{figure}
\plotone{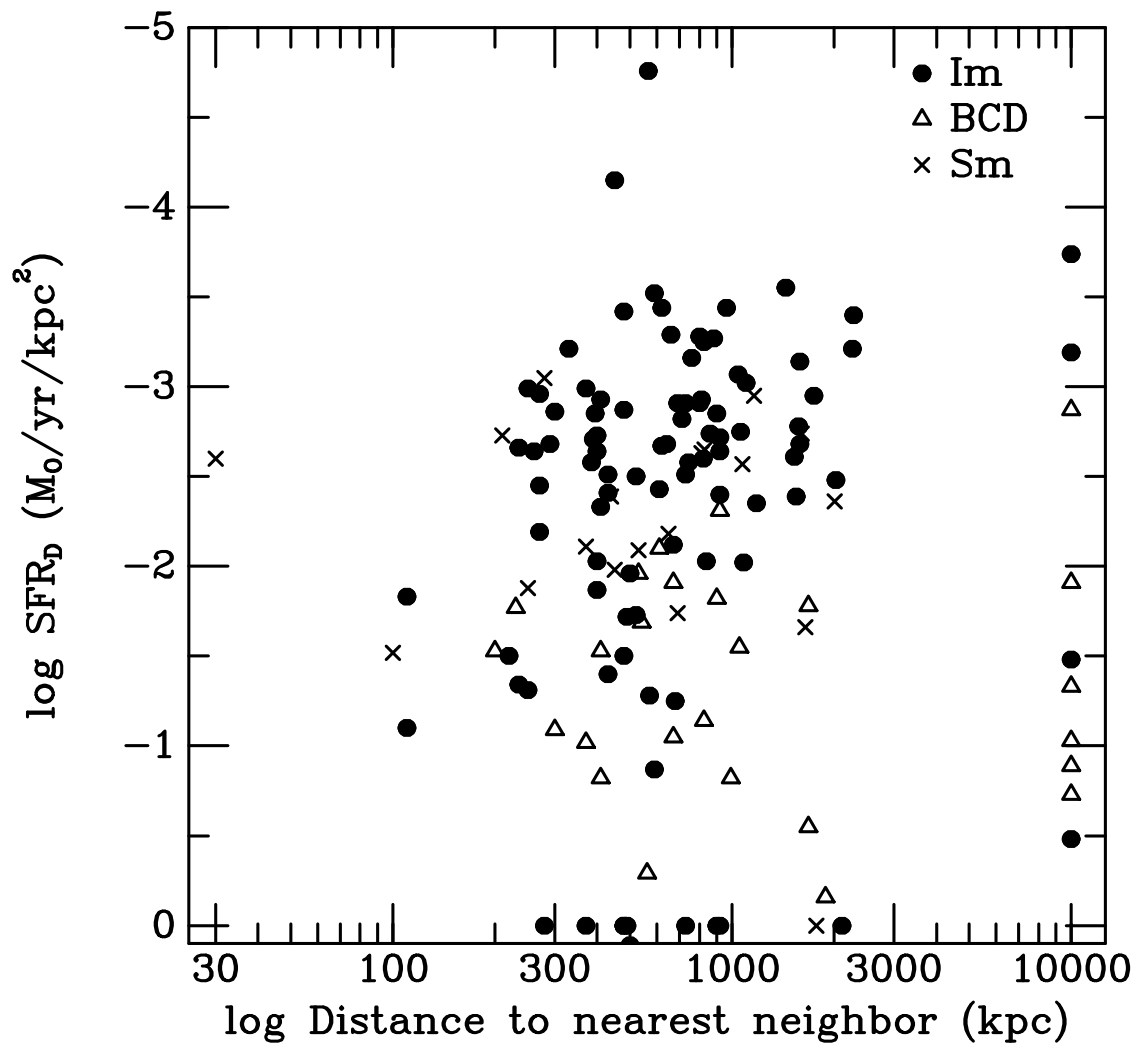}
\caption{
Normalized star-formation rate \protect\sfrd\ plotted against
the log of the distance to the nearest neighboring galaxy.
Those plotted at a distance of 10000 kpc are galaxies for which
no neighbor was found within the search parameters: 1 Mpc and
150 \protect\kms.
The Im, BCD, and Sm samples
are shown with different symbols.
Galaxies plotted at \protect\logsfrd$=0$ have a star formation
rate of 0.
\label{fig-sfrdist}}
\end{figure}

\begin{figure}
\plotone{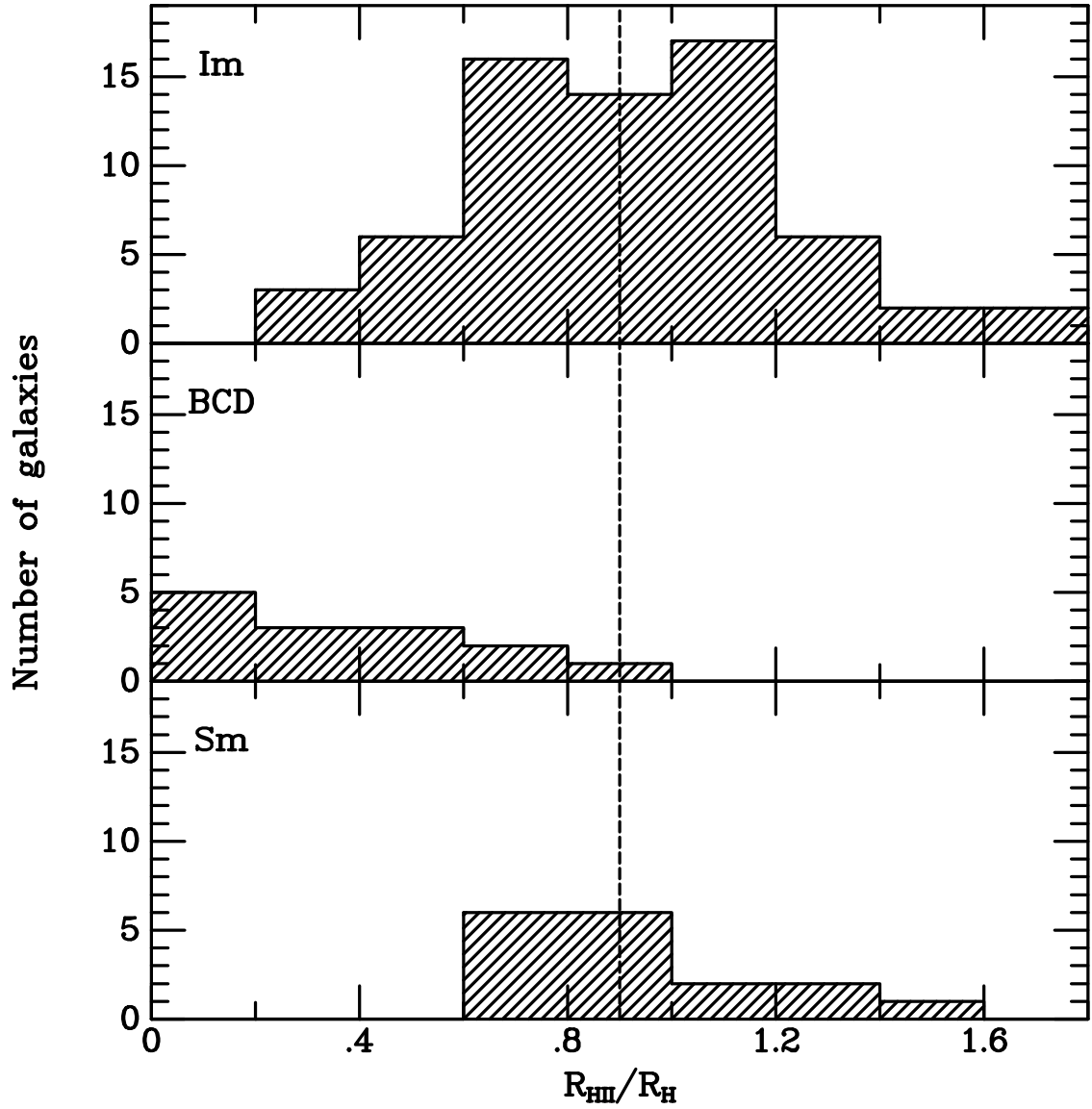}
\caption{Number distribution of the extent of discrete \protect\HII\ regions
relative to the Holmberg radius \protect\rh.
The vertical dashed line marks the median value of \protect\rhah\
for the Im galaxies.
\label{fig-histrharh}}
\end{figure}

\begin{figure}
\plotone{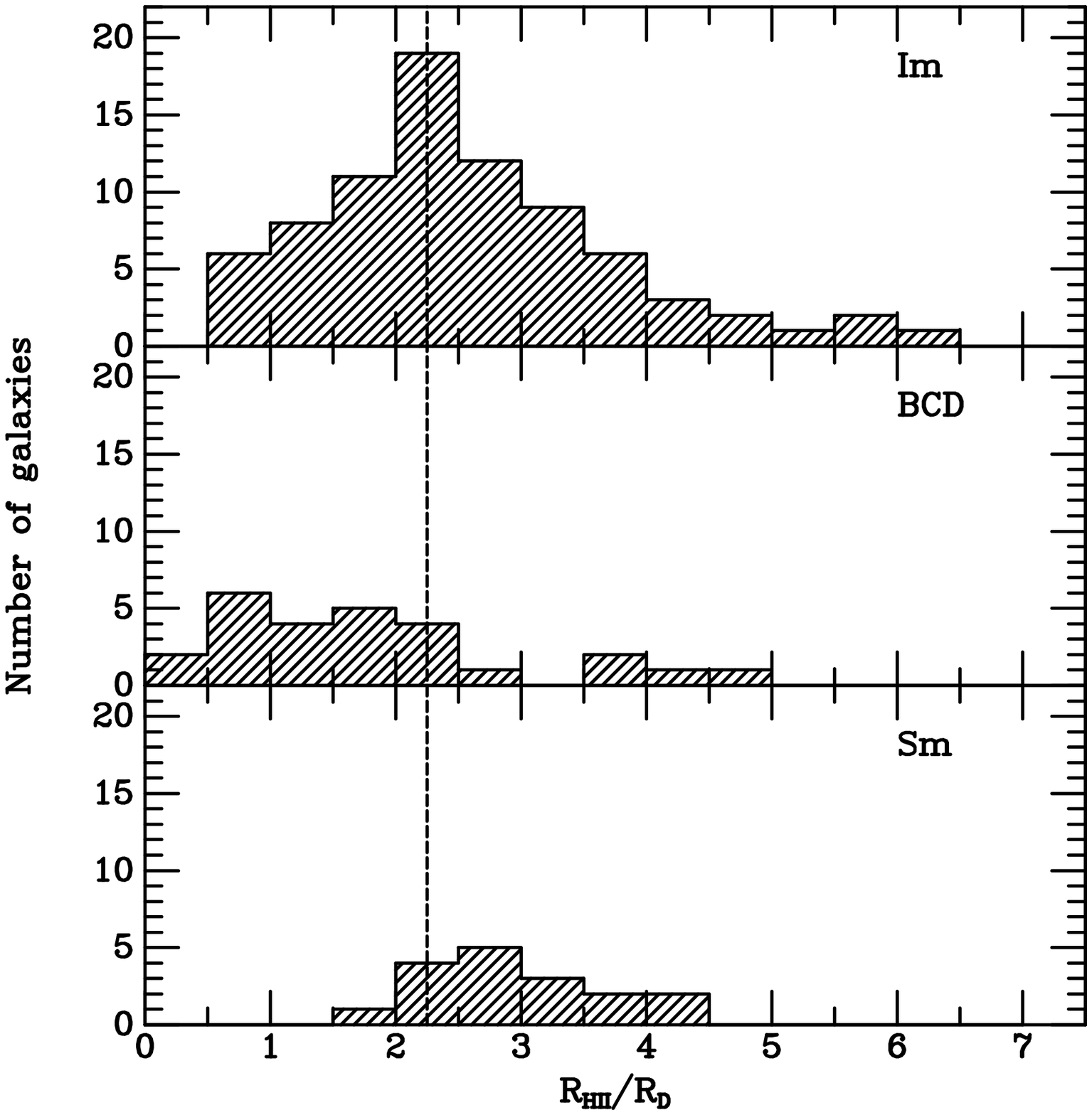}
\caption{Number distribution of the extent of discrete \protect\HII\ regions
relative to the scale length \protect\rd.
The vertical dashed line marks the median value of \protect\rhad\
for the Im galaxies.
\label{fig-histrhard}}
\end{figure}

\begin{figure}
\plotone{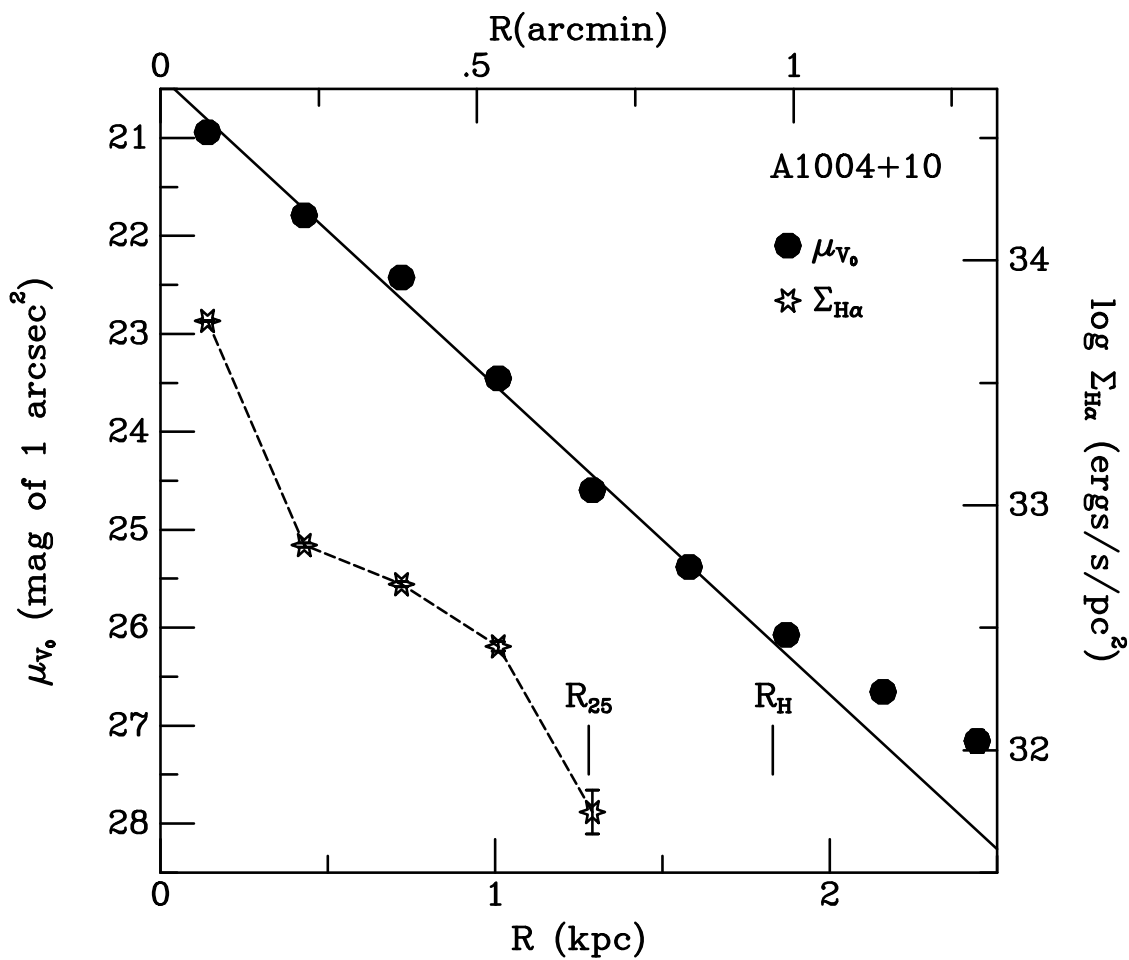} \caption{Azimuthually-averaged
H$\alpha$ and V-band surface photometry. Both are corrected for
reddening. The scales for $\Sigma_{H\alpha}$ and $\mu_{V_0}$ have
been set so that they cover the same logarithmic interval. The
solid line is a fit to the V-band surface photometry. The radii
corresponding to \protect\rtf\ and \protect\rh\ are marked with
vertical lines near the bottom of the plot. This plot is the
surface photometry for A1004$+$10, the first galaxy in our sample.
For a few galaxies V-band images were not available; these are
denoted ``off'' or ``DSS'' for use of the H$\alpha$ off-band image
or Digitized Sky Survey. The galaxy A1004$+$10 is also an example
of a good correspondence between $\mu_V$ and \protect\sigha.
\label{fig-sb}}
\end{figure}

\begin{figure}
\plotone{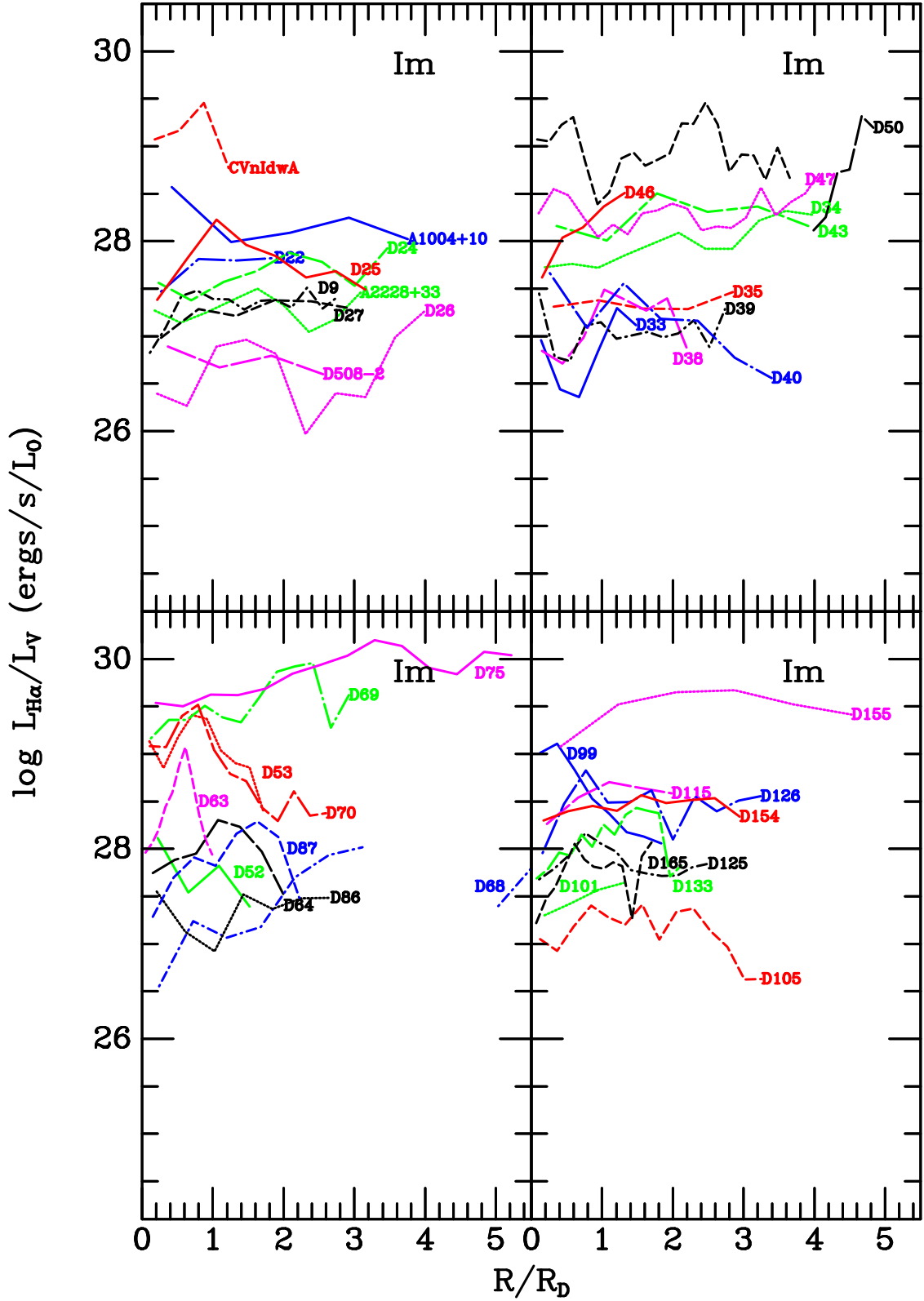}
\caption{Azimuthally-averaged ratio of L$_{H\alpha}$/L$_V$ as a
function of radius relative
to the scale-length in the
individual sample galaxies.
\label{fig-hadivv}}
\end{figure}

\clearpage
\plotone{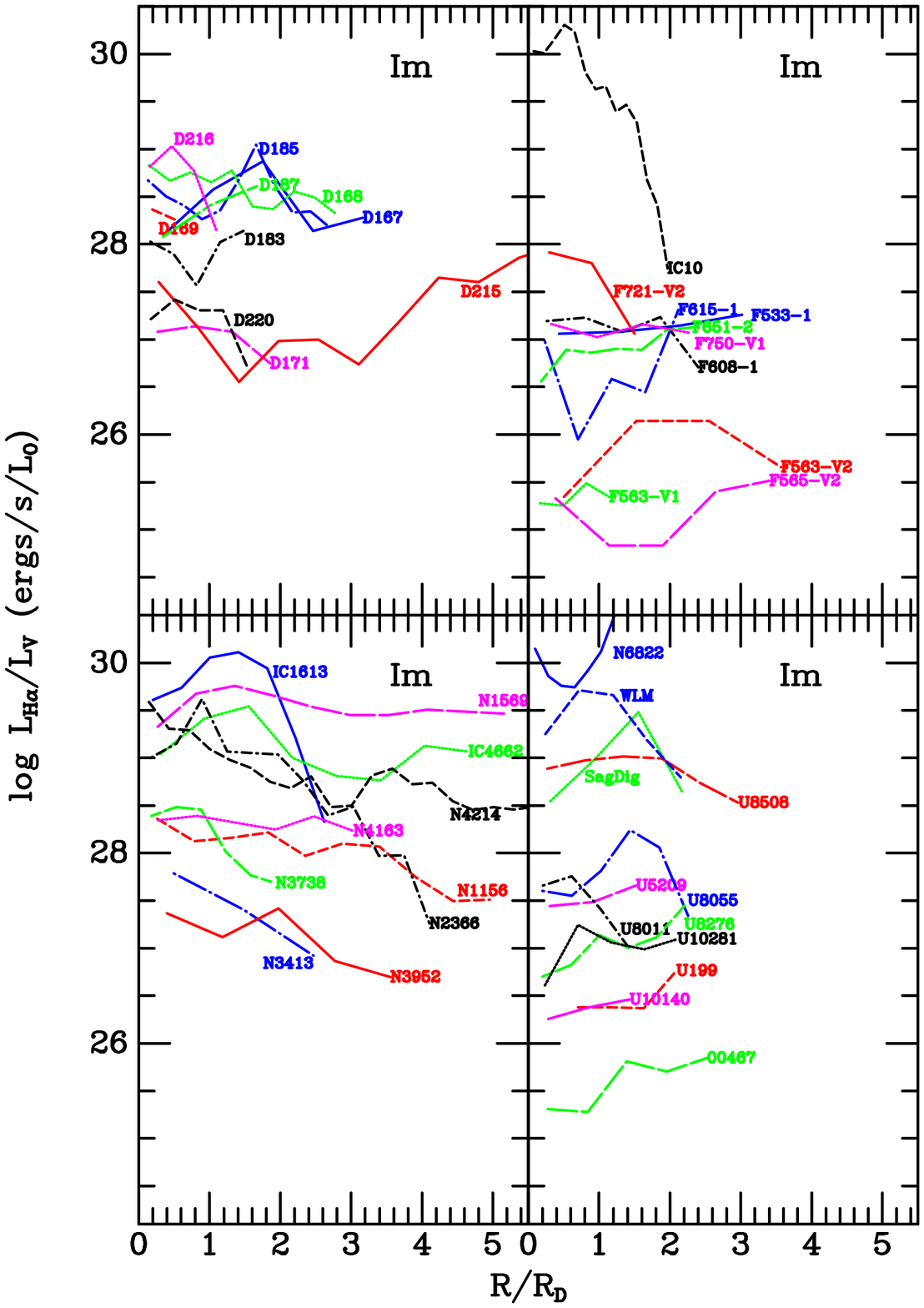}
\clearpage
\plotone{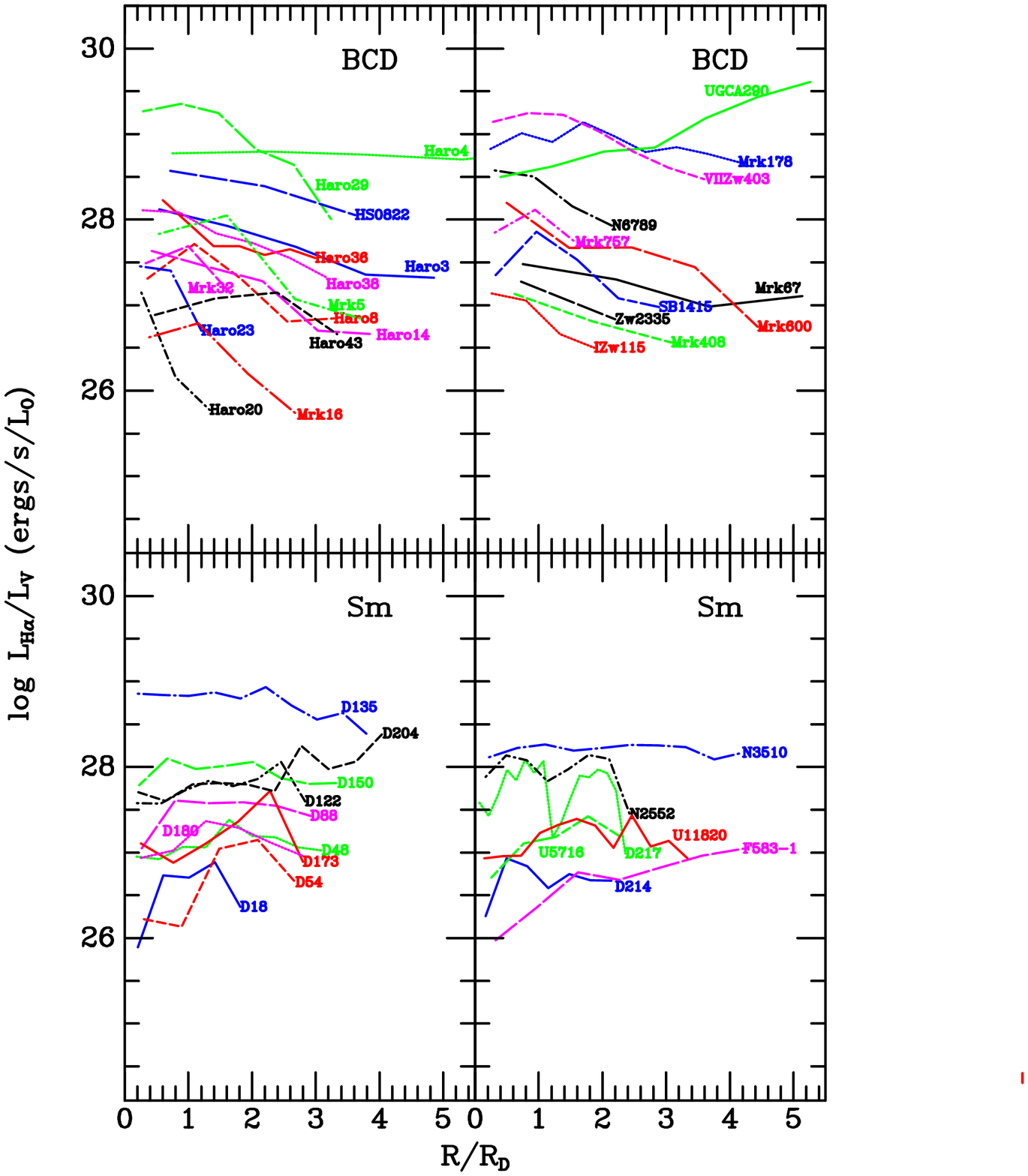}

\clearpage
\begin{figure}
\epsscale{0.6}
\plotone{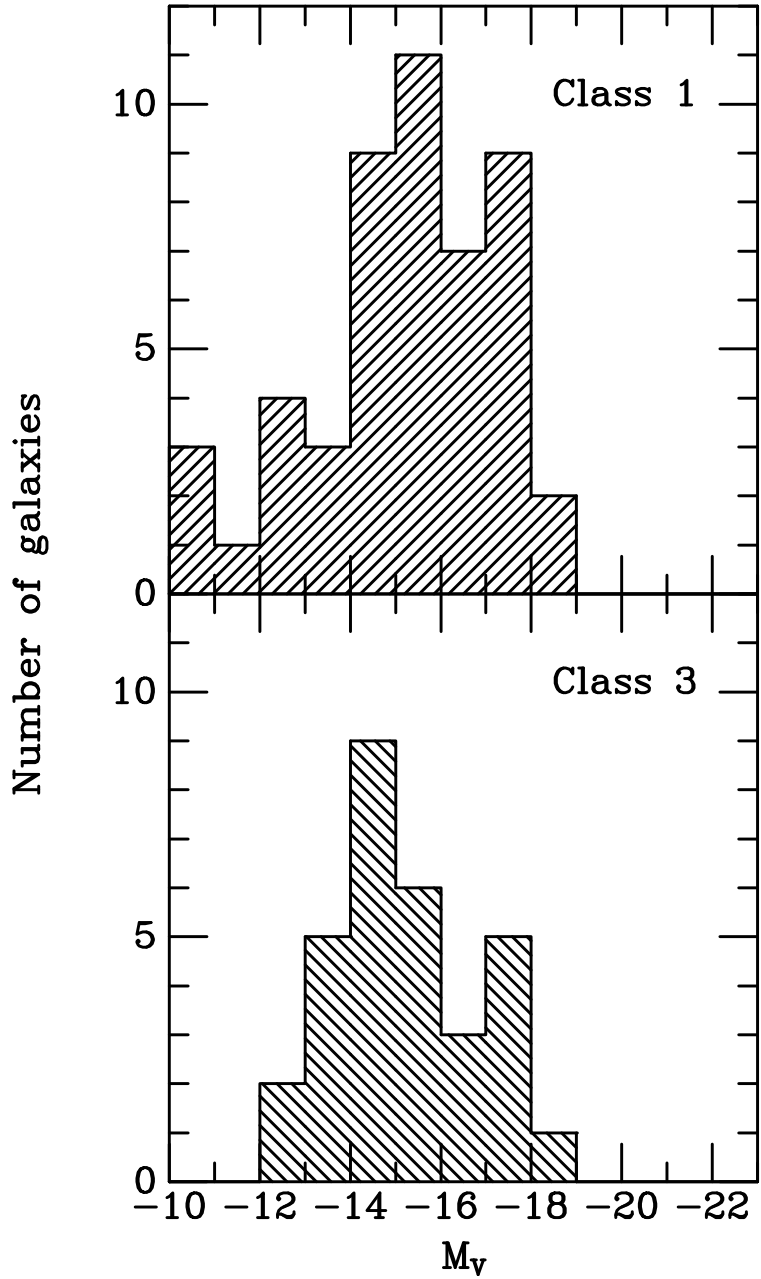}
\caption{Number distribution of L$_{H\alpha}$/L$_V$ classes 1 and 3
as a function of
M$_V$. Class 1 are galaxies in which the ratio L$_{H\alpha}$/L$_V$ is
roughly constant with radius, and class 3 are those in which the
ratio shows large-scale trends with radius.
\label{fig-havclass_mv}}
\end{figure}

\begin{figure}
\epsscale{1.0}
\plotone{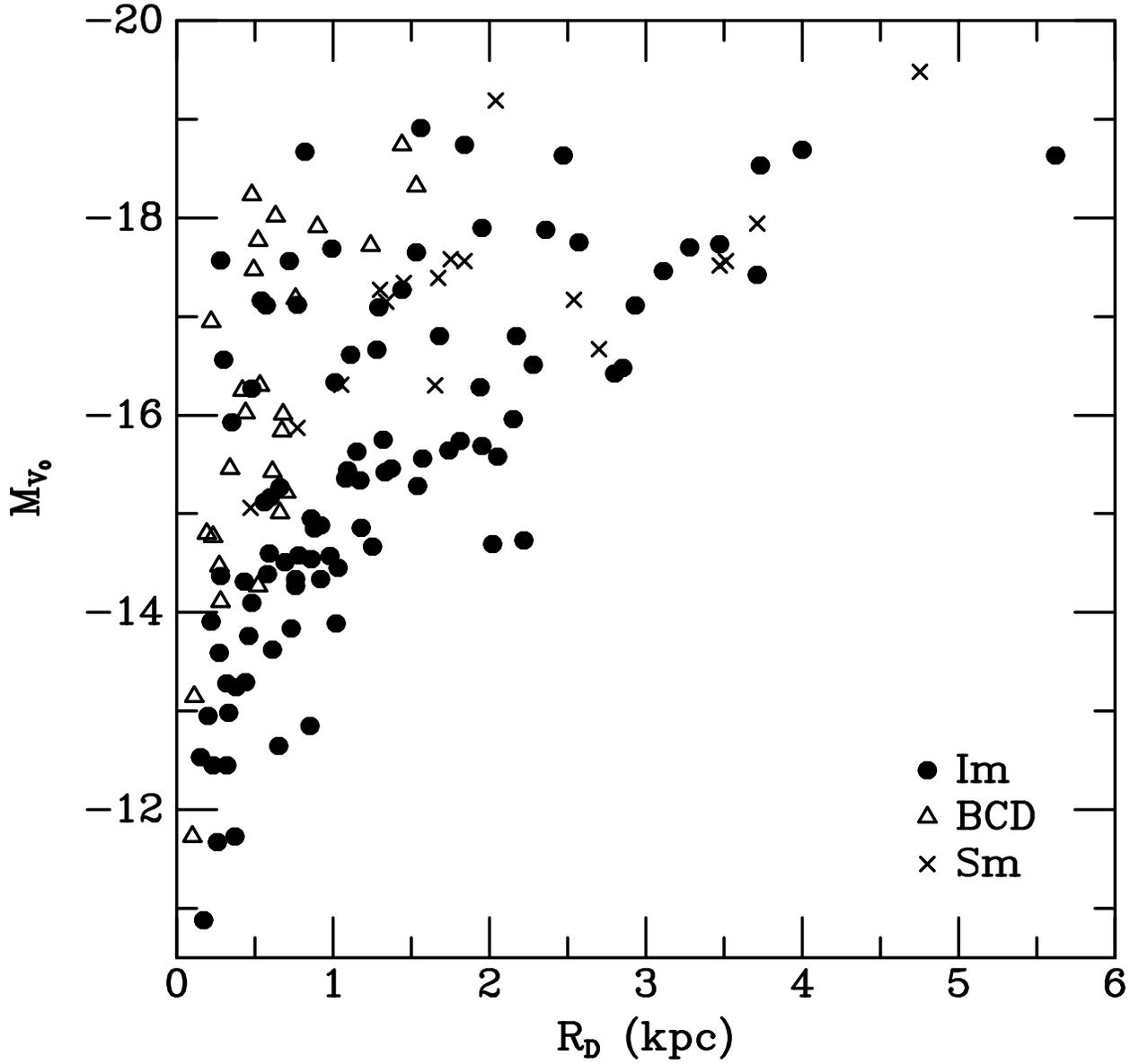}
\caption{M$_V$, corrected for reddening,
as a function of the scale-length R$_D$ for the
three types of galaxies.
\label{fig-mvrd}}
\end{figure}

\begin{figure}
\epsscale{0.8}
\plotone{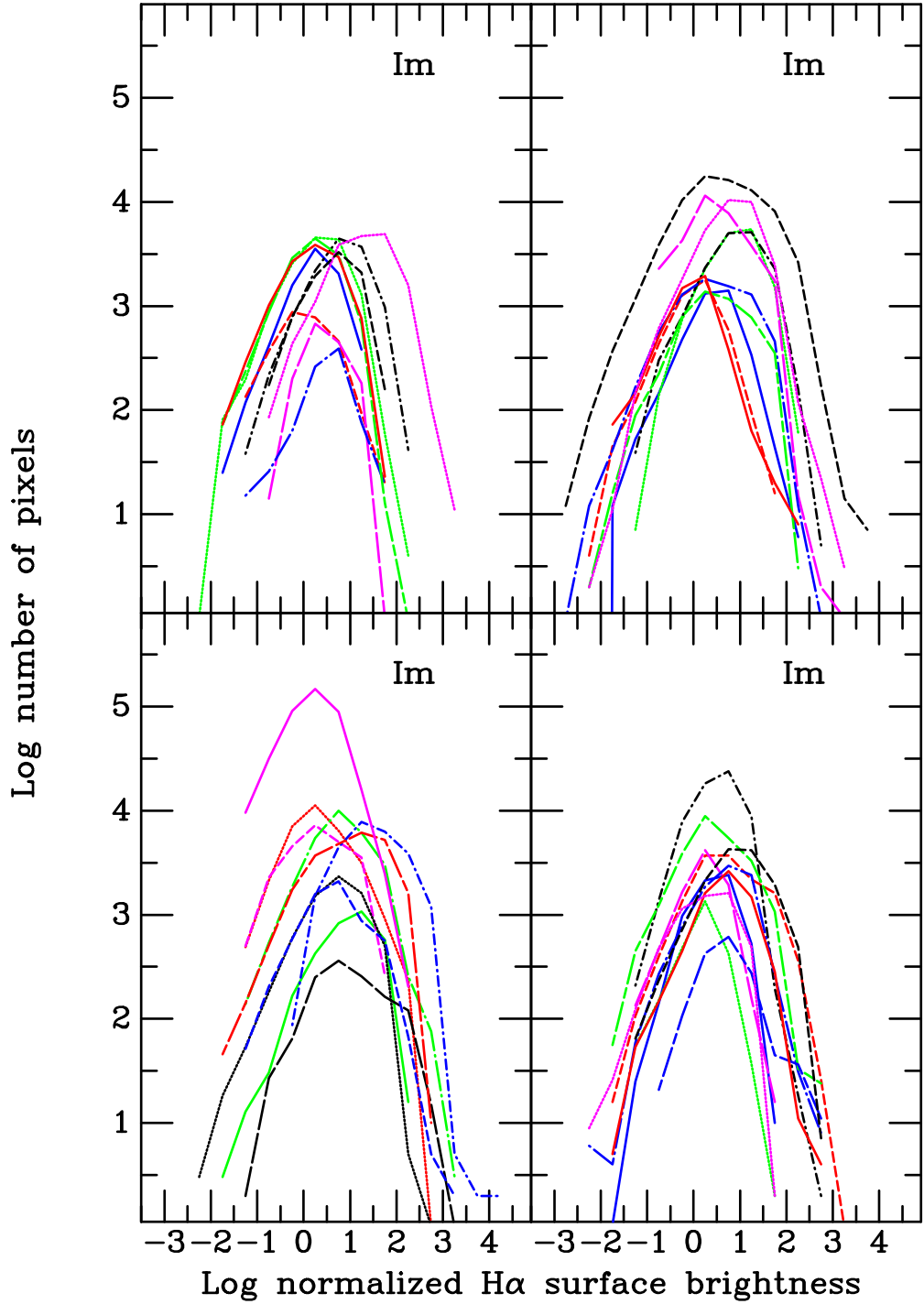}
\caption{Probability distribution functions (pdfs) for the H$\alpha$
images of each galaxy. The logarithm of the number of pixels
is plotted as a function of the logarithm of the normalized
H$\alpha$ surface brightness of the pixel. The H$\alpha$ surface
brightness of each pixel has been normalized to the azimuthally-averaged
surface brightness profile at that radius.
\label{fig-pdf}}
\end{figure}

\clearpage
\epsscale{0.8}
\plotone{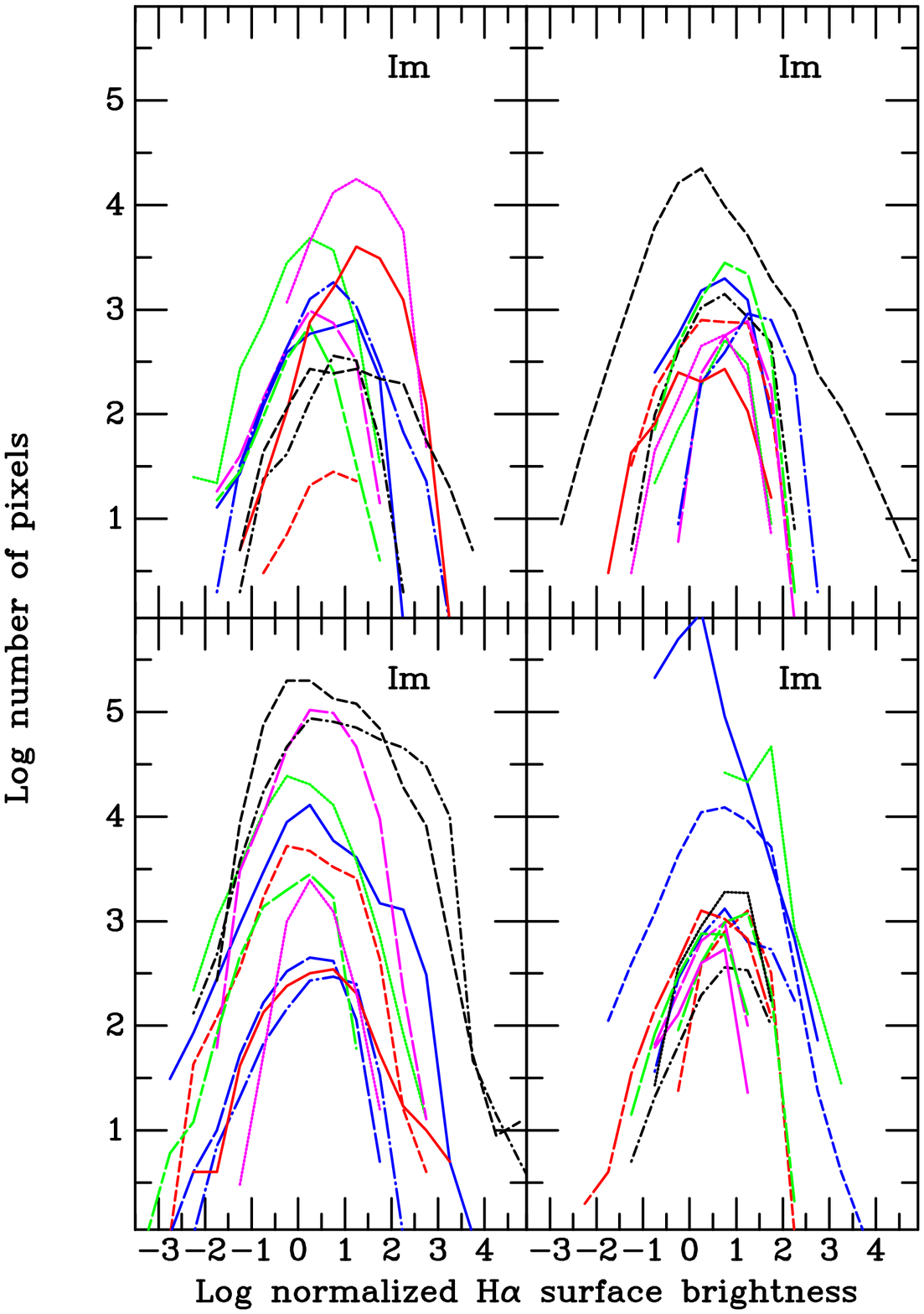}
\clearpage
\epsscale{0.8}
\plotone{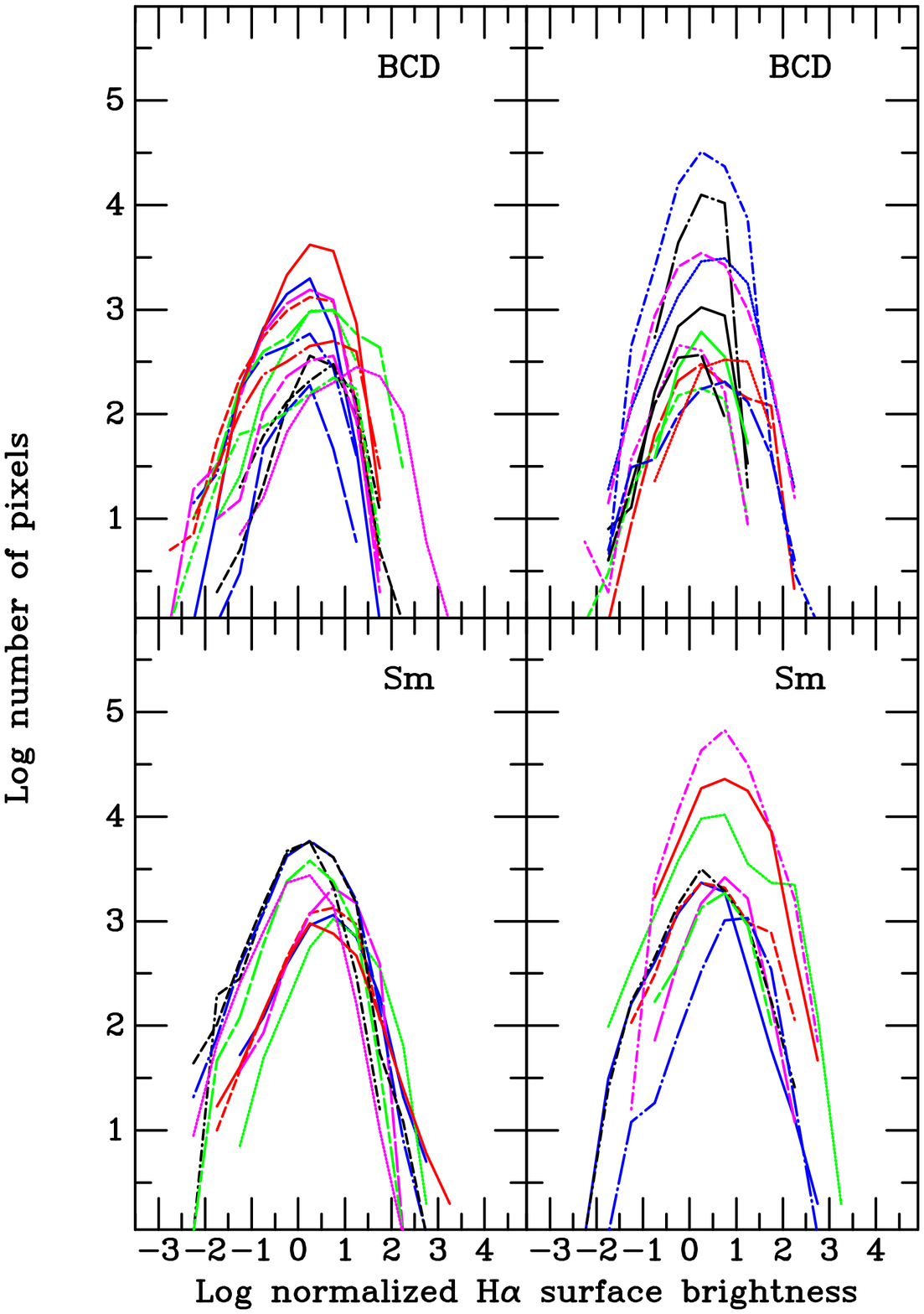}

\end{document}